\documentclass[10pt,preprint]{aastex}

\begin{document}

\shortauthors{Welty et al.}
\shorttitle{Magellanic Clouds Molecules and DIBs}


\title{VLT/UVES Observations of Interstellar Molecules and Diffuse Bands in the Magellanic Clouds\footnotemark}

\footnotetext{Based on observations collected at the European Southern Observatory, Chile, under programs 67.C-0281, 70.D-0164, 72.C-0064, 72.C-0682, and 74.D-0109.}

\author{D. E. Welty\altaffilmark{2},
S. R. Federman\altaffilmark{3},
R. Gredel\altaffilmark{4},
J. A. Thorburn\altaffilmark{5}, and
D. L. Lambert\altaffilmark{6}}

\altaffiltext{2}{University of Chicago, Astronomy and Astrophysics Center, 5640 S. Ellis Ave., Chicago, IL  60637; welty@oddjob.uchicago.edu}

\altaffiltext{3}{University of Toledo, Dept. of Physics \& Astronomy, Toledo, OH  43606; steven.federman@utoledo.edu}

\altaffiltext{4}{Max-Planck-Institut fur Astronomie, Koenigstuhl 17, Heidelberg, D69117, Germany; gredel@caha.es}

\altaffiltext{5}{Yerkes Observatory, University of Chicago, Williams Bay, WI 53191; thorburn@yerkes.uchicago.edu}

\altaffiltext{6}{University of Texas, Dept. of Astronomy, Austin, TX  78712-1083; dll@astro.as.utexas.edu}

\begin{abstract}

We discuss the abundances of interstellar CH, CH$^+$, and CN in the Magellanic Clouds, derived from spectra of 7 SMC and 13 LMC stars obtained (mostly) with the VLT/UVES.
CH and/or CH$^+$ have now been detected toward 3 SMC and 9 LMC stars; CN is detected toward Sk~143 (SMC) and Sk$-67\arcdeg$2 (LMC).  
These data represent nearly all the optical detections of these molecular species in interstellar media beyond the Milky Way.
In the LMC, the CH/H$_2$ ratio is comparable to that found for diffuse Galactic molecular clouds in four sight lines, but is lower by factors of 2.5--4.0 in two others.
In the SMC, the CH/H$_2$ ratio is comparable to the local Galactic value in one sight line, but is lower by factors of 10--15 in two others.  
The abundance of CH in the Magellanic Clouds thus appears to depend on local physical conditions --- and not just on metallicity.
In both the SMC and LMC, the observed relationships between the column density of CH and those of CN, CH$^+$, \ion{Na}{1}, and \ion{K}{1} are generally consistent with the trends observed in our Galaxy.

Using existing data for the rotational populations of H$_2$ in these sight lines, we estimate temperatures, radiation field strengths, and local hydrogen densities for the diffuse molecular gas.
The inferred temperatures range from about 45 to 90 K, the radiation fields range from about 1 to 900 times the typical local Galactic field, and the densities (in most cases) lie between 100 and 600 cm$^{-3}$. 
Densities estimated from the observed $N$(CH), under the assumption that CH is produced via steady-state gas-phase reactions, are considerably higher than those derived from H$_2$.
Much better agreement is found by assuming that the CH is made via the (still undetermined) process(es) responsible for the observed CH$^+$.
A significant fraction of the CH and CH$^+$ in diffuse molecular material in the SMC and LMC may be produced in photon-dominated regions.
The excitation temperature obtained from the populations of the two lowest CN rotational levels toward Sk$-67\arcdeg$2 is quite consistent with the temperature of the cosmic microwave background radiation measured with {\it COBE}.

Toward most of our targets, the UVES spectra also reveal absorption at velocities corresponding to the Magellanic Clouds ISM from several of the strongest of the diffuse interstellar bands (at 5780, 5797, and 6284 \AA).
On average, the three DIBs are weaker by factors of 7 to 9 (LMC) and about 20 (SMC), compared to those typically observed in Galactic sight lines with similar $N$(\ion{H}{1}) --- presumably due to the lower metallicities and stronger radiation fields in the LMC and SMC. 
The three DIBs are also weaker (on average, but with some exceptions) by factors of order 2 to 6, relative to $E(B-V)$, $N$(\ion{Na}{1}), and $N$(\ion{K}{1}) in the Magellanic Clouds.
The detection of several of the so-called ``C$_2$ DIBs'' toward Sk~143 and Sk$-67\arcdeg$2, with strengths similar to those in comparable Galactic sight lines, however, indicates that no single, uniform scaling factor (e.g., one related to metallicity) applies to all DIBs (or for all sight lines) in the Magellanic Clouds. 

\end{abstract}

\keywords{galaxies: ISM --- ISM: abundances --- ISM: lines and bands --- ISM: molecules --- Magellanic Clouds}

\section{Introduction}
\label{sec-intro}

Studies of the interstellar medium (ISM) in the Large and Small Magellanic Clouds (LMC and SMC) explore somewhat different environmental conditions from those typically probed in our own Galactic ISM.
The Magellanic Clouds are characterized by lower overall metallicities ($-$0.3 dex for the LMC, $-$0.6 to $-$0.7 dex for the SMC), lower dust-to-gas ratios, generally stronger ambient radiation fields, and significant differences in UV extinction (especially in the SMC).
These differences are predicted to affect the structure and properties of interstellar clouds in the LMC and SMC (Wolfire et al. 1995; Pak et al. 1998) and, in principle, they may also affect the relative abundances of various constituents of those clouds.
Comparisons of the relative interstellar abundances of different atomic and molecular species --- for the diverse environments in the Milky Way, the LMC, and the SMC --- may thus give insights into the physical and chemical processes shaping the clouds.

Improvements in instrumentation (in several wavelength domains) over the past 10 to 15 years have begun to provide more detailed and accurate information on the properties of diffuse atomic gas in the Magellanic Clouds.
By combining 21 cm data from both the Australia Telescope Compact Array (ATCA) and the Parkes 64m radio telescope, Stanimirovic et al. (1999) and Staveley-Smith et al. (2003) have investigated the structure and kinematics of \ion{H}{1} in the SMC and LMC (respectively) on spatial scales ranging from 15 pc to 10 kpc.
The ATCA interferometric data have revealed complex structure on the smallest scales, with numerous shells and filaments in both galaxies.
Relatively high-resolution UV spectra obtained with {\it HST} for a handful of stars in the Magellanic Clouds have revealed some differences in the patterns of gas-phase elemental abundances (and thus, presumably, in the depletions of those elements into dust grains) and also have yielded estimates for the local thermal pressures and densities in several sight lines (Welty et al. 1997, 2001, and in prep.).
While earlier studies of the optical interstellar absorption lines of \ion{Na}{1} and \ion{Ca}{2} focused largely on the kinematics of the gas in the Magellanic Clouds (e.g., Songaila et al. 1986; Wayte 1990), more recent surveys, incorporating higher resolution spectra of those and other species, have begun to explore the abundances and physical conditions in the clouds.
Relative to hydrogen, \ion{Na}{1} and \ion{K}{1} generally are less abundant than in our Galaxy --- due, presumably, to the lower metallicities and stronger radiation fields (Welty, in prep.); \ion{Ti}{2} generally appears to be less severely depleted in both the SMC and LMC (Welty \& Crowther, in prep.).

Surveys of the $^{12}$CO $J$ = 1--0 emission line at 115 GHz have provided the most complete inventories of relatively dense molecular gas in the Magellanic Clouds.
The initial surveys, undertaken at a spatial resolution of 8.8 arcmin ($\sim$140 pc for the LMC; $\sim$160 pc for the SMC), found CO emission over about 10\% of the LMC (Cohen et al. 1988) and over about 20\% of the main SMC bar (Rubio et al. 1991); a number of distinct molecular clouds (or cloud complexes) were identified in the two galaxies.
Subsequent observations with the SEST telescope, at much higher resolution (45 arcsec, corresponding to 12 pc for the LMC and 14 pc for the SMC), explored the regions around known far-IR sources, star-forming regions, and prior CO detections in much greater detail (Israel et al. 1993, 2003, and references therein).
The higher resolution SEST data indicate that the CO emission in the Magellanic Clouds is (1) weaker than that seen in Galactic molecular clouds of comparable size, and (2) concentrated in fairly small, discrete clumps of relatively dense molecular gas, with little emission from the lower density ``interclump'' gas (e.g., Lequeux et al. 1994).
The more recent large-scale surveys performed with the NANTEN telescope (resolution 2.6 arcmin) have confirmed the clumpy nature of the CO emission, with about 8\% of the LMC pointings and about 10\% of the (more limited) SMC pointings yielding detections (Mizuno et al. 2001; Yamaguchi et al. 2001a).
The molecular clouds seen in CO are well correlated with young stellar clusters and \ion{H}{2} regions, but poorly correlated with older clusters and supernova remnants --- consistent with relatively rapid dispersal of the molecular gas (Fukui et al. 1999; Yamaguchi et al. 2001b).
Observations of a number of other molecular species, near peaks in the CO emission in the N159 region of the LMC and in the LIRS~36 star-forming region in the SMC, have provided some information on molecular abundances, isotopic ratios, and cloud structure (Johansson et al. 1994; Chin et al. 1998).
In the N159 region, the molecular abundances are about a factor of 10 lower (on average) than the corresponding values found for Orion KL and TMC-1.
Similar abundances are found for a number of molecular species in LIRS~36, though the abundances of several nitrogen-bearing species (including CN) appear to be lower by an additional factor of 10.

Unfortunately, very little is known regarding optical/UV absorption from molecules in the Magellanic Clouds, apart from the recent {\it FUSE} H$_2$ survey of Tumlinson et al. (2002). 
Detections of the optical lines of CH and CH$^+$ have been reported only toward SN 1987A (Magain \& Gillet 1987); far-UV lines of CO and/or HD have been measured in several other lines of sight (Bluhm \& de Boer 2001; Andr\'{e} et al. 2004).
Because the optical molecular lines evidently are fairly weak in the Magellanic Clouds and because the stellar targets typically are relatively faint ($V$ $\sim$ 11--14 mag), efficient detection of the lines requires both a large telescope and a spectrograph that can achieve fairly high spectral resolution.
We have therefore used the ESO VLT/UVES to obtain high S/N, moderately high resolution optical spectra of a number of stars in the LMC and SMC, in an attempt to expand the sample of detections of CH, CH$^+$, and CN --- so that we may then compare the relative abundances of those species with the values and trends found in our Galaxy.

The UVES spectra also cover the wavelengths of a number of the enigmatic diffuse interstellar bands --- broader interstellar absorption features first recognized in the spectra of Galactic stars more than 80 years ago (Heger 1922).
While the DIBs are generally thought to be due to large carbon-based molecules, no specific carrier has been securely identified as yet for any of the several hundred features now known (e.g., Jenniskens \& D\'{e}sert 1994; Herbig 1995; Tuairisg et al. 2000; Snow 2001a; York et al., in prep.; Snow \& McCall, in prep.; and references therein).
In view of the different environmental factors present in the ISM of the Magellanic Clouds, observations of the DIBs there may yield useful constraints on the DIB carriers --- but (again) the available data are scarce (Snow 2001b).
The recent detections of several of the stronger DIBs toward a handful of Magellanic Clouds stars seem to indicate that the DIBs are generally weaker in the LMC and SMC (Ehrenfreund et al. 2002; Sollerman et al. 2005; Cox et al. 2006), but additional data are needed to determine the detailed behavior of the DIBs under different physical conditions.

In this paper, we discuss new observations of CH, CH$^+$, CN, and DIBs in the ISM of the Magellanic Clouds.
In \S~\ref{sec-data}, we describe the sets of sight lines observed, the spectra obtained, and the various steps taken to analyze those spectra.
In \S~\ref{sec-res}, we compare the column densities of various atomic and molecular features in the Magellanic Clouds with the corresponding relationships seen in our Galaxy and we characterize the physical conditions in the observed diffuse molecular gas.
In \S~\ref{sec-disc}, we compare our results to the predictions of theoretical models for clouds in environments like those found in the Magellanic Clouds and we discuss possible implications for the behavior and carriers of the DIBs.
In \S~\ref{sec-summ}, we summarize our conclusions.

\section{Data}
\label{sec-data}

\subsection{Stellar Sample}
\label{sec-stel}

Table~\ref{tab:los} lists the 20 Magellanic Clouds sight lines --- seven SMC and thirteen LMC --- discussed in this mini-survey.
(In this and subsequent tables, the SMC sight lines are listed first, then the LMC sight lines.)
As most of the sight lines were selected for having strong observed absorption from H$_2$ [with $N$(H$_2$) $>$ 10$^{19}$ cm$^{-2}$; Tumlinson et al. 2002; Cartledge et al. 2005] and/or \ion{Na}{1} [with $N$(\ion{Na}{1}) $>$ 3 $\times$ 10$^{12}$ cm$^{-2}$; Welty, in prep.], it is certainly not an unbiased sample.
The sight lines do, however, sample a variety of regions and environments in the two galaxies --- as indicated by location (relative to \ion{H}{2} regions and molecular clouds), by studies of UV extinction (e.g., Gordon et al. 2003), and by the local physical conditions estimated below (\S~\ref{sec-chem}).
In the SMC:  Sk~13, Sk~18, Sk~40, and AV~80 lie in the southwest part of the main ``bar''; Sk~143, AV~476 and Sk~155 are in the ``wing'' region.
In the LMC:  LH~10-3061, Sk$-67\arcdeg$2, and Sk$-67\arcdeg$5 lie in the northwest; BI~237 is in the northeast; Sk$-68\arcdeg$52 and Sk$-68\arcdeg$73 are near the center; Sk$-68\arcdeg$135, Sk$-69\arcdeg$202 (SN 1987A), BI~253, Melnick 42, and Sk$-69\arcdeg$246 are near 30 Dor (at the southeast end of the main bar); Sk$-69\arcdeg$191 is southwest of 30 Dor; and Sk$-70\arcdeg$115 is in the LMC2 region (southeast of 30 Dor).
For each sight line, the table gives equatorial coordinates (J2000), spectral type, $V$, $E(B-V)$ (total and MC), total exposure time, and the observing run(s) in which the observations were obtained.
Most of the spectral types, photoelectric $V$ magnitudes, and $B-V$ colors were taken from similar compilations in the recent {\it FUSE} surveys by Danforth et al. (2002) and Tumlinson et al. (2002) and in a forthcoming survey of atomic lines (Welty, in prep.), where the original references are given; sources for the rest are given in footnote (a) to the table.
The targets have $V$ magnitudes between 10.9 and 13.9 (except for SN 1987A), with those in the LMC generally somewhat brighter than those in the SMC.
The total $E(B-V)$ color excesses were obtained using the intrinsic colors for LMC stars adopted by Fitzpatrick \& Garmany (1990).
For the Galactic contributions to those total $E(B-V)$, a constant 0.04 mag was adopted for all the SMC sight lines (Schlegel, Finkbeiner, \& Davis 1998); Galactic values ranging from 0.02 to 0.06 mag were estimated from Figure 13 of Staveley-Smith et al. (2003) for the LMC sight lines.
The second $E(B-V)$ value listed for each sight line represents the portion due to dust in the SMC or LMC (total minus Galactic); the values range from 0.07 to 0.34 mag for the SMC sight lines and from 0.08 to 0.51 for the LMC sight lines.\footnotemark
\footnotetext{Our values for $E(B-V)_{\rm MC}$ toward Sk$-67\arcdeg$2, Sk$-67\arcdeg$5, and Sk$-68\arcdeg$135 are somewhat higher than those determined by Cox et al. (2006), due to differences in the adopted spectral types, intrinsic colors, and/or Galactic contributions.
While the $E(B-V)$ = 0.51 toward Sk$-68\arcdeg$73 (type Ofpe/WN9) is somewhat uncertain (and may be too high), it is clear, from observations of various atomic and molecular species, that there is a substantial amount of interstellar material along that sight line.}

\subsection{Observations and Data Processing}
\label{sec-obs}

All the spectroscopic data discussed in this paper were obtained with the ESO/VLT UT2 telescope and UVES spectrograph (Dekker et al. 2000)\footnotemark, under several different observing programs executed in 2001--2004.
\footnotetext{See also http://www.eso.org/instruments/uves.}
Spectra of five SMC and seven LMC stars were obtained during three nights in 2003 November (run V03) for program 72.C-0682, which was aimed at detecting weak interstellar atomic and molecular absorption lines toward stars in the Magellanic Clouds.  
The standard dichroic \#1 390/564 setting and a slit width corresponding to 0.7 arcsec were used to obtain nearly complete coverage of the wavelength range from 3260 to 6680 \AA\ (on three CCDs) at a resolution of 4.5--4.9 km~s$^{-1}$.
This setup included lines from \ion{Na}{1} (U and D doublets at 3302 and 5889/5895~\AA), \ion{Ca}{1} (4226~\AA), \ion{Ca}{2} (3933/3968~\AA), and \ion{Ti}{2} (3383~\AA); the strongest lines from CH (4300~\AA), CH$^+$ (4232~\AA), and CN (3874~\AA); and a number of the diffuse interstellar bands.
Multiple exposures, generally of length 20 to 45 minutes, were obtained for most targets over the three nights, with the camera tilted by $\pm$50 units on the second and third nights to ensure that the spectra fell on slightly different parts of the three CCDs.
The run was characterized by very good observing conditions:  low humidity, good transparency, and seeing generally better than 0.8 arcsec.

Standard routines within IRAF were used to remove the bias from the CCD frames and to divide sections of the 2-D images containing the spectral order(s) of interest by a normalized flat-field derived from quartz lamp exposures.
The 1-D spectra then were extracted from the flat-fielded image segments via the apextract routines, using variance weighting (with the appropriate values for read noise and gain for each detector).
Wavelength calibration was accomplished via Th-Ar lamp exposures, which were obtained at the beginning and end of each night, using the thorium rest wavelengths tabulated by Palmer \& Engelman (1983).
The widths of unblended thorium lines in those lamp exposures indicate that the spectral resolution is about 4.5$\pm$0.1 km~s$^{-1}$ for the blue spectra (3260--4515~\AA) and is about 4.9 km~s$^{-1}$ near the \ion{Na}{1} D lines.
For the first night's data, obtained using the standard camera tilt, our IRAF-based reductions yielded extracted spectra very similar to those produced by the UVES pipeline reduction package.
Multiple spectra for a given target were sinc interpolated to a common, approximately optimally sampled heliocentric wavelength grid, then summed.
The narrow telluric absorption lines present near the \ion{Na}{1} D lines and near the 6284~\AA\ DIB were removed by dividing by appropriately scaled spectra of the lightly reddened Galactic star $\psi^2$~Aqr.
The summed spectra were then normalized via Legendre polynomial fits to the continuum regions surrounding the interstellar (and stellar) absorption lines.
Determination of the appropriate continuum regions was fairly unambiguous for the narrower atomic and molecular features.
For the DIB at 6284~\AA\ (which exhibits weak, broad wings), higher S/N spectra of reddened Galactic stars (Thorburn et al. 2003), shifted to the velocities of the strongest Galactic and Magellanic Clouds \ion{Na}{1} absorption, provided guidance on setting the continuum regions.
The S/N ratios in the normalized UVES spectra, obtained from the continuum fits, are typically $\sim$ 200--350 per half resolution element near the various molecular lines.

Spectra of four LMC stars were obtained via service observing with the VLT/UVES during several nights in 2004 January--March (run V04) under program 72.C-0064, which also was aimed at detecting molecular absorption in the ISM of the Magellanic Clouds.
The standard dichroic \#2 437/860 setting employed for these observations covered the wavelength ranges from 3740 to 4990~\AA\ and from 6670 to 10610~\AA, again at a resolution of about 5 km~s$^{-1}$.
This setup included the lines from \ion{Ca}{1}, \ion{Ca}{2}, \ion{K}{1} (7668/7698~\AA), CN, CH, and CH$^+$, but few of the stronger diffuse bands.
The spectra were processed through the UVES pipeline reduction package (Ballester et al. 2000), which yielded a single combined, extracted spectrum for each CCD detector segment.
For the three stars also observed in run V03, the shorter exposure times in run V04 yielded somewhat lower S/N ratios for the corresponding regions in the extracted spectra. 

UVES spectra of 9 SMC and 22 LMC OB stars, originally obtained for studies of stellar abundances and stellar wind properties [programs 70.D-0164 (run C02) and 74.D-0109 (run C04); e.g., Crowther et al. 2002; Evans et al. 2004], were graciously made available for this study of interstellar molecular abundances by P. A. Crowther.
While several different instrumental setups were used to acquire those spectra, the observed spectral ranges included the lines of \ion{Na}{1}, \ion{Ca}{1}, \ion{Ca}{2}, CN, CH, CH$^+$, and the stronger DIBs in all cases.
The spectra were processed through the UVES pipeline reduction package.
The stars observed in those programs are typically less reddened and have lower \ion{Na}{1} column densities than the ones in our sample, however, and the spectra generally are characterized by lower S/N ratios.
The typically very weak lines from CN and CH thus were not detected in any of the sight lines, and CH$^+$ was detected only toward the LMC star LH~10-3061.  
Several of the stronger diffuse bands were detected toward one of the SMC stars and five of the LMC stars (including Sk$-70\arcdeg$115, observed at higher S/N in run V03).

UVES spectra of the SMC star Sk~143, obtained under program 67.C-0281 for studies of the diffuse interstellar bands (run E01; Ehrenfreund et al. 2002) and with the same instrumental setup used for our V03 data, were obtained from the UVES archive and processed through our IRAF-based procedure.
This sight line has a very high fraction of hydrogen in molecular form [$f$(H$_2$) = 2$N$(H$_2$)/($N$(\ion{H}{1})+2$N$(H$_2$)) = 0.52; Cartledge et al. 2005] and a higher than average (for the SMC) dust-to-gas ratio, and is the only known SMC sight line with a Milky Way-like extinction curve (exhibiting the 2175 \AA\ bump; Gordon et al. 2003).

This paper will focus on the molecular absorption lines and some of the diffuse interstellar bands found in the UVES spectra of 20 Magellanic Clouds targets (the set of 13 stars observed in runs V03 and V04, plus the 5 additional stars observed in runs C02 and C04 with detected diffuse bands, plus Sk~143 and SN 1987A).
Several other papers will discuss the absorption from \ion{Na}{1}, \ion{K}{1}, \ion{Ca}{1}, \ion{Ca}{2}, and \ion{Ti}{2} observed in a larger set of SMC and LMC sight lines (Welty, in prep.; Welty \& Crowther, in prep.).

\subsection{Spectra and Equivalent Widths}
\label{sec-ew}

\subsubsection{Molecular Lines}
\label{sec:ewmol}

\begin{figure}
\epsscale{1.0}
\plottwo{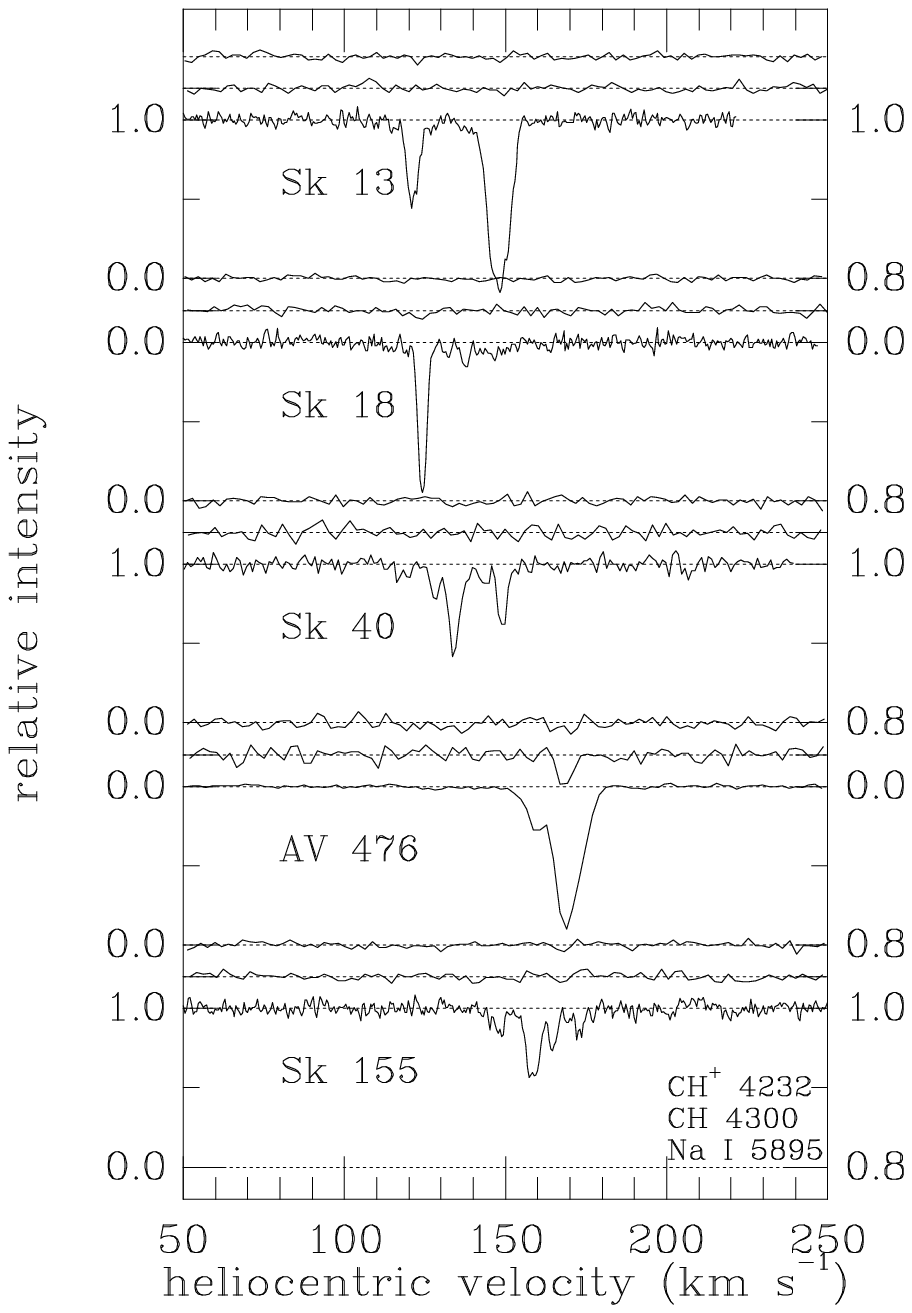}{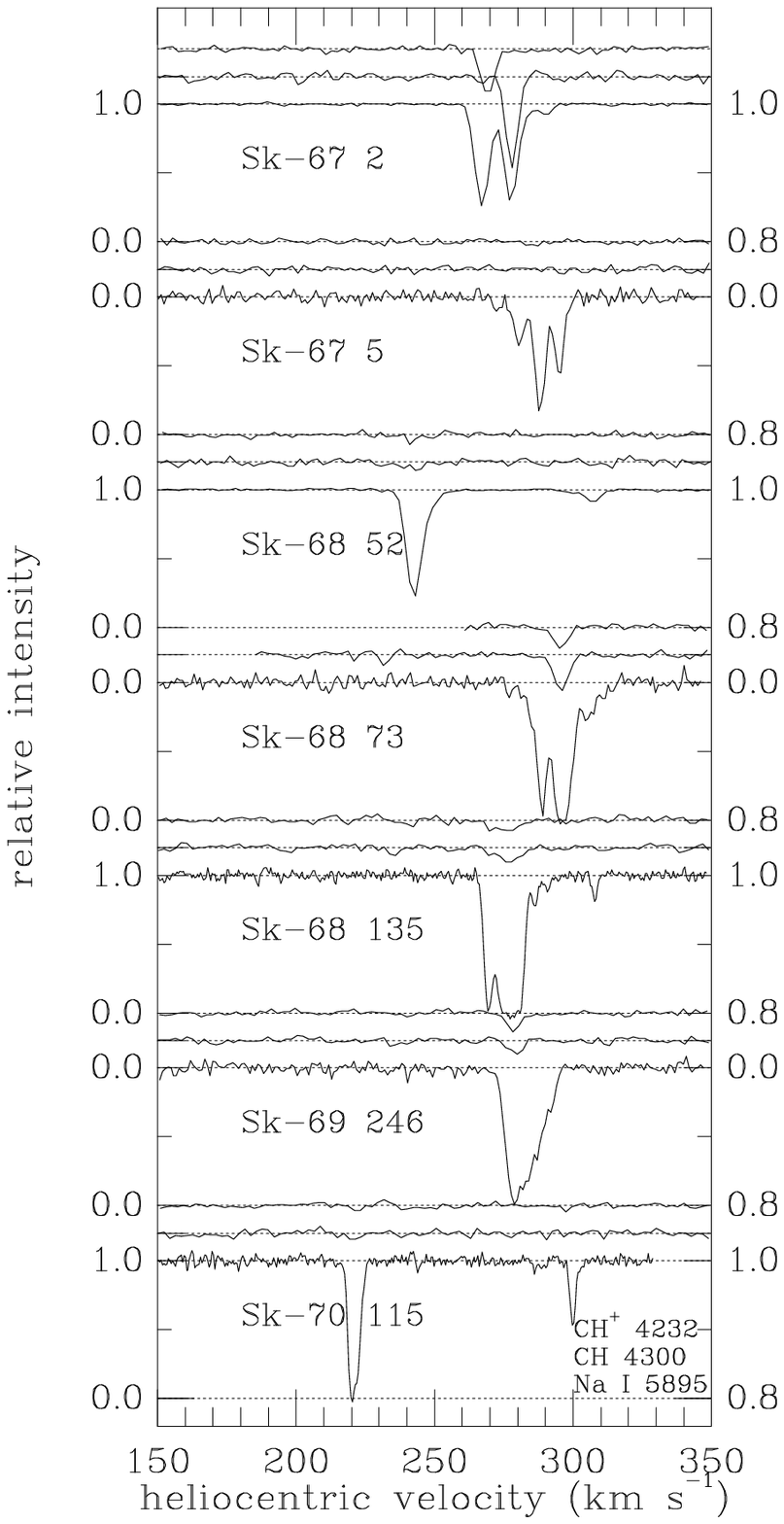}
\caption{Absorption lines of CH$^+$ ($\lambda$4232), CH ($\lambda$4300), and Na~I ($\lambda$5895) toward SMC ({\it left}) and LMC ({\it right}) stars observed under program V03.
Only the Magellanic Clouds absorption is shown.
The vertical scale is expanded by a factor of 5 for the molecular lines ({\it right-hand scales}).
The molecular lines were observed with UVES at resolutions of 4.5--4.9 km~s$^{-1}$; most of the Na~I lines (except those toward AV 476, Sk$-67\arcdeg$2, and Sk$-68\arcdeg$52) were observed with the ESO 3.6m CES, at resolutions of 1.2--2.0 km~s$^{-1}$.}
\label{fig:specmol}
\end{figure}

\begin{figure}
\plottwo{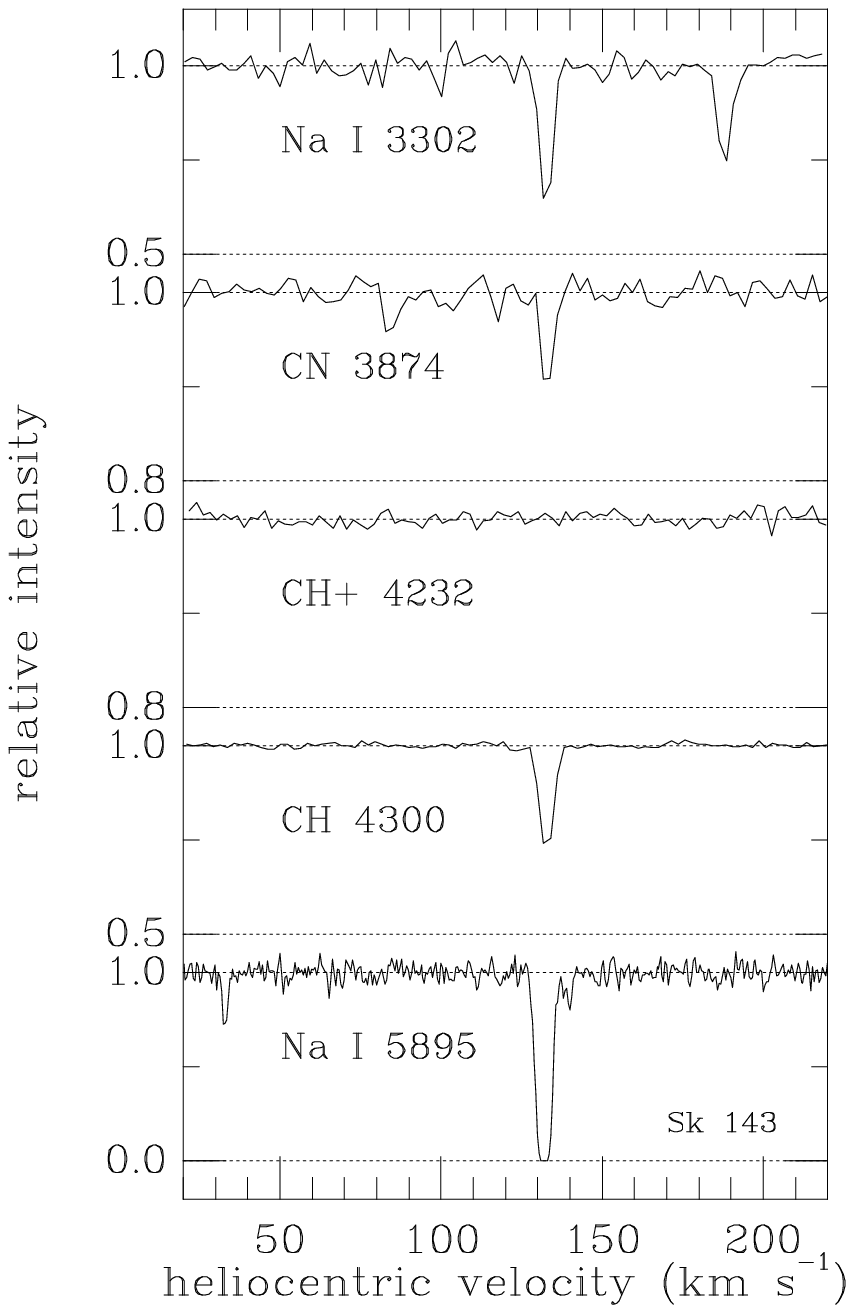}{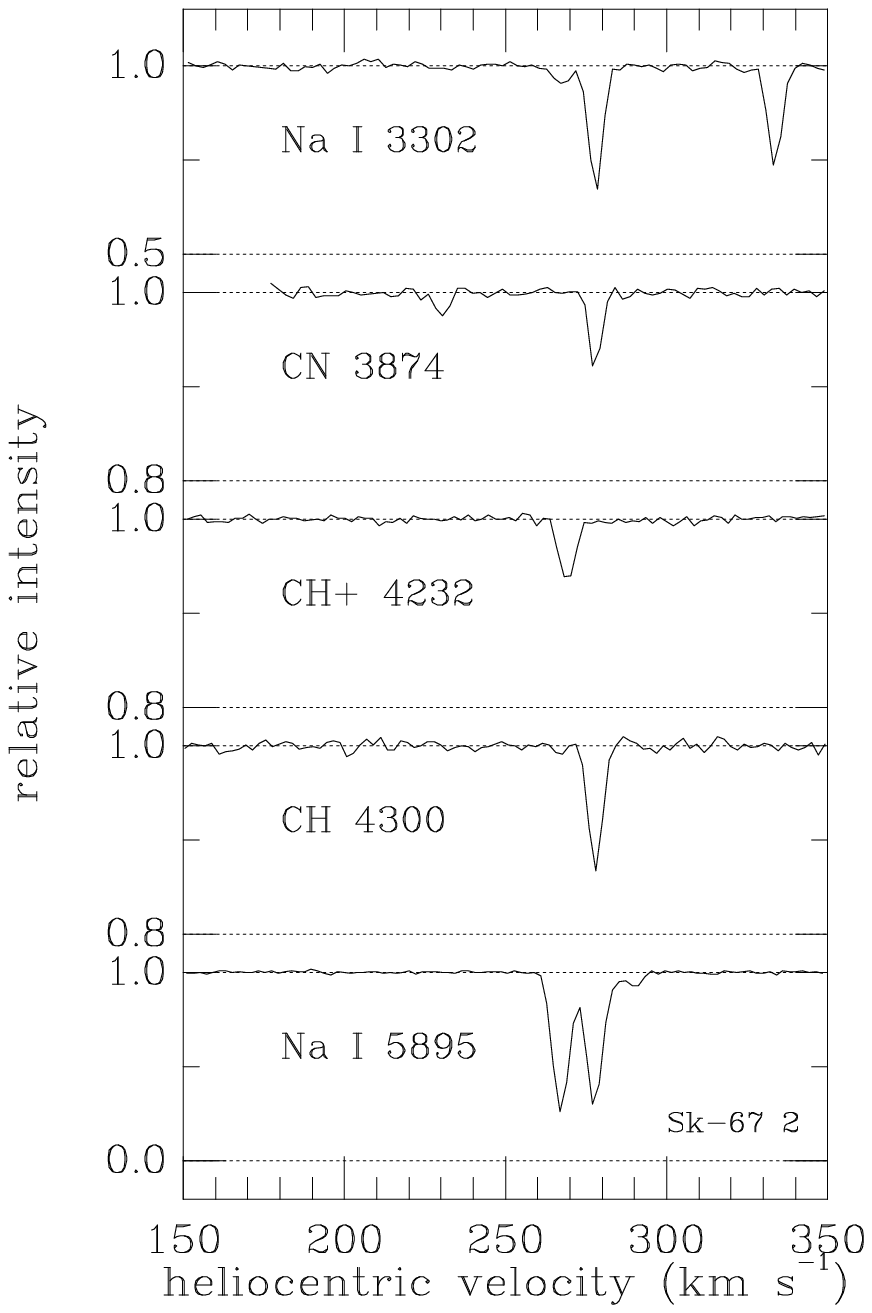}
\caption{Absorption lines of CN, CH$^+$, CH, and Na~I toward the SMC star Sk~143 and the LMC star Sk$-67\arcdeg$2.
Only the Magellanic Clouds absorption is shown.
The vertical scale is expanded by a factor of 2 or 5 for the molecular lines.
The weaker member of the Na~I 3302 doublet appears at about +55 km~s$^{-1}$, relative to the stronger line; the R(1) line of CN appears at about $-$47 km~s$^{-1}$, relative to the R(0) line.
For Sk$-67\arcdeg$2, note the velocity differences between the stronger component in Na~I, CN, and CH (278 km~s$^{-1}$), the weaker component in Na~I and CH (267 km~s$^{-1}$), and the stronger component in CH$^+$ (268 km~s$^{-1}$).}
\label{fig:sk67d2}
\end{figure}

\begin{figure}
\epsscale{0.5}
\plotone{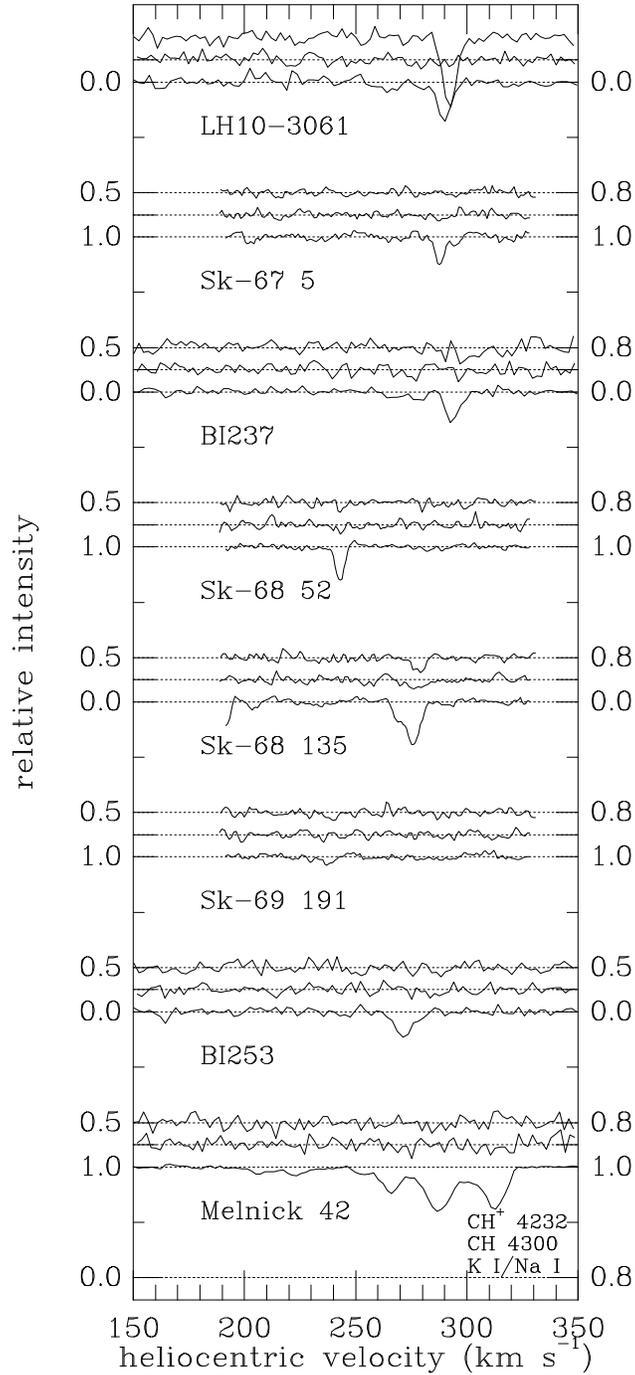}
\caption{Absorption lines of CH$^+$ ($\lambda$4232), CH ($\lambda$4300), and either K~I ($\lambda$7698) or Na~I ($\lambda$5895) (Melnick 42 only) toward LMC stars observed with UVES under programs V04, C02, and C04.
Only the Magellanic Clouds absorption is shown.
Note the strong CH$^+$ absorption toward the LMC star LH~10-3061.
The vertical scale is expanded by a factor of 2 for K~I ({\it left-hand scales}) and by factors of 2 to 5 for the molecular lines ({\it right-hand scales}).}
\label{fig:v04}
\end{figure}

The normalized CH$^+$ $\lambda$4232 and CH $\lambda$4300 profiles from run V03 are displayed together with the corresponding \ion{Na}{1} $\lambda$5895 profiles in Figure~\ref{fig:specmol}; the CH$^+$, CH, CN $\lambda$3874, and \ion{Na}{1} profiles observed toward the SMC star Sk~143 (run E01) and the LMC star Sk$-67\arcdeg$2 (run V03) are shown in Figure~\ref{fig:sk67d2}.
The profiles of CH$^+$, CH, and \ion{K}{1} $\lambda$7698 or \ion{Na}{1} $\lambda$5895 for eight of the LMC stars observed in runs V04, C02, and C04 are shown in Figure~\ref{fig:v04}.
Only the Magellanic Clouds absorption --- from 100 to 180 km~s$^{-1}$ for the SMC sight lines and from 210 to 320 km~s$^{-1}$ for the LMC sight lines --- is included in these figures; note the expanded vertical scales for the molecular lines.
In most cases, the molecular lines lie at the same velocity as the strongest \ion{Na}{1} (and \ion{K}{1}) absorption; the difference in velocity between the strongest component of CH$^+$ (268 km~s$^{-1}$) and those of \ion{Na}{1}, CN, and CH (278 km~s$^{-1}$) toward Sk$-67\arcdeg$2 is a notable exception, however.
The higher resolution (FWHM $\sim$ 1.2--2.0 km~s$^{-1}$) \ion{Na}{1} spectra [observed with the ESO 3.6m telescope and coud\'{e} echelle spectrograph (Welty, in prep.)] shown in Figs.~\ref{fig:specmol} and \ref{fig:sk67d2} reveal both more complex structure than is discernible in the UVES spectra and additional lower column density components --- suggesting that some of the molecular absorption features observed with UVES may be unresolved blends of several narrower components.

Equivalent widths for the Magellanic Clouds molecular absorption features measured from the normalized spectra are given in Table~\ref{tab:ewmol}, together with the rest wavelengths and oscillator strengths of the transitions.
For detected lines, the listed 1$\sigma$ uncertainties include contributions from both photon noise and continuum fitting (Jenkins et al. 1973; Sembach \& Savage 1992); the 3-$\sigma$ detection limits for weak, unresolved absorption lines are typically $\sim$ 1.2--1.8~m\AA\ for spectra obtained in run V03 and $\sim$ 3--5 m\AA\ for the lower S/N spectra obtained in runs C02 and C04.
Even at the fairly sensitive limits achieved in runs V03, V04, and E01 --- and for sight lines generally selected for having relatively strong absorption from H$_2$ and/or \ion{Na}{1} --- CN is detected only toward the SMC star Sk~143 and toward the LMC star Sk$-67\arcdeg$2, at 6.3$\pm$0.8 m\AA\ and 4.9$\pm$0.4 m\AA, respectively, for the B-X(0,0) R(0) line at 3874.6~\AA\ (Fig.~\ref{fig:sk67d2}). 
CH and/or CH$^+$ are detected with greater than 2-$\sigma$ significance toward most of the LMC stars listed in Table~\ref{tab:ewmol}, but only toward two stars (Sk~143, AV~476) in the SMC.  
Typical (detected) equivalent widths for the CH A-X(0,0) line at 4300.3~\AA\ and for the CH$^+$ A-X(0,0) line at 4232.5~\AA\ are both $\sim$ 1--5~m\AA.
Agreement in velocity with the stronger absorption from \ion{Na}{1} and/or \ion{K}{1} (\S~\ref{sec-coldens}) suggests, however, that the weaker possible molecular features listed in Table~\ref{tab:ewmol} are likely to be real as well.
The relative strengths of the possible molecular features at 3878.8, 3886.4, and 3890.2~\AA\ (CH) and at 3957.7 and 3745.3~\AA\ (CH$^+$) are consistent with those of the stronger CH and CH$^+$ lines, given the respective $f$-values.
While the red spectra obtained in run V04 also cover several of the C$_2$ A-X (Phillips) bands, the equivalent width limits [e.g., $W(3\sigma)$ $<$ 5 m\AA\ for individual rotational lines in the (3-0) band toward Sk$-68\arcdeg$135] do not yield very interesting limits on the total C$_2$ column densities toward the four stars observed in that run.
The equivalent widths measured for various atomic and molecular lines toward several Galactic stars (23 Ori, HD~62542, HD~73882, $\psi^2$~Aqr) also observed during run V03 show very good agreement with previously reported values (Cardelli et al. 1990; Gredel, van Dishoeck, \& Black 1991, 1993; Albert et al. 1993; Welty et al. 1999b).

To our knowledge, these are the first reported optical detections of CH, CH$^+$, and CN in the ISM of the Magellanic Clouds, apart from the detections of CH $\lambda$4300 (0.9$\pm$0.2~m\AA) and CH$^+$ $\lambda$4232 (0.4$\pm$0.2~m\AA) toward SN 1987A by Magain \& Gillet (1987).
The only other extragalactic detections of CH ($\lambda$4300) and CH$^+$ ($\lambda$3957, $\lambda$4232) were toward SN~1986G in NGC~5128 (D'Odorico et al. 1989).
While extragalactic CN has been detected in both emission and absorption at millimeter wavelengths (e.g., Henkel, Mauersberger, \& Schilke 1988; Eckart et al. 1990), the present detections of CN in the Magellanic Clouds appear to be the first, in optical absorption, beyond our Galaxy.

\subsubsection{Diffuse Bands}
\label{sec-ewdibs}

\begin{figure}
\epsscale{1.0}
\plottwo{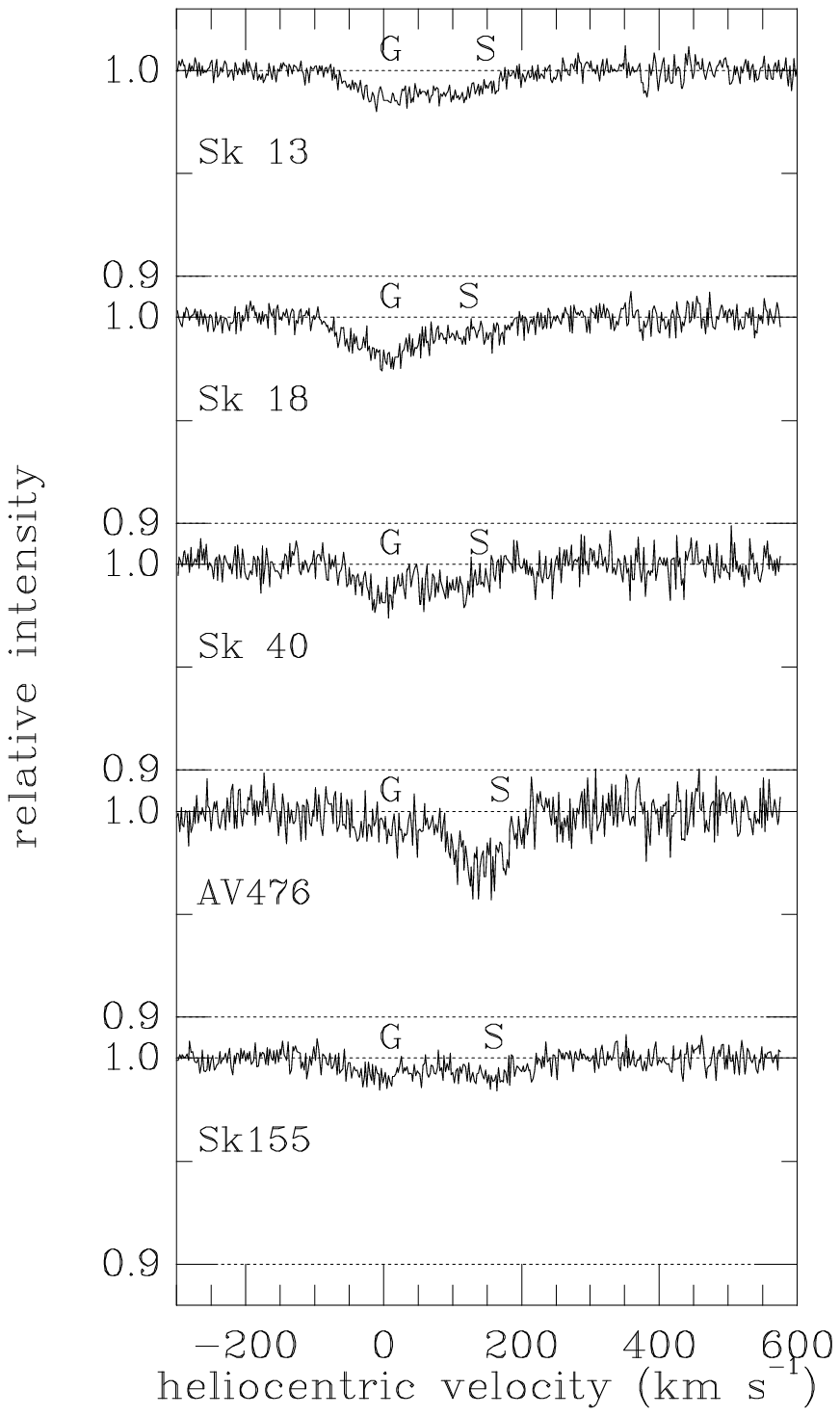}{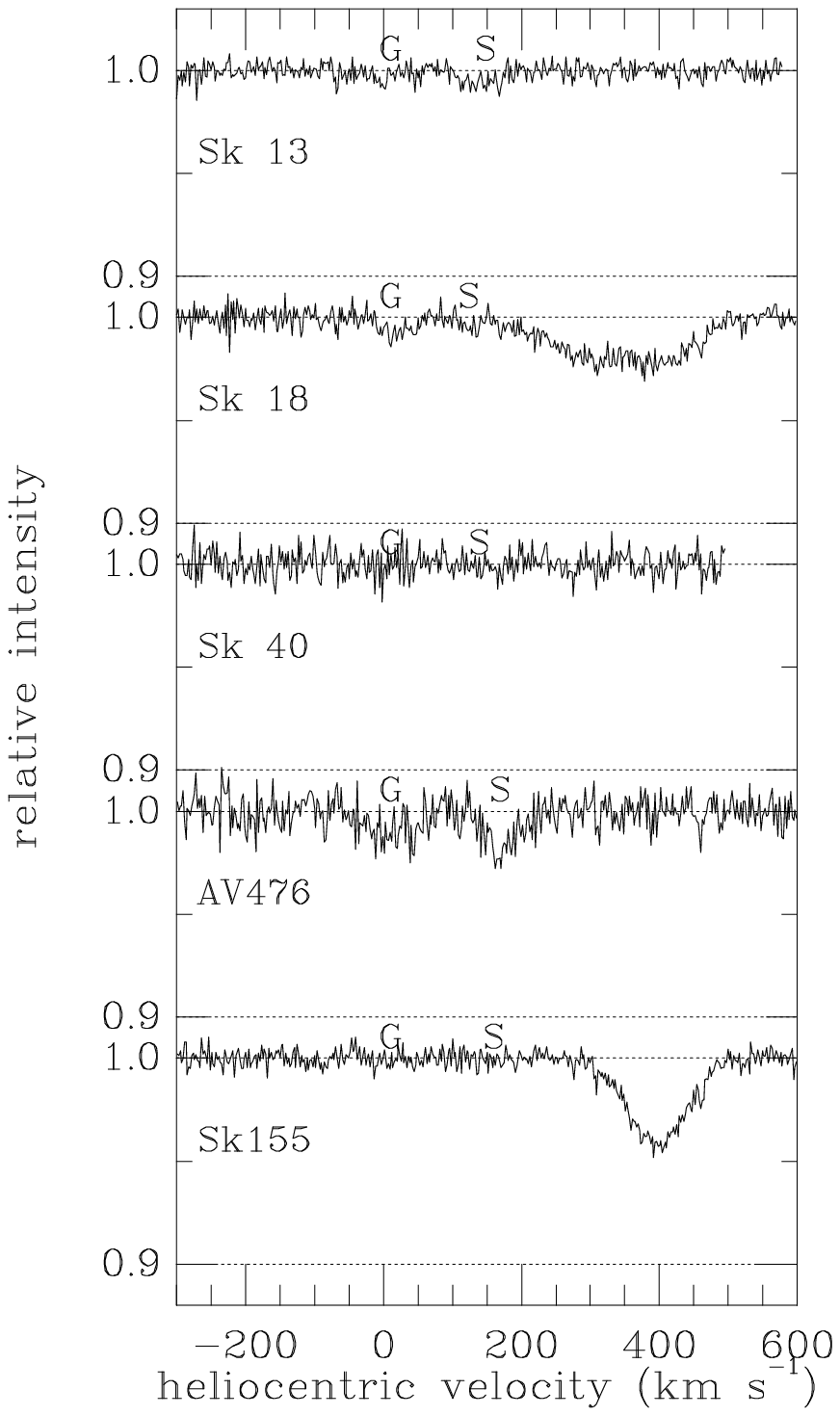}
\caption{UVES spectra of DIBs at 5780 ({\it left}) and 5797 \AA\ ({\it right}) toward SMC stars observed under program V03.
The vertical scale is expanded by a factor of 10.
Letters G and S mark velocities of strongest Galactic and SMC Na~I absorption; unmarked features red-ward of the 5797 \AA\ DIB are stellar lines.}
\label{fig:d57smc}
\end{figure}

\begin{figure}
\epsscale{1.0}
\plottwo{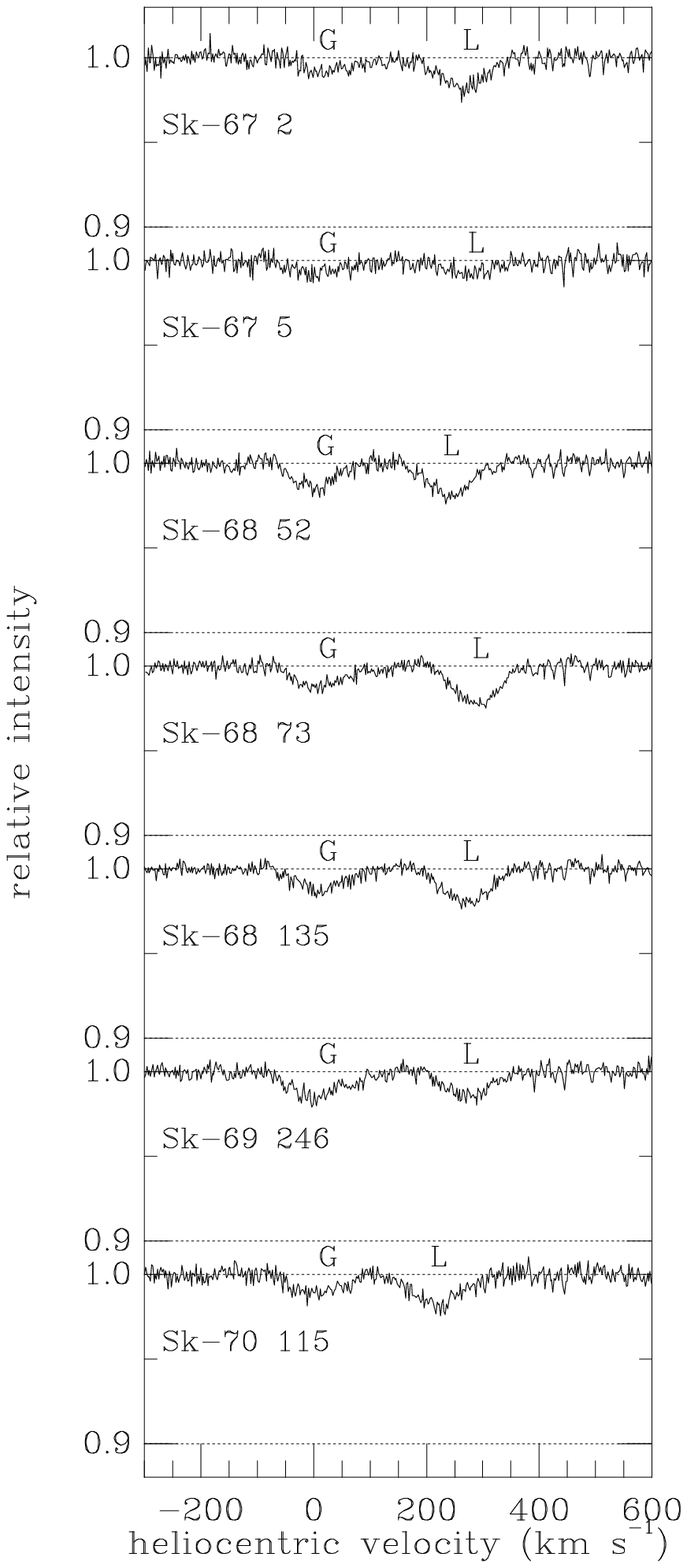}{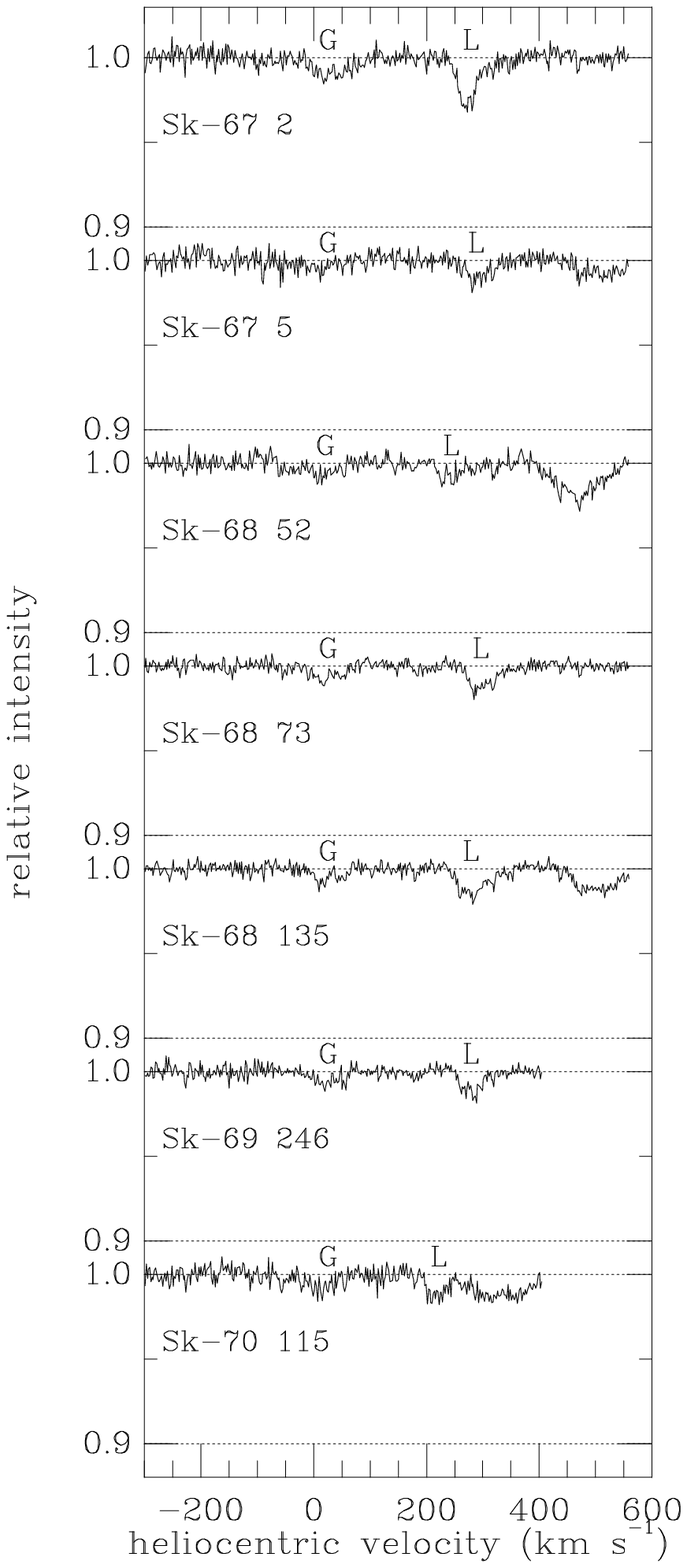}
\caption{UVES spectra of DIBs at 5780 ({\it left}) and 5797 \AA\ ({\it right}) toward LMC stars observed under program V03.
The vertical scale is expanded by a factor of 10.
Letters G and L mark velocities of strongest Galactic and LMC Na~I absorption; unmarked features red-ward of the 5797 \AA\ DIB are stellar lines.}
\label{fig:d57lmc}
\end{figure}

\begin{figure}
\epsscale{1.0}
\plottwo{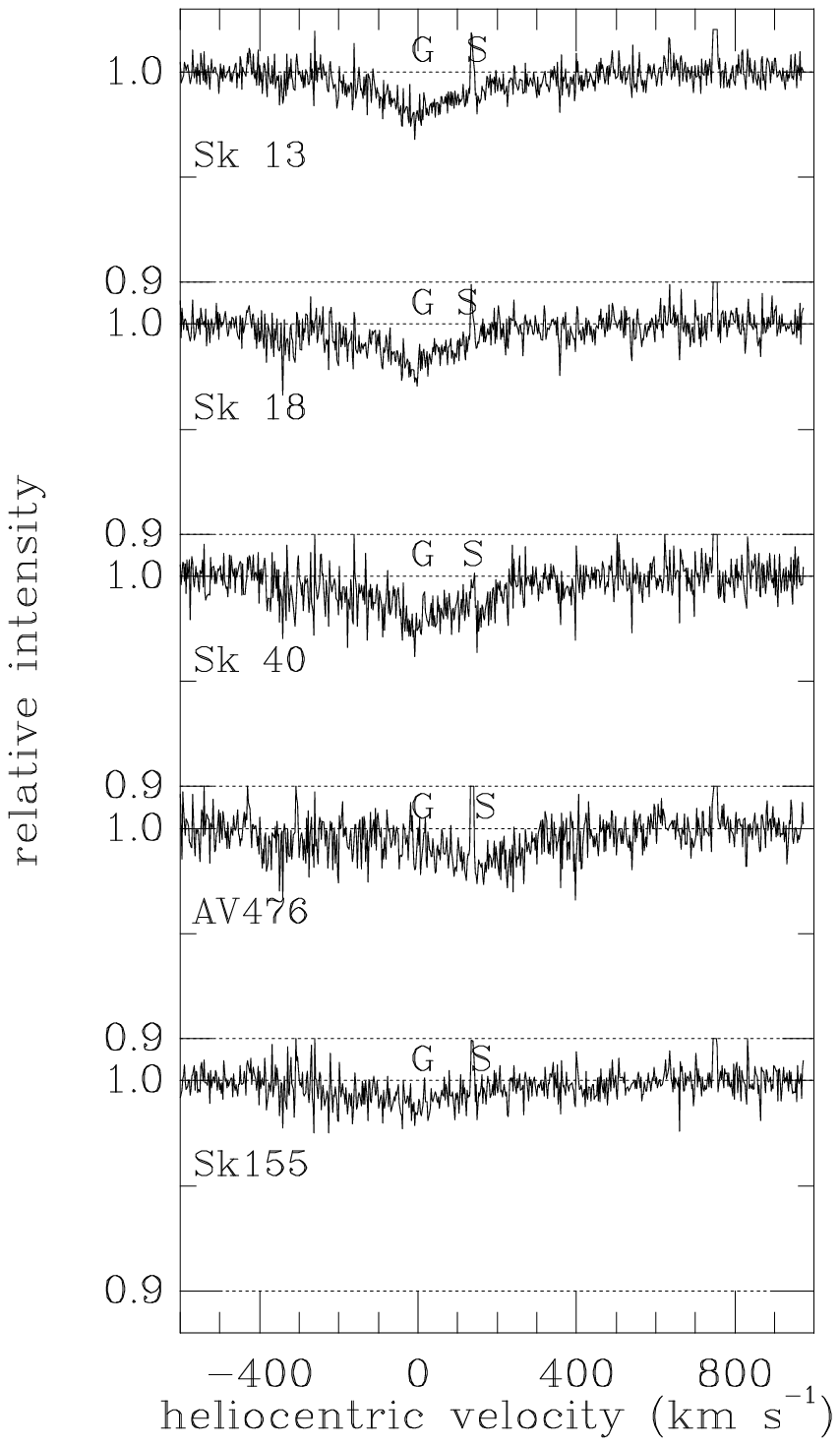}{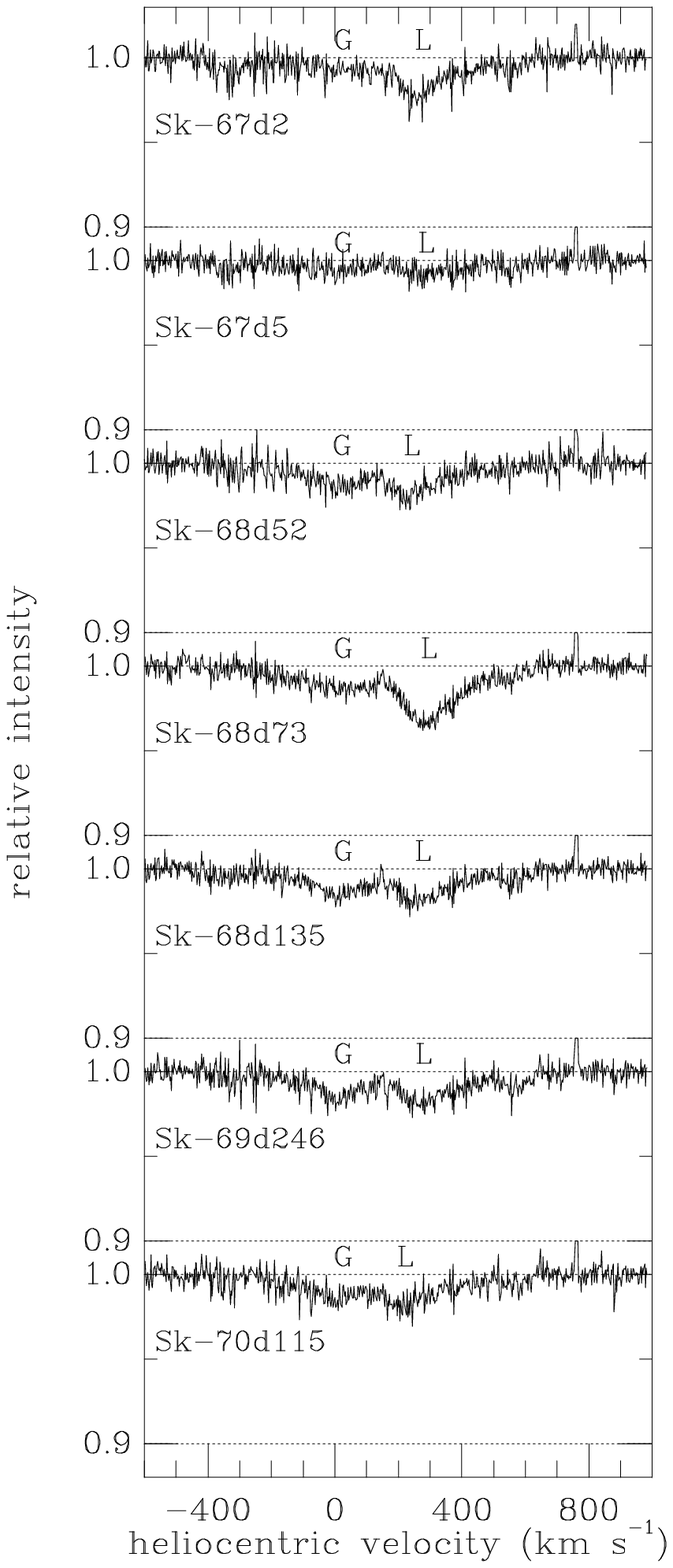}
\caption{UVES spectra of DIB at 6284 \AA\ toward SMC ({\it left}) and LMC ({\it right}) stars observed under program V03.
The vertical scale is expanded by a factor of 10.
Letters G, S, and L mark velocities of strongest Galactic, SMC, and LMC Na~I absorption.
The weak feature near 580 km~s$^{-1}$ is a residual instrumental artefact; telluric emission lines may be noted near 150 and 760 km~s$^{-1}$ (truncated).}
\label{fig:d62mc}
\end{figure}

Apart from the very strong, broad diffuse interstellar band near 4428~\AA, those at 5780, 5797, and 6284~\AA\ are typically among the strongest and most frequently observed DIBs in the Galactic ISM.
Figures~\ref{fig:d57smc}~and~\ref{fig:d57lmc} show the absorption from the DIBs at 5780 and 5797~\AA\ toward the SMC and LMC targets (respectively) observed in program V03; Figure~\ref{fig:d62mc} shows the corresponding profiles for the DIB at 6284~\AA.
The rest wavelengths for these DIBs were taken from Thorburn et al. (2003).
The letters G, S, and L above the spectra indicate the velocities of the strongest Galactic, SMC, and LMC components seen in \ion{Na}{1}; stellar lines are also present (red-ward of the 5797~\AA\ DIB) in a number of cases.
For all three of these DIBs, weak, broad absorption features --- centered at velocities similar to those of the strongest Magellanic Clouds \ion{Na}{1} components and with central depths generally $\la$~2\% below the local continuum --- appear to be present toward nearly all of the SMC and LMC targets. 
The weaker DIBs observed adjacent to (and perhaps blended with) the 5780~\AA\ DIB in more heavily reddened Galactic sight lines and the weak, broad wings of the 6284~\AA\ DIB (e.g., Jenniskens \& D\'{e}sert 1994; Tuairisg et al. 2000; Thorburn et al. 2003) can be difficult to discern in these spectra, given the relative weakness of the features and the blending between the Galactic and Magellanic Clouds absorption. 

Table~\ref{tab:ewdib} lists the equivalent widths determined for the 5780, 5797, and 6284~\AA\ DIBs in the SMC and LMC.
Values for the few Magellanic Clouds sight lines reported by Vladilo et al. (1987), Ehrenfreund et al. (2002), Sollerman et al. (2005), and Cox et al. (2006) --- and for several Galactic sight lines toward the Perseus and Sco-Oph regions (Thorburn et al. 2003) --- are also included in the table.
Because the velocity difference between the Galactic and SMC absorption is sometimes insufficient to completely separate the respective 5780~\AA\ DIB features, equivalent widths for the two galaxies were estimated by fitting two Gaussian profiles to the observed absorption. 
For the sight lines in this study, the equivalent widths for the 5780~\AA\ DIB in the Magellanic Clouds range from 14 to 111~m\AA, while the values for the 5797~\AA\ DIB range from 4 to 37~m\AA.
The uncertainties in the equivalent widths of the 5780 and 5797 \AA\ DIB features range from about 2--10~m\AA\ for the spectra from run V03 and from about 15--20~m\AA\ for the lower S/N spectra from runs C02 and C04; limits on $W$(5780) for the C02/C04 sight lines in which the DIBs were not detected are typically $\sim$40 m\AA\ (3$\sigma$). 
In view of the weak, broad wings of the 6284~\AA\ DIB --- and of the consequent blending between Galactic and Magellanic Clouds absorption --- the equivalent widths and corresponding uncertainties for that DIB were estimated (to the nearest 5 m\AA) by comparing the stronger, more well defined absorption in the cores of the DIB features to scaled, shifted high S/N spectra of the Galactic star $\rho$~Oph.
(Such estimates assume that the intrinsic profile of the 6284~\AA\ DIB is the same in all three galaxies.)
The equivalent widths obtained for the 6284~\AA\ DIB range from about 50 to 180~m\AA\ for the stars observed in run V03, with uncertainties typically $\sim$ 20~m\AA; somewhat larger $W$(6284) are found toward several of the stars observed in run C04.

For the DIBs at 5780 and 5797 \AA, a number of the equivalent widths in our sample are smaller than those previously found for the more reddened sight lines toward Sk~143 (Ehrenfreund et al. 2002) and Sk$-69\arcdeg$223 (Sollerman et al. 2005), but they are very similar to those seen toward the comparably reddened Sk$-69\arcdeg$202 (SN 1987A; Vladilo et al. 1987).
Our values for Sk~143 are consistent with those reported by Ehrenfreund et al. (2002).
For the 6284 \AA\ DIB, the equivalent width obtained toward Sk$-67\arcdeg$2 is consistent with those quoted by Ehrenfreund et al. (2002), Sollerman et al. (2005), and Cox et al. (2006) (which are all based on the same UVES data); the value for Sk$-68\arcdeg$135 is somewhat higher than the ones listed in those three references, however.

\begin{figure}
\plottwo{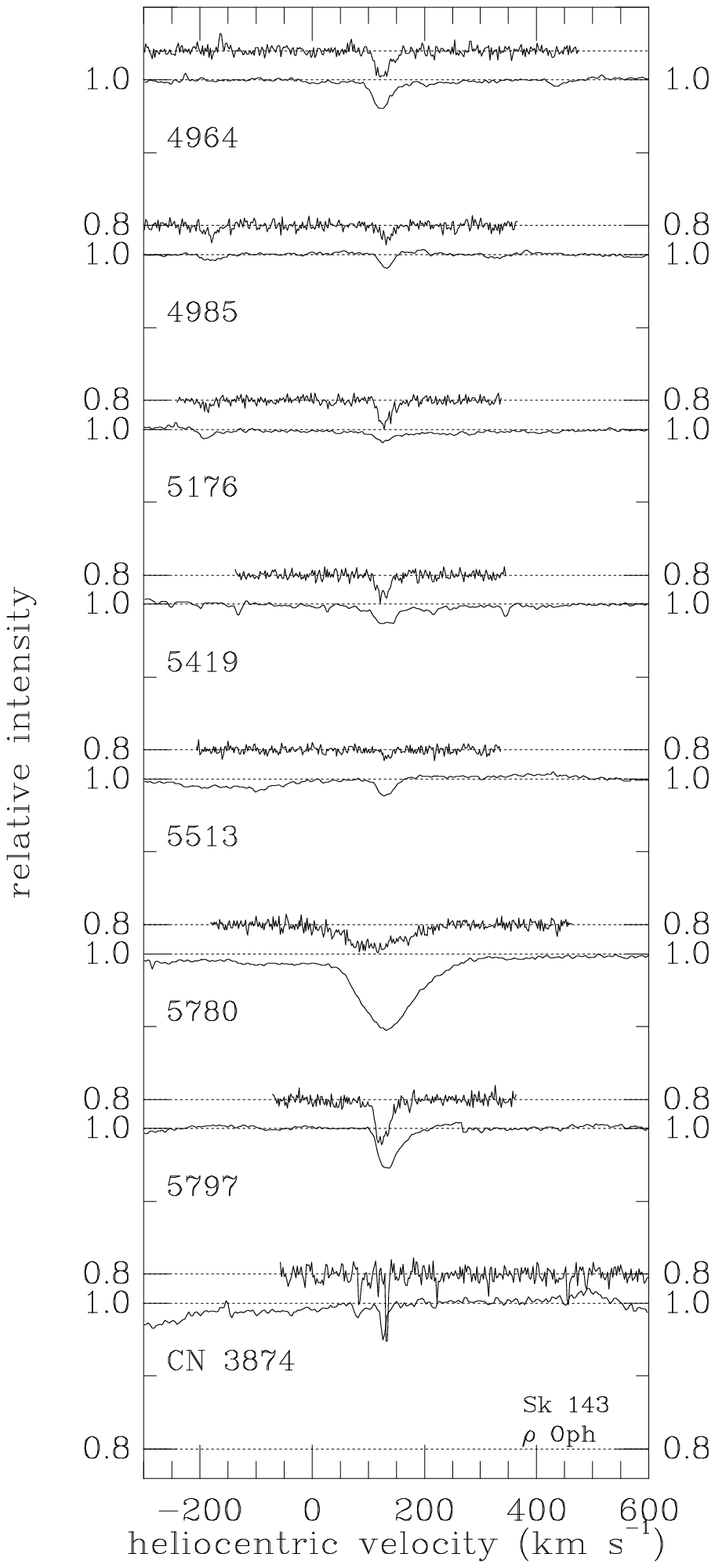}{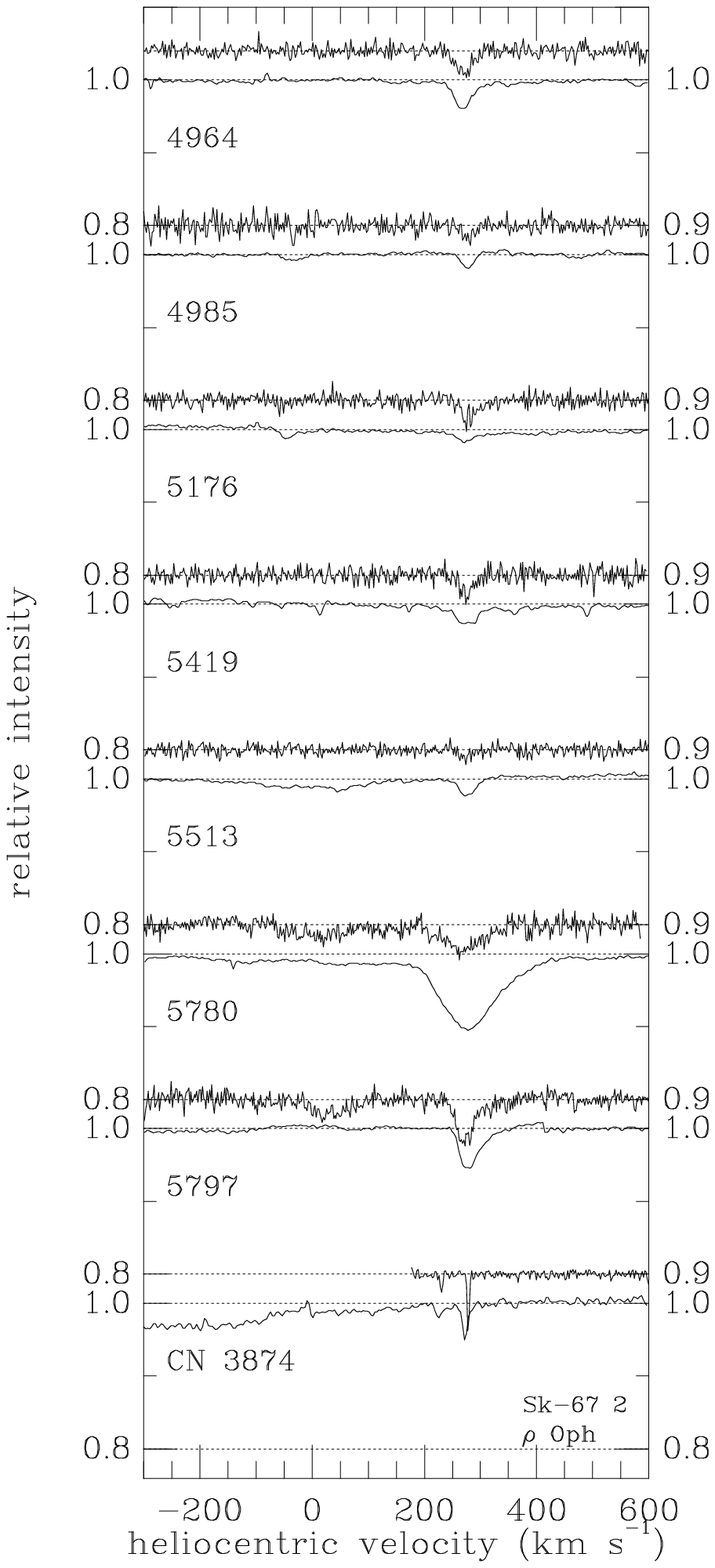}
\caption{Five strongest ``C$_2$ DIBs'' toward Sk 143 (SMC) ({\it left}) and Sk$-67\arcdeg$2 (LMC) ({\it right}), compared with those toward Galactic star $\rho$ Oph ({\it lower spectra}; shifted to same velocity).
The vertical scale is expanded by a factor of 5 for Sk~143, 10 for Sk$-67\arcdeg$2, and 5 for $\rho$ Oph.
Note also the possible absorption from the weaker C$_2$ DIBs at 4980 and 5170 \AA\ in the spectra of the 4985 and 5176 \AA\ DIBs (offset by about $-$315 km~s$^{-1}$ in each case).
Absorption from DIBs at 5780 and 5797 \AA\ and from CN is shown in the lower three pairs; Galactic absorption from those two DIBs is present near 25 km~s$^{-1}$ toward Sk$-67\arcdeg$2.}
\label{fig:c2dibs}
\end{figure}

Thorburn et al. (2003) found a set of relatively weak, narrow DIBs whose strengths, for a given range in $E(B-V)$, showed significant secondary correlations with the column densities of CN and C$_2$.
Figure~\ref{fig:c2dibs} shows the profiles of several of the strongest of those so-called ``C$_2$ DIBs'' observed toward Sk~143 and Sk$-67\arcdeg$2 (the lone SMC and LMC stars toward which CN was detected, with the highest fractions of hydrogen in molecular form in our sample) and toward the Galactic star $\rho$~Oph~A.
Equivalent widths measured for the DIB at 4963 \AA\ (which is generally the strongest of the C$_2$ DIBs) are listed in Table~\ref{tab:ewdib}; this DIB will be used to represent the C$_2$ DIBs in the comparisons with other species discussed below.
Table~\ref{tab:c2dibs} compares the equivalent widths of five of the C$_2$ DIBs (and of several other of the typically stronger DIBs) toward Sk$-67\arcdeg$2 and Sk~143 with those for six diverse Galactic sight lines which exhibit different molecular ratios and relative DIB strengths (Thorburn et al. 2003).
Of the six Galactic sight lines in Table~\ref{tab:c2dibs}, $o$ Per is a well-studied diffuse molecular sight line; HD~37061 is near the Orion Trapezium, with very low column densities of \ion{Na}{1}, \ion{K}{1}, CH, and CN; $\rho$~Oph~A is in the Sco-Oph region, where the various trace neutral species are slightly less abundant than usual; HD~183143 is often used as a ``standard'' sight line for DIB measurements (e.g., Herbig 1995); HD~204827 has relatively strong absorption from C$_2$, C$_3$, and the C$_2$ DIBs (Oka et al. 2003; Thorburn et al. 2003); and HD~210121 has UV extinction somewhat reminiscent of that seen in some Magellanic Clouds sight lines, with a very steep far-UV rise and a relatively weak 2175 \AA\ bump (Welty \& Fowler 1992).
On average, the C$_2$ DIBs toward Sk$-67\arcdeg$2 are roughly two-thirds as strong as those toward $o$~Per and about one-sixth as strong as those toward HD~204827 --- though some band-to-band variations in relative strength may be noted.
The C$_2$ DIBs toward Sk 143 are about twice as strong as those toward Sk$-67\arcdeg$2.

\subsection{Column Densities}
\label{sec-coldens}

Because the molecular lines observed toward our Magellanic Clouds targets are weak, fairly accurate total column densities may be obtained directly from the equivalent widths (assuming the lines to be optically thin) and/or by integrating the ``apparent'' optical depth (AOD) over the line profiles.
Several of the sight lines, however, appear to contain more than one component in CH and/or CH$^+$, and even more complex structure is seen in the corresponding higher resolution spectra of \ion{Na}{1} (Fig.~\ref{fig:specmol}; Welty, in prep.).
We have therefore also used the method of profile fitting to estimate column densities ($N$), line widths ($b$ $\sim$ FWHM/1.665), and velocities ($v$) for the individual components discernible in the spectra (e.g., Welty, Hobbs, \& Morton 2003).  
For each line profile, we initially adopted the minimum number of components needed to achieve a ``satisfactory'' fit, given the resolution and S/N characterizing the spectrum and assuming that each component may be represented by a (symmetric) Voigt profile.  
Given the more complex structure seen in \ion{Na}{1} and the generally good correlation between the column densities of \ion{Na}{1}, \ion{K}{1}, and CH found in the Galactic ISM (Welty \& Hobbs 2001; see below), we also used the component structures (i.e., number of components, $b$-values, relative velocities) found for \ion{Na}{1} to model the molecular lines.
The total column densities obtained from these more complex fits agree well with the values determined from the simpler fits and (in most cases) from the AOD integrations and equivalent widths.

The interstellar component parameters for \ion{Na}{1}, \ion{K}{1}, CH, and CH$^+$ obtained from the fits to the UVES line profiles are listed in Tables~\ref{tab:comps} and \ref{tab:compf}.
For each line of sight, the first line gives the sources of the spectra and the approximate 3-$\sigma$ column density limits for a single, unresolved component with ``typical'' width; subsequent lines give the heliocentric velocity, column density, and $b$-value of each component for each of those three species.  
Note that the column densities of some weak components are smaller than the listed 3-$\sigma$ limits.
Component $b$-values in square brackets were fixed, but relatively well determined in the fits; values in parentheses also were fixed, but are less well determined.  
There are strong similarities between the velocities and relative strengths of the components seen in \ion{Na}{1} $\lambda$3302, \ion{K}{1} $\lambda$7698, and CH $\lambda$4300 for sight lines with clear detections of those species.
In view of the statistics of component width and separation obtained from higher resolution, higher S/N spectra of atomic and molecular species in Galactic sight lines (Welty, Hobbs, \& Kulkarni 1994; Welty, Morton, \& Hobbs 1996; Crane, Lambert, \& Sheffer 1995; Crawford 1995; Welty \& Hobbs 2001) and of the additional structure seen in the higher resolution \ion{Na}{1} spectra (Fig.~\ref{fig:specmol}; Welty, in prep.), however, it is likely that the ``true'' component structures in the SMC and LMC sight lines are even more complex than those given in the tables.
Many of the components with $b$ $\ga$ 2--3 km~s$^{-1}$ (for example) may well be unresolved blends of narrower components.

Because of the likelihood of unresolved structure in the line profiles, we choose not to list specific uncertainties for the component parameters in Tables~\ref{tab:comps} and \ref{tab:compf}, but instead to briefly characterize typical values.
The relatively weak molecular and \ion{Na}{1} $\lambda$3302 lines (with $W_{\lambda}$ $\la$ 5 m\AA) appear to be essentially unsaturated.
For such weak lines, the total column densities derived from the equivalent widths, from integration over the apparent optical depth (AOD), and from the profile fits generally agree within 10\% (0.04 dex).
The strongest \ion{Na}{1} lines are somewhat saturated, however, with doublet ratios less than 2.
For those stronger lines, the column densities derived from the profile fits (which account for the saturation) are somewhat larger than those estimated from the equivalent widths and AOD integrations.
For each sight line, the estimated (3$\sigma$) uncertainties in $N$ for a single relatively weak, unresolved component are given in the Tables; the uncertainties in the total column densities, which are due primarily to the corresponding uncertainties in the total equivalent widths (Table~\ref{tab:ewmol}), range from about 0.03 to 0.3 dex.

The formal uncertainties in $b$ obtained in the profile fits are typically of order 20\% for reasonably isolated, well-defined lines, but can be somewhat larger for weak, broad, and/or blended lines. 
For the narrowest lines, the $b$-values are also sensitive to the exact value adopted for the instrumental resolution.
For example, if FWHM = 4.5$\pm$0.1 km~s$^{-1}$ (for the V03 UVES blue spectra), then there is an additional 15-20\% uncertainty for $b$ $\sim$ 1.0 km~s$^{-1}$ due to the uncertainty in the FWHM.
The $b$-values determined for the CH $\lambda$4300 and CH$^+$ $\lambda$4232 lines toward 23 Ori (2.2$\pm$0.2 and 2.7$\pm$0.2 km~s$^{-1}$; observed in run V03) are in excellent agreement with those determined from higher resolution (FWHM $\sim$ 1.3--1.4 km~s$^{-1}$) spectra obtained with the KPNO coud\'{e} feed (2.4$\pm$0.2 and 2.7$\pm$0.1 km~s$^{-1}$; Welty et al. 1999b) --- confirming the accuracy of the instrumental width adopted for the UVES spectra.

The formal uncertainties in $v$ are typically several tenths of a km~s$^{-1}$ for well-defined lines, but can be larger for weak, broad, and/or blended lines.
In most cases, the component velocities for the three LMC sight lines observed in runs V03 and V04 agree within about 0.5 km~s$^{-1}$; the velocities of the \ion{Na}{1} lines observed both with UVES and (at higher resolution) with the ESO 3.6m+CES (Welty, in prep.) are also in good agreement. 

While the \ion{Na}{1} D-line data for these Magellanic Clouds sight lines are discussed in more detail elsewhere (Welty, in prep.), some comments regarding the reliability of the total \ion{Na}{1} column densities are in order.
In general, analyses of \ion{Na}{1} D-line spectra obtained at resolutions $\sim$ 3--5 km~s$^{-1}$ appear to yield reasonably accurate total column densities for $N$(\ion{Na}{1}) $\la$ 10$^{12}$ cm$^{-2}$.  
For $N$(\ion{Na}{1}) $\ga$ 10$^{12}$ cm$^{-2}$, fits to higher resolution (0.6--2.0 km~s$^{-1}$) D-line spectra generally require more complex component structures and yield somewhat higher total column densities.
Observations of the weaker $\lambda$3302 lines, however, indicate that even those higher values can underestimate the true total column densities --- by more than an order of magnitude in some cases.
The \ion{Na}{1} lines toward Sk$-67\arcdeg$2 (Fig.~\ref{fig:sk67d2}) provide an example of the care needed in interpreting even moderately strong absorption lines in spectra obtained at a resolution of 5 km~s$^{-1}$.
While the component at 268 km~s$^{-1}$ appears to be stronger than the component at 278 km~s$^{-1}$ in the $\lambda$5895 (D1) line, the intrinsically much weaker $\lambda$3302 lines clearly indicate that the component at 278 km~s$^{-1}$ has a much higher column density (Table~\ref{tab:comps}).
With $b$ $\sim$ 0.65 km~s$^{-1}$ and a doublet ratio $<$ 2, even the $\lambda$3302 lines are somewhat saturated. 

Our \ion{Na}{1} component structures and total column densities, derived from both higher resolution spectra of the D lines and UVES spectra of the $\lambda$3302 doublet, thus differ somewhat from some previously published values derived solely from UVES D-line spectra.
The \ion{Na}{1} component structures derived for Sk$-67\arcdeg$5, Sk$-68\arcdeg$135, and Sk$-69\arcdeg$246 (Welty, in prep.) are more complex than those adopted by Andr\'{e} et al. (2004), and some of the component column densities are very different.
Our characterization of the properties of the individual components and association of molecular (H$_2$) gas with particular components in those sight lines thus are also somewhat different (e.g., toward Sk$-68\arcdeg$135) --- but are more consistent with the observed CH profiles.
While our total \ion{Na}{1} and \ion{K}{1} column densities toward Sk$-67\arcdeg$5 and Sk$-68\arcdeg$135 are consistent with those adopted by Sollerman et al. (2005) and by Cox et al. (2006), our values for Sk$-67\arcdeg$2 are significantly higher. 

Because the specific carriers of the DIBs are as yet unidentified (and the corresponding $f$-values are thus unknown), their column densities cannot be determined, and so the DIB strengths typically are characterized either by their equivalent widths or by their central depths.
Comparisons between DIB equivalent widths (or central depths) and the corresponding equivalent widths or central depths of other atomic or molecular features can be somewhat misleading, however, if the narrower (and unresolved) atomic or molecular features are affected by saturation --- e.g., for the \ion{Na}{1} D lines, the $\lambda$7698 line of \ion{K}{1}, and (for higher column density sight lines) the strongest lines of CH and CN.
While saturation in the atomic and molecular lines can be difficult to gauge in the moderate resolution spectra usually employed for DIB surveys, the DIBs themselves generally seem to be unsaturated.
Thorburn et al. (2003) noted that the ratio of the equivalent widths of the DIBs at 5780 and 5512 \AA\ is essentially constant, with a value near 4, for $W$(5780) up to (at least) 800 m\AA\ --- suggesting that the stronger 5780 \AA\ DIB is not saturated even at such large equivalent widths.
If even fairly strong DIBs (like 5780) are not affected by saturation, then the column densities of the (unknown) DIB carriers can be assumed to be proportional to the equivalent widths of the corresponding DIBs over the entire measured range --- enabling meaningful comparisons between DIB equivalent widths and the column densities of various atomic and molecular species even for high column density lines of sight.

\section{Results}
\label{sec-res}

\subsection{Comparisons with CO Emission}
\label{sec-coem}

Models constructed to interpret the existing CO emission data in the Magellanic Clouds have suggested that the molecular gas is concentrated in relatively small, dense clumps and that the ``interclump'' gas should be overwhelmingly atomic (e.g., Lequeux et al. 1994; Israel et al. 2003). 
The recent observations of absorption from H$_2$ (Tumlinson et al. 2002; Cartledge et al. 2005), CO and HD (Bluhm \& de Boer 2001; Andr\'{e} et al. 2004), and the optical lines of CH, CH$^+$ and CN (this paper) indicate, however, that there is some more diffuse molecular gas located outside the CO contours.
Comparisons between the diffuse gas and any nearby denser gas may provide clues to the processes responsible for the formation and/or dissipation of the denser molecular clouds.

For each line of sight in this optical survey, we have compared the heliocentric velocity of the strongest molecular (or atomic) absorption (Tables~\ref{tab:comps} and \ref{tab:compf}) to the velocity and FWHM of the CO emission from the nearest molecular cloud identified in each of the three main CO surveys (CTIO, SEST, NANTEN; see the references listed above in \S~\ref{sec-intro}).
Note that several sight lines (Sk 13 and Sk 18 in the SMC, Sk$-69\arcdeg$246 in the LMC) fall within the CO contours in the lower resolution surveys, but outside the contours of the smaller CO clumps discerned in the higher resolution SEST data.
For some of the sight lines (Sk~18, Sk~40, AV~476, LH~10-3061, Sk$-68\arcdeg$73, Sk$-68\arcdeg$135, Sk$-69\arcdeg$202), the optical and CO velocities agree within 3 km~s$^{-1}$ --- suggesting that the dense and diffuse molecular gas may be related.
For Sk~155 and Sk$-70\arcdeg$115, however, the optical and CO velocities differ by $\sim$ 20 km~s$^{-1}$; Sk~155 (and thus any gas seen in absorption toward it) appears to be located in front of the dense molecular gas in that direction (Welty et al., in prep.).
The situation is not as clear-cut for the other sight lines, where the optical and CO velocities differ by 5 to 10 km~s$^{-1}$.
No such comparisons are currently possible for Sk~143 or Sk$-67\arcdeg$2 (the only sight lines with detected absorption from CN). 
The region around Sk$-67\arcdeg$2 was only observed in the latest NANTEN survey (Yamaguchi et al. 2001a; Mizuno et al. 2002), and an updated list of molecular clouds has not yet been published.
The new maps do seem to indicate reasonably strong CO emission close to that line of sight, however.
The region around Sk~143 has not been observed in any published CO survey of the SMC.

\subsection{Column Density Comparisons}
\label{sec-cdcomp}

In this section, we compare the column densities of several atomic and molecular species with each other and with the strengths of several of the diffuse interstellar bands.  
In each case, we first display and describe the general relationships found in the local Galactic ISM and in two more restricted regions [``Sco-Oph'' and ``Orion Trapezium'' --- where the behavior of \ion{Na}{1} and \ion{K}{1} differs somewhat from the general Galactic trends (e.g., Welty \& Hobbs 2001)], then examine the corresponding relationships found for the SMC and LMC.
Such comparisons, for the different environments probed in the three galaxies, may reveal useful diagnostics of the local physical conditions and may provide clues to the (still unknown) carriers of the DIBs.
Because individual component column densities cannot be determined for several of the species, the comparisons discussed below are based on integrated sight line values --- though it should be kept in mind that significant component-to-component variations may be hidden in the integrated values, as seen (for example) for CH and CH$^+$ toward Sk$-67\arcdeg$2.
Table~\ref{tab:coldens} lists the total column densities for the SMC and LMC sight lines, together with values for several Galactic sight lines (included for comparison).
The column densities and DIB strengths for the Galactic sight lines --- both those listed in Tables~\ref{tab:coldens} and \ref{tab:ewdib} and the larger set used in various figures below --- are taken primarily from the compilations in Welty \& Hobbs (2001), Rachford et al. (2002), and Thorburn et al. (2003); most of the SMC and LMC column densities and DIB strengths are from Tumlinson et al. (2002), Welty (in prep.), and this paper.\footnotemark
\footnotetext{Compilations of atomic and molecular column densities and diffuse band strengths (with uncertainties and sources) for both Galactic and Magellanic Clouds sight lines are maintained at http://astro.uchicago.edu/$\sim$welty/coldens.html and http://astro.uchicago.edu/$\sim$welty/coldens\_mc.html.} 

As noted above, the thirteen stars observed in runs V03 and V04, chosen because they exhibit strong absorption from H$_2$ and/or \ion{Na}{1}, may not probe an entirely representative sample of the ISM in the Magellanic Clouds.
While the 31 stars observed in runs C02 and C04 were selected instead to sample a range of stellar properties, most are only lightly reddened [$E(B-V)$ $<$ 0.1], and the S/N ratios achieved in the spectra are typically factors of 2 to 4 lower than those obtained in run V03.
As a result, the 5780 \AA\ DIB was detected in only six of those 31 sight lines, and molecular absorption (from CH$^+$) was detected only toward LH~10-3061 (Tables~\ref{tab:ewmol} and \ref{tab:ewdib}).
While the upper limits on the molecular lines and DIBs are not very restrictive for many of the other C02/C04 sight lines [given the low reddening and $N$(\ion{Na}{1})], they are nonetheless generally consistent with the trends established by the data from runs V03 and V04.

The overall Galactic trends are summarized in Table~\ref{tab:corr}, which lists for each comparison [log(Y) vs. log(X)] the number of sight lines with detections of both quantities (N) included in the regression fits, the linear correlation coefficient ($r$), the slope of the line fitted to the ensemble of points, and the rms scatter of the points about that best-fit line.\footnotemark\
\footnotetext{We used a slightly modified version of the subroutine regrwt.f, obtained from the statistical software archive maintained at Penn State (http://www.astro.psu.edu/statcodes), to perform the fits (see Welty \& Hobbs 2001).}
In each case, both weighted and unweighted fits were performed; ``discrepant'' sight lines (e.g., those in the Sco-Oph, Pleiades, and Orion Trapezium regions) were not included.
While the fits allow for uncertainties in both Y and X, the unweighted fits assume that the uncertainties in the two quantities are equal; limits are not included, in either case.

For many of the relationships, the slope of the best-fit line is close to either 1.0 or 0.5 --- suggesting that Y is roughly proportional either to X or to X$^{1/2}$.
The corresponding average Galactic ratios [Y/X or Y/(X$^{1/2}$); not including the values for any ``discrepant'' sight lines] are listed in Table~\ref{tab:ratgal}, together with the average ratios for the smaller Sco-Oph, Trapezium, LMC, and SMC samples.
The second line for each ratio gives the differences between the average values for the latter four samples and the overall Galactic average value. 
While slight differences in the individual subsets of sight lines with detections of both X and Y (for small N) can in principle lead to differences in the average ratios, some significant, systematic differences do appear to be present among the various sight line samples. 
The overall average differences in the behavior of the DIBs at 5780, 5797, and 6284 \AA\ in the four subsamples/regions are summarized at the bottom of the table.

\subsubsection{CN, CH, and CH$^+$}
\label{sec-cdmol}

\begin{figure}
\epsscale{1.0}
\plotone{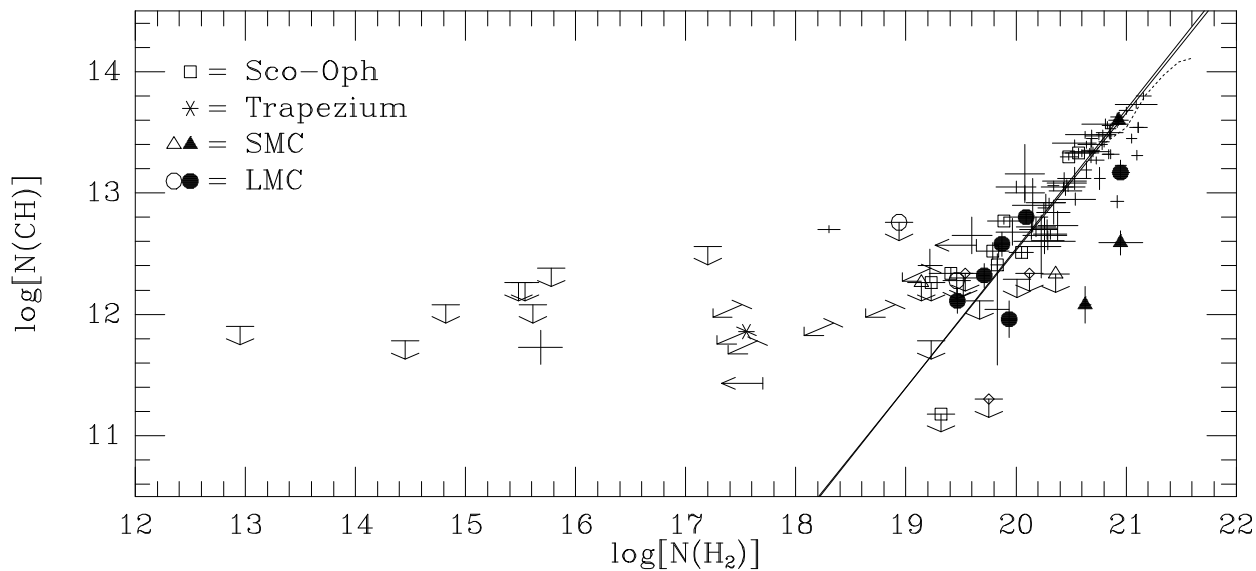}
\caption{CH vs. H$_2$.
Circles and triangles denote LMC and SMC sight lines, respectively; filled symbols indicate detections of both species; open symbols indicate limits for one (or both) of the species.
All other symbols denote Galactic sight lines (open squares are for Sco-Oph; asterisks are for Orion Trapezium; smaller open diamonds are for other ``discrepant'' sight lines).
The solid lines show weighted and unweighted fits to the Galactic data (with slopes $\sim$ 1.1), for $N$(H$_2$) $\ge$ 10$^{19}$ cm$^{-2}$ and omitting various ``discrepant'' sight lines.
The dotted line shows the values predicted by ``translucent cloud'' models T1--T6 of van Dishoeck \& Black (1989).}
\label{fig:chh2}
\end{figure}

The roughly linear relationship between the column densities of CH and H$_2$, found for diffuse sight lines by Federman (1982) and Danks, Federman, \& Lambert (1984) and for more reddened (``translucent'') sight lines by Mattila (1986) and Rachford et al. (2002), is displayed in Figure~\ref{fig:chh2}.
In this figure (and in subsequent similar figures), the Galactic sight lines are represented by the crosses, with the lengths of the lines indicating the 1-$\sigma$ measurement uncertainties.
The open squares and asterisks denote sight lines in Sco-Oph and the Orion Trapezium, respectively; the smaller open diamonds denote other Galactic sight lines (e.g., in the Pleiades) which have ``discrepant'' column densities of \ion{Na}{1} and/or \ion{K}{1} (Welty \& Hobbs 2001). 
Triangles and circles denote sight lines in the SMC and LMC, respectively; for these, the filled symbols represent sight lines with detections of both species, while the open symbols indicate sight lines with limits for one (or both) of the species.
The solid line, fitted to the points with $N$(H$_2$) $\ge$ 10$^{19}$ cm$^{-2}$ (where the H$_2$ is fully self-shielded), has slope 1.13$\pm$0.08; the average column density ratio log[$N$(CH)/$N$(H$_2$)], for 52 Galactic sight lines, is $-$7.38$\pm$0.22.
As discussed below, the observed nearly linear relationship between the column densities of CH and H$_2$ can be understood via models of steady-state gas-phase chemistry in moderately dense gas.
For example, the dotted line in Fig.~\ref{fig:chh2} shows the column densities predicted by the ``translucent cloud'' models T1--T6 of van Dishoeck \& Black (1989), which have $n_{\rm H}$ = 500--1000 cm$^{-3}$.

The general Galactic relationship between CH and H$_2$ seems to hold as well for the seven Sco-Oph sight lines with detections of both species (though not for $\pi$ Sco), for the six LMC sight lines (with the possible exception of those toward Sk$-67\arcdeg$2 and Sk$-70\arcdeg$115), and for Sk~143 (SMC).
The other two SMC sight lines with detections of H$_2$ and CH (Sk~18, AV~476), however, have $N$(CH)/$N$(H$_2$) ratios at least an order of magnitude smaller than the Galactic average value, and the upper limit toward Sk~13 is a factor of 4 below that average value.
The C02/C04 sight lines with only upper limits for $N$(CH) and $W$(5780) (not plotted in this and subsequent figures) have log[$N$(H$_2$)] $\la$ 19.4 and log[$N$(CH)] $\la$ 12.6 --- not inconsistent with the observed trends.
The SMC and LMC sight lines with CH detections sample different regions, which apparently are characterized by somewhat different physical conditions (see below), and it is not yet clear (given the small number of detections) whether the $N$(CH)/$N$(H$_2$) ratios observed for these few cases are characteristic of the two galaxies.

\begin{figure}
\epsscale{1.0}
\plotone{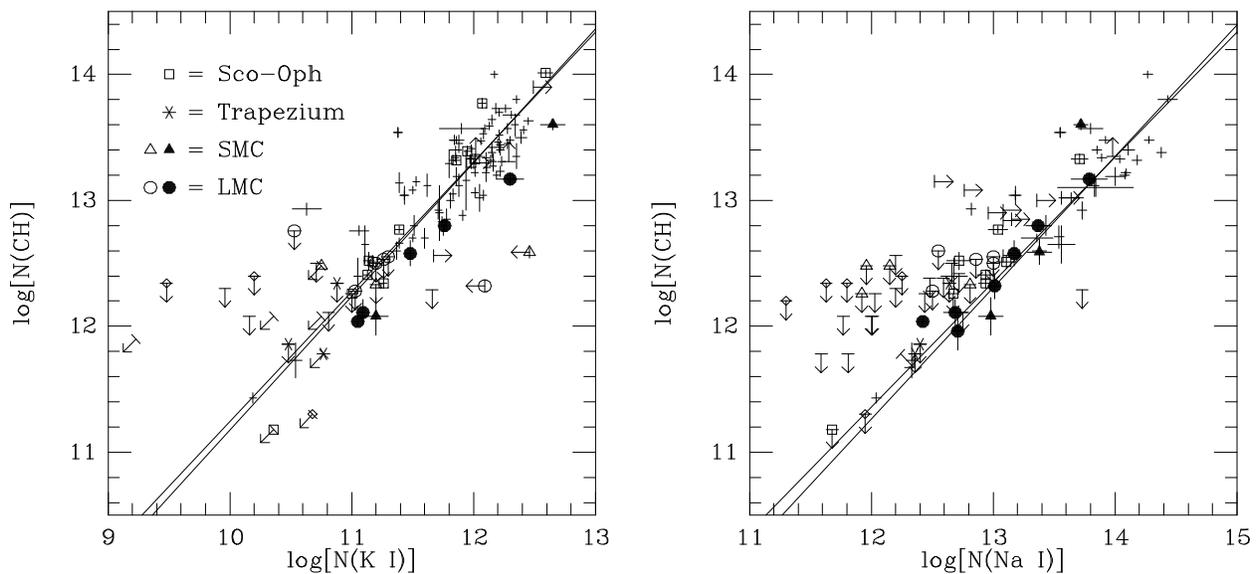}
\caption{CH vs. K~I ({\it left}) and CH vs. Na~I ({\it right}).
Circles denote LMC sight lines; triangles denote SMC sight lines; all others are for Galactic sight lines (see caption to Fig.~8).
The solid lines show weighted and unweighted fits to the Galactic data (with slopes $\sim$ 1.0; omitting ``discrepant'' sight lines).}
\label{fig:chk1na1}
\end{figure}

Welty \& Hobbs (2001) noted both a fairly tight, nearly linear relationship between the column densities of CH and \ion{K}{1} and striking similarities in the respective absorption-line profiles in a number of Galactic sight lines --- suggestive of a generally close association between the two species (see also Pan et al. 2004, 2005).
Figure~\ref{fig:chk1na1} shows the relationships between $N$(CH) and the column densities of both \ion{K}{1} (with twice the number of sight lines plotted by Welty \& Hobbs) and \ion{Na}{1} (which is also tightly correlated with \ion{K}{1}).
In both cases, the column densities for the Galactic sight lines follow an essentially linear relationship, with slopes $\sim$ 0.99--1.06, very small rms scatter ($\le$ 0.15 dex), and very few ``discrepant'' points.
Similar relationships appear to hold for both the LMC and SMC.
The column density ratios log[$N$(CH)/$N$(\ion{K}{1})] and log[$N$(CH)/$N$(\ion{Na}{1})] for the Sco-Oph region (6--10 sight lines) and the LMC (4--7 sight lines) appear to be consistent with the overall Galactic values 1.30$\pm$0.20 and $-$0.65$\pm$0.19, respectively; the slightly lower than average $N$(\ion{Na}{1})/$N$(\ion{K}{1}) ratios in those two regions are reflected in the slight differences from the average Galactic values.
While the average $N$(CH)/$N$(\ion{Na}{1}) ratio for the three SMC sight lines with detections of CH and \ion{Na}{1} is also consistent with the Galactic value, the ratio is higher by a factor of about 3 toward Sk~143.
The other C02/C04 sight lines have both log[$N$(\ion{Na}{1})] and log[$N$(CH)] $\la$ 12.6 --- again not inconsistent with the observed trends.

\begin{figure}
\epsscale{1.0}
\plotone{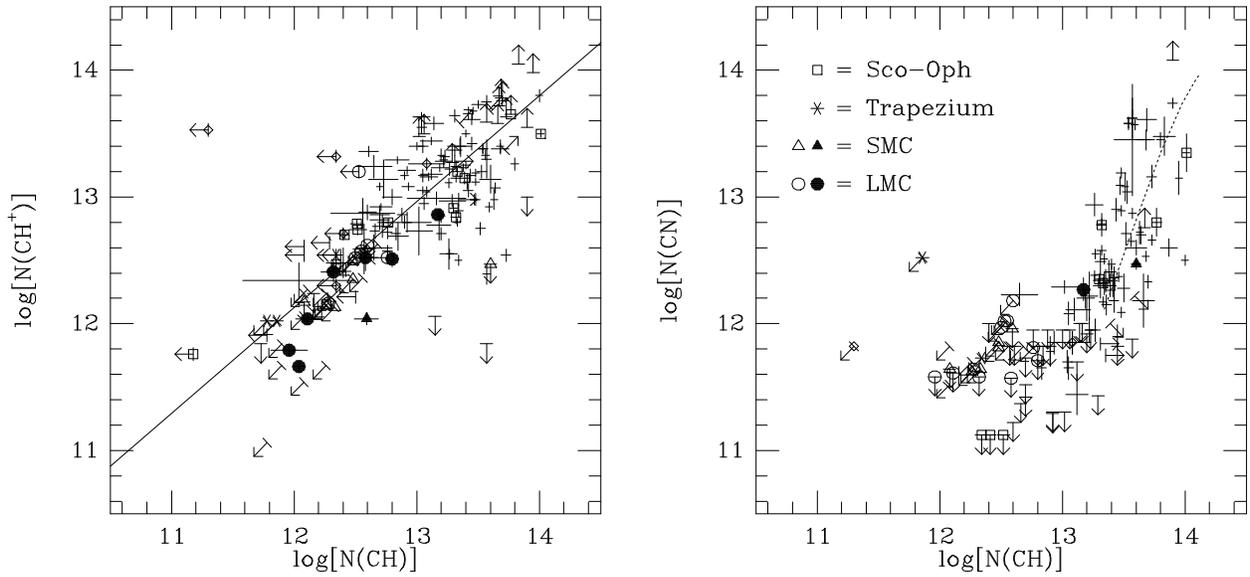}
\caption{CH$^+$ vs. CH ({\it left}) and CN vs. CH ({\it right}).
Circles denote LMC sight lines; triangles denote SMC sight lines; all others are for Galactic sight lines (see caption to Fig.~8).
The solid lines show weighted and unweighted fits to the Galactic data (omitting ``discrepant'' sight lines).
The dotted line in the right-hand plot shows the values predicted by ``translucent cloud'' models T1--T6 of van Dishoeck \& Black (1989).}
\label{fig:chpcn}
\end{figure}

The left-hand panel of Figure~\ref{fig:chpcn} shows the relationship between $N$(CH$^+$) and $N$(CH) (with slope = 0.84$\pm$0.09), including values for the sight lines in the Magellanic Clouds.
There is more scatter (rms = 0.23 dex) than for the relationships between CH and H$_2$, \ion{K}{1}, and \ion{Na}{1} --- and there would be even more scatter if individual component values were plotted instead of total sight line values (e.g., Pan et al. 2005).
For the Galactic sight lines, the mean $N$(CH$^+$)/$N$(CH) ratio is about 0.87 (Table~\ref{tab:ratgal}).
The only systematic regional difference seems to be for the Pleiades, where the $N$(CH$^+$)/$N$(CH) ratio is unusually high (White 1984), but there also are a number of sight lines with very low $N$(CH$^+$)/$N$(CH) ratios [often together with fairly high $N$(CN)/$N$(CH); e.g., Cardelli et al. (1990); Gredel 2004].
In general, the total sight line ratios for the SMC and LMC seem consistent with the observed Galactic relationship (for relatively low column densities) --- though the $N$(CH$^+$)/$N$(CH) ratio is high toward LH~10-3061, very low toward Sk~143, and the individual components toward Sk$-67\arcdeg$2 exhibit quite different $N$(CH$^+$)/$N$(CH) ratios (9.3 at +268 km~s$^{-1}$, 0.03 at +278 km~s$^{-1}$).

The right-hand panel of Figure~\ref{fig:chpcn} shows the fairly steep relationship between the column densities of CN and CH.
There is considerable scatter, and most of the sight lines with $N$(CH) $\la$ 10$^{13}$ cm$^{-2}$ (both Galactic and Magellanic Clouds) have only upper limits for $N$(CN).
There are no apparent systematic regional differences; the values toward Sk$-67\arcdeg$2 and Sk~143 are similar to those found in the local Galactic ISM.
The dotted line, showing the column densities predicted by models T1--T6 of van Dishoeck \& Black (1989), is in general agreement with the observed column densities.

\subsubsection{Diffuse Bands}
\label{sec-cddibs}

\begin{figure}
\epsscale{0.80}
\plotone{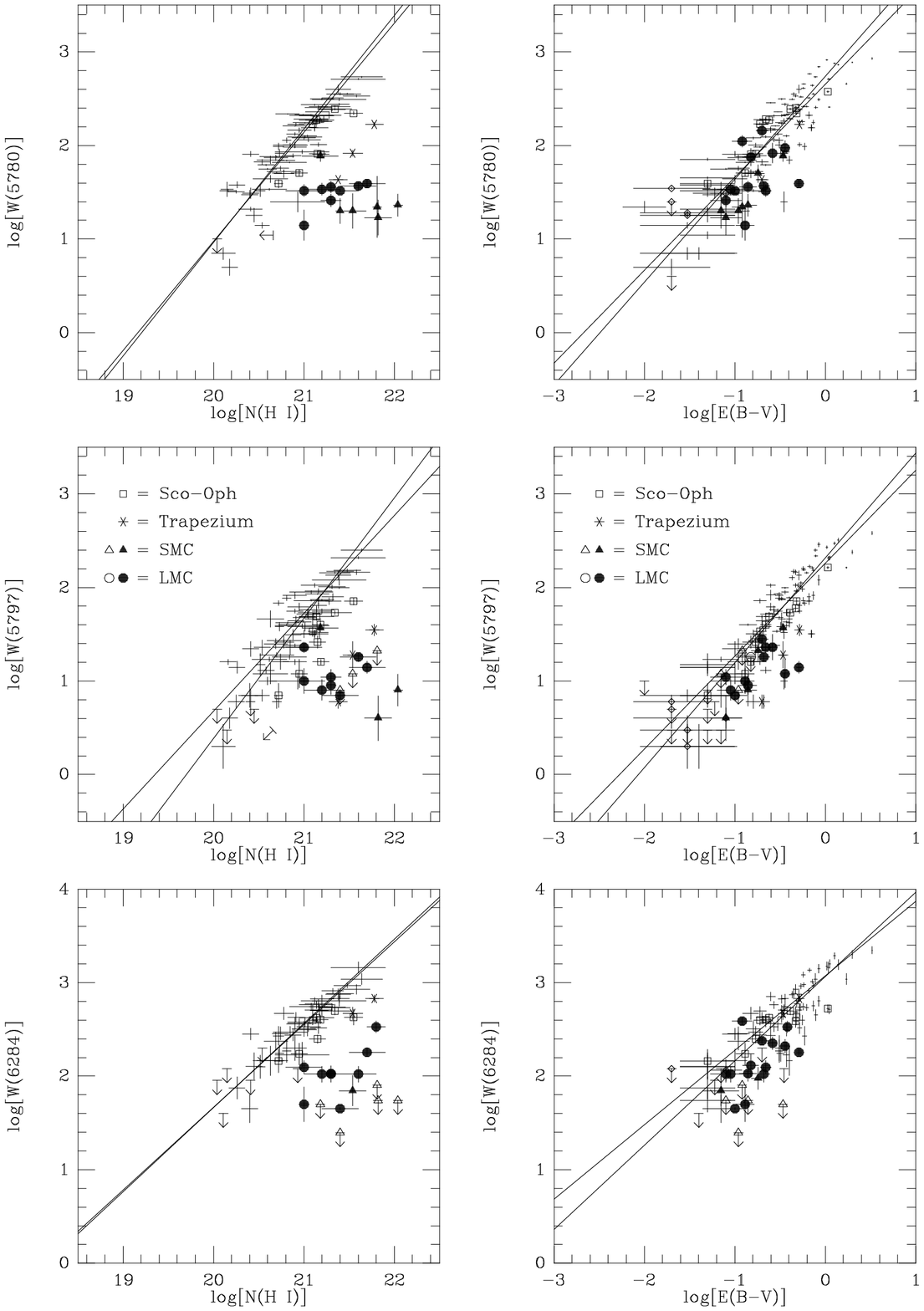}
\caption{5780 DIB ({\it top}), 5797 DIB ({\it middle}), and 6284 DIB ({\it bottom}) vs. $N$(H~I) ({\it left}) and vs. $E(B-V)$ ({\it right}).
Circles denote LMC sight lines; triangles denote SMC sight lines; all others are for Galactic sight lines (see caption to Fig.~8).
The solid lines show weighted and unweighted fits to the Galactic data (with slopes $\sim$ 0.9--1.3 vs. $N$(H~I) and $\sim$ 0.8--1.1 vs. $E(B-V)$; omitting ``discrepant'' sight lines).}
\label{fig:h1ebv}
\end{figure}

Figure~\ref{fig:h1ebv} shows the well-known correlations between the strengths of the 5780, 5797, and 6284 \AA\ DIBs, the column density of \ion{H}{1}, and the color excess $E(B-V)$ (e.g., Herbig 1993, 1995, and references therein; York et al., in prep.; see also Cox et al. 2006).
The slopes of the general Galactic relationships with both $N$(\ion{H}{1}) and $E(B-V)$ are 1.0--1.3 for the 5780 and 5797 \AA\ DIBs, and are 0.8--0.9 for the 6284 \AA\ DIB; the rms scatter is $\le$ 0.15 dex in all cases (Table~\ref{tab:corr}).
Some regional differences are apparent, however, for the DIBs in the Galactic ISM (Table~\ref{tab:ratgal}).
Toward the Sco-Oph stars, all three DIBs are weaker by a factor of about 2, relative to $N$(\ion{H}{1}), but are consistent with the general trends versus $E(B-V)$.
Toward the three Trapezium region stars, the three DIBs are weaker by factors of 3 to 10, relative to $N$(\ion{H}{1}); the 5780 and 5797 \AA\ DIBs are weaker by factors of about 2 and 4, respectively, relative to $E(B-V)$ (see also Herbig 1993; Jenniskens, Ehrenfreund, \& Foing 1994).
The weakness of the DIBs versus $N$(\ion{H}{1}) is reminiscent of the lower column densities of \ion{Na}{1} and \ion{K}{1}, for a given $N$(H$_{\rm tot}$) [= $N$(\ion{H}{1}) + 2$N$(H$_2$)], noted for these two regions by Welty \& Hobbs (2001).

In the Magellanic Clouds, the 5780, 5797, and 6284 \AA\ DIBs are (on average) significantly weaker than their Galactic counterparts --- by factors of 7 to 9 (LMC) and factors of $\sim$ 20 (SMC), relative to $N$(\ion{H}{1}), and by factors of about 2 (both LMC and SMC), relative to $E(B-V)$. 
The column densities of \ion{Na}{1} and \ion{K}{1} also are significantly lower in the LMC and SMC, for a given $N$(H$_{\rm tot}$), than in the local Galactic ISM (Welty, in prep.).
The 6284 \AA\ DIB may generally be even weaker in the SMC, as several sight lines (Sk~13, Sk~18, Sk~143) have upper limits for $W$(6284)/$N$(\ion{H}{1}) that are factors of 30--70 below the average Galactic value.
The range in the ratios $W$(DIB)/$N$(\ion{H}{1}) and $W$(DIB)/$E(B-V)$, however, is often larger for the set of sight lines in the Magellanic Clouds than for those in our Galaxy --- so that some of the ratios in some individual LMC and SMC sight lines (e.g., for 5780 and 5797 toward Sk~143) are closer to the Galactic values.
The sight lines observed in runs C02 and C04 with upper limits for CH and the 5780 \AA\ DIB have log[$N$(\ion{H}{1})] $\la$ 21.6 (21.1) for the SMC (LMC), $E(B-V)$ $<$ 0.1, and $W$(5780) $\la$ 40 m\AA\ --- consistent with the values shown for the SMC and LMC sight lines in the upper panels of Figure~\ref{fig:h1ebv}.
Curiously, the Magellanic Clouds DIBs (for the present small sample) do not exhibit the same clear linear relationships with $N$(\ion{H}{1}) found for most Galactic sight lines.

\begin{figure}
\plotone{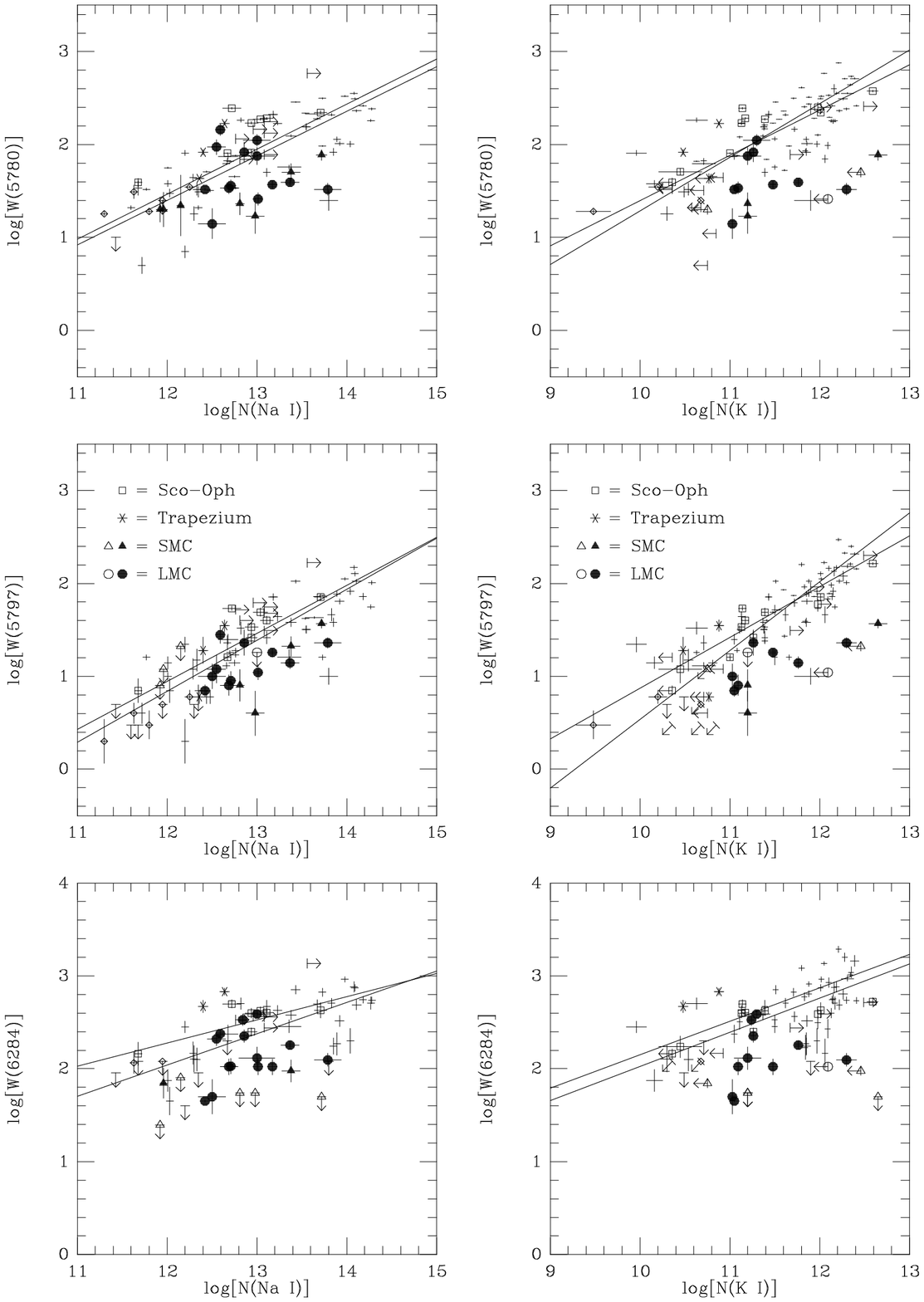}
\caption{5780 DIB ({\it top}), 5797 DIB ({\it middle}), and 6284 DIB ({\it bottom}) vs. $N$(Na~I) ({\it left}) and vs. $N$(K~I) ({\it right}).
Circles denote LMC sight lines; triangles denote SMC sight lines; all others are for Galactic sight lines (see caption to Fig.~8).
The solid lines show weighted and unweighted fits to the Galactic data (with slopes
$\sim$ 0.3--0.5 vs. $N$(Na~I) and $\sim$ 0.4--0.7 vs. $N$(K~I); omitting ``discrepant'' sight lines).}
\label{fig:nak}
\end{figure}

Figure~\ref{fig:nak} indicates that the strengths of the three DIBs also appear to be correlated with the column densities of both \ion{Na}{1} and \ion{K}{1} (Herbig 1993; see also Krelowski, Galazutdinov, \& Musaev 1998; Galazutdinov et al. 2004; Cox et al. 2006), though with somewhat larger rms scatter (0.15--0.20 dex for 5797, 0.22--0.26 dex for 5780 and 6284) and correspondingly smaller correlation coefficients than those characterizing the relationships with $N$(\ion{H}{1}) and $E(B-V)$ (Table~\ref{tab:corr}).
Moreover, the slopes of the general Galactic relationships with $N$(\ion{Na}{1}) and $N$(\ion{K}{1}) are smaller --- 0.5--0.7 for the 5780 and 5797 \AA\ DIBs and 0.3--0.4 for the 6284 \AA\ DIB; the slopes versus $N$(\ion{C}{1}) are slightly smaller in each case (Table~\ref{tab:corr}).
The DIB strengths thus may be roughly proportional to the square roots of the column densities of the trace neutral species --- consistent with the roughly linear relationships between the DIBs and $N$(\ion{H}{1}) and the roughly quadratic relationships between the trace species and $N$(H$_{\rm tot}$) (Welty \& Hobbs 2001).
As noted by Welty \& Hobbs (2001), these slopes are somewhat smaller than those found by Herbig (1993) from similar plots of DIB strength versus $N$(\ion{X}{1}).  
Most of the \ion{Na}{1} column densities used in Figure~\ref{fig:nak} have been derived from fits to relatively high resolution spectra of the \ion{Na}{1} D lines, using additional constraints from spectra of the much weaker (and thus less saturated) \ion{Na}{1} doublet at 3302 \AA\ and the \ion{K}{1} line at 7698 \AA\ for many of the higher column density sight lines (e.g., Welty \& Hobbs 2001).
For those higher $N$(\ion{Na}{1}) sight lines, the derived \ion{Na}{1} column densities are systematically higher than those estimated by Herbig (1993) from considerations of the D-line doublet ratios --- thus leading to smaller overall slopes.

There are also some apparent regional differences in average DIB strength versus both [$N$(\ion{Na}{1})]$^{1/2}$ and [$N$(\ion{K}{1})]$^{1/2}$ (Table~\ref{tab:ratgal}).
Toward the Sco-Oph stars, the three DIBs are consistent with the general Galactic values, relative to [$N$(\ion{K}{1})]$^{1/2}$, but are stronger by a factor of about 2, relative to [$N$(\ion{Na}{1})]$^{1/2}$ --- due, presumably, to the slight relative weakness of \ion{Na}{1} in that region (as noted above).
Toward the Trapezium stars, the strength of the 5797 \AA\ DIB, relative to both [$N$(\ion{K}{1})]$^{1/2}$ and [$N$(\ion{Na}{1})]$^{1/2}$, is consistent with the general Galactic values, but the corresponding ratios for the 5780 and 6284 \AA\ DIBs are factors of about 2 and 4 higher, respectively (see also Herbig 1993; Jenniskens et al. 1994).
In both the LMC and SMC, the three DIBs are all weaker, by average factors of about 1.3--2.6 and 3--6, relative to [$N$(\ion{Na}{1})]$^{1/2}$ and [$N$(\ion{K}{1})]$^{1/2}$, respectively.
Again, however, there are individual sight lines in the SMC and LMC for which some of the ratios are consistent with typical Galactic values (e.g., for the 5780 \AA\ DIB toward Sk 40, Sk~155, BI 237, BI 253, and LH~10-3061).
The additional C02/C04 sight lines have log[$N$(\ion{Na}{1})] $\la$ 12.6 and $W$(5780) $\la$ 40 m\AA\ --- not inconsistent with the values obtained for the other Magellanic Clouds sight lines.

\begin{figure}
\plotone{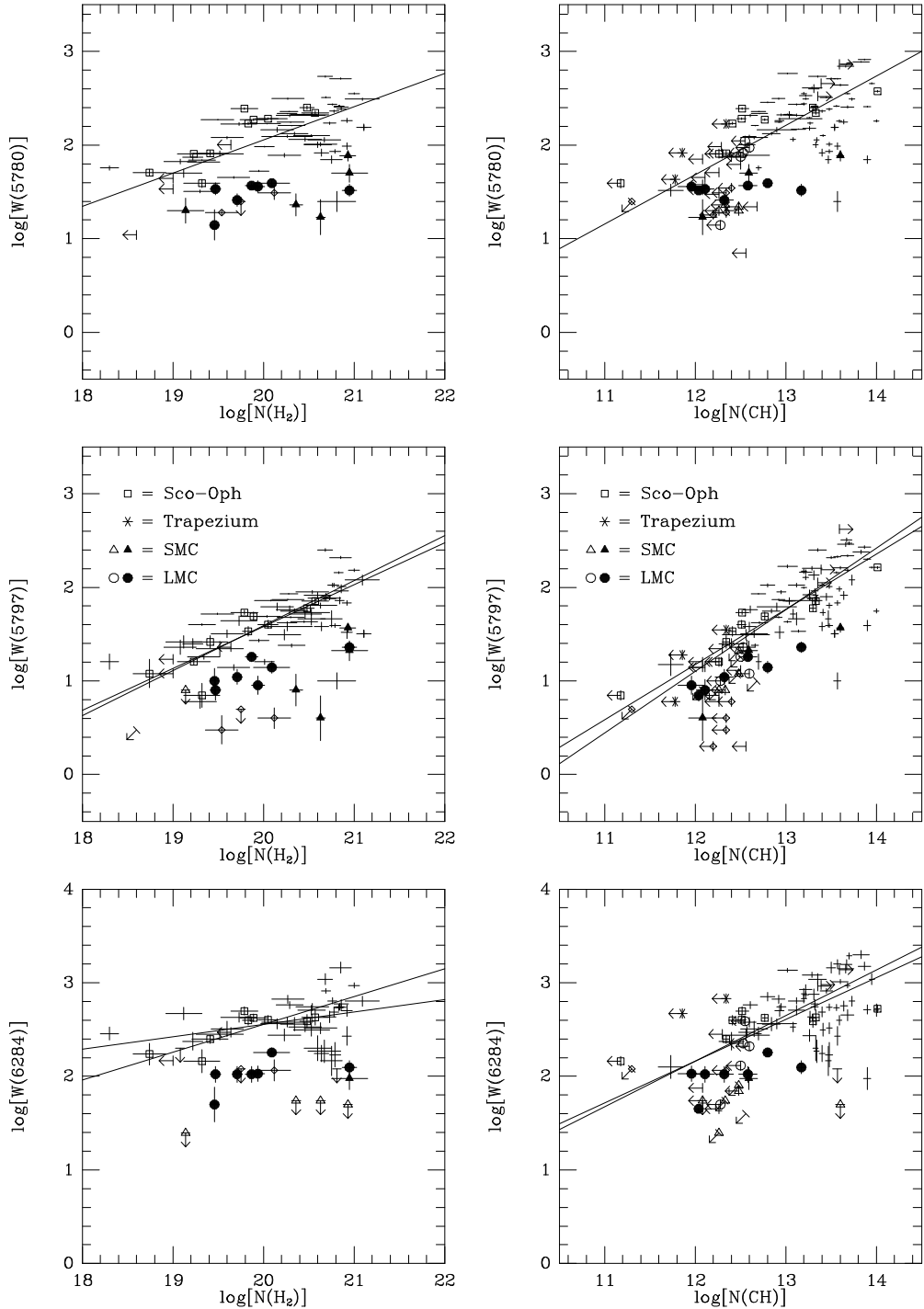}
\caption{5780 DIB ({\it top}), 5797 DIB ({\it middle}), and 6284 DIB ({\it bottom}) vs. $N$(H$_2$) ({\it left}) and vs. $N$(CH) ({\it right}).
Circles denote LMC sight lines; triangles denote SMC sight lines; all others are for Galactic sight lines (see caption to Fig.~7).
The solid lines show weighted and/or unweighted fits to the Galactic data (with slopes
$\sim$ 0.1--0.7; omitting ``discrepant'' sight lines).}
\label{fig:dibmol}
\end{figure}

As the DIBs have sometimes been ascribed to large, carbon-based molecules, comparisons between the strengths of various DIBs and the column densities of known molecules might provide some clues toward the behavior and identification of the DIB carriers (e.g., Thorburn et al. 2003).
Figure~\ref{fig:dibmol} compares the equivalent widths of the three DIBs to the column densities of H$_2$ and CH (see also Herbig 1993; Krelowski et al. 1999; Snow et al. 2002).
While all three DIBs appear to strengthen, on average, with increasing $N$(H$_2$) and $N$(CH), the correlation coefficients are all smaller than those for the relationships with $N$(\ion{H}{1}), $E(B-V)$, $N$(\ion{Na}{1}), and $N$(\ion{K}{1}) (Table~\ref{tab:corr}).
There are also sight lines with $N$(H$_2$) much lower than 10$^{18}$ cm$^{-2}$ (where the H$_2$ is not self-shielded) that have DIB equivalent widths comparable to those observed for $N$(H$_2$) $\sim$ 10$^{19}$ cm$^{-2}$ (Herbig 1993) --- similar to the behavior of the column densities of \ion{Na}{1} and \ion{K}{1} versus $N$(H$_2$) (Welty \& Hobbs 2001).
Moreover, there appears to be considerable scatter in the DIB equivalent widths for the sight lines with the highest column densities of H$_2$ and CH, particularly for the DIBs at 5780 and 6284 \AA.
Herbig (1993) ascribed the seeming correlation between $W$(5780) and $N$(H$_2$) (for higher column densities) to the underlying strong correlation between $W$(5780) and $N$(\ion{H}{1}) and the tendency for $N$(H$_2$) to increase together with $N$(\ion{H}{1}). 
After removing the trend with $N$(\ion{H}{1}), Herbig found no residual correlation between $W$(5780) and $N$(H$_2$) and concluded that the 5780 \AA\ DIB (at least) is not related to H$_2$ [see also York et al. (in prep.)]. 
While the DIBs at 5780 and 6284 \AA\ thus appear to be unrelated to H$_2$ (and to CH, given the observed close relationship between H$_2$ and CH), the equivalent width of the 5797 \AA\ DIB exhibits somewhat stronger correlations with $N$(H$_2$) and $N$(CH), and the strength of the 4963 \AA\ C$_2$ DIB also appears to be fairly well correlated with $N$(CH) (Table~\ref{tab:corr}) --- suggesting that those two narrower DIBs might be related to the diffuse molecular component of the clouds.

\begin{figure}
\plotone{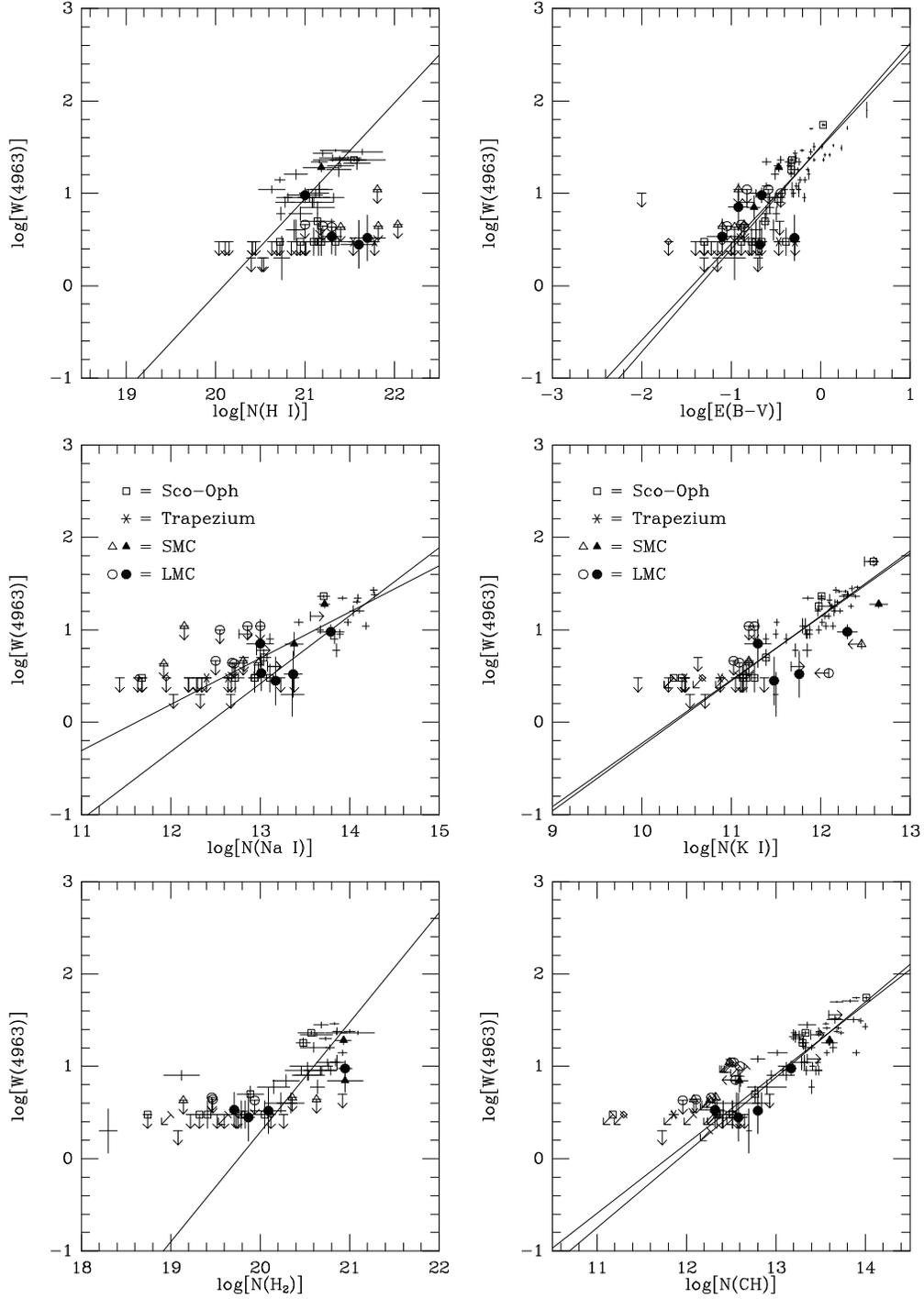}
\caption{C$_2$ DIB at 4963 \AA\ vs. $N$(H~I), $E(B-V)$, $N$(Na~I), $N$(K~I), $N$(H$_2$), and $N$(CH).
Circles denote LMC sight lines; triangles denote SMC sight lines; all others are for Galactic sight lines (see caption to Fig.~7).
The solid lines show weighted and/or unweighted fits to the Galactic data
(omitting ``discrepant'' sight lines).
In most cases, the 4963 \AA\ DIB is as strong toward Sk~143 (SMC) and Sk$-67\arcdeg$2 (the two Magellanic Clouds stars with detected CN) as it is in comparable Galactic sight lines.}
\label{fig:d4963}
\end{figure}

The equivalent widths of the C$_2$ DIBs measured for several of the SMC and LMC sight lines appear to signal some intriguing differences in the behavior among the DIBs in the Magellanic Clouds.
As discussed above, the typically stronger, more well studied DIBs at 5780, 5797, and 6284 \AA\ are generally weaker (on average) in the Magellanic Clouds than in the Galactic ISM (Figures~\ref{fig:h1ebv}--\ref{fig:dibmol} and Table~\ref{tab:ratgal}) --- especially relative to the column densities of \ion{H}{1} and H$_2$. 
The comparisons shown in Figure~\ref{fig:d4963} and listed in Table~\ref{tab:c2dibs}, however, indicate that the 4963 \AA\ DIB (typically the strongest of the C$_2$ DIBs) is just as strong toward Sk~143 and Sk$-67\arcdeg$2 (the two Magellanic Clouds sight lines in which CN is detected) as it is in Galactic sight lines with comparable values of $N$(\ion{H}{1}) and $N$(H$_2$).
Moreover, while the 4963 \AA\ DIB is significantly weaker, relative to $N$(\ion{H}{1}), in the SMC and LMC sight lines in which CN is not detected, it is still more comparable in strength to the average Galactic values --- relative to $E(B-V)$ and the column densities of \ion{Na}{1}, \ion{K}{1}, H$_2$, and CH --- than the DIBs at 5780, 5790, and 6284 \AA. 
Similar results are obtained if the sum of the equivalent widths of the five C$_2$ DIBs listed in Table~\ref{tab:c2dibs} is substituted for $W$(4963).

\subsection{Chemical Analysis}
\label{sec-chem}

The column densities of H$_2$, CN, CH, and CH$^+$ may be used to estimate the local densities, temperatures, and radiation fields characterizing the molecule-containing gas in the Magellanic Clouds.  
While these local physical conditions may be derived solely from the abundance and rotational excitation of H$_2$, complementary estimates may also be obtained from the abundances of CH, CH$^+$, and CN by using simple models for the chemistry in diffuse clouds (e.g., Federman \& Huntress 1989; Federman et al. 1994, 1997).
Because the chemical analyses often employ column densities and/or abundances of different species, it is important to determine to what extent those species coexist.
In general, CN traces relatively cold, dense regions, while CH$^+$ traces warmer, more diffuse regions. 
The CH molecule, however, can be produced either via steady-state processes in denser gas or via the non-thermal process(es) responsible for the formation of CH$^+$ in more diffuse gas.
Comparisons of the physical parameters derived under different assumptions and/or using different diagnostics thus can provide information on cloud structure, the distribution of different species, and the importance of various possible processes.

In an extensive study based on relatively high resolution spectra of 29 sight lines toward the $\rho$ Oph, Cep OB2, and Cep OB3 associations, Pan et al. (2005) identified 78 ``CH$^+$-like CH'' components and 12 ``CN-like CH'' components, out of 125 total components seen in CH (see also Crane, Lambert, \& Sheffer 1990; Gredel et al. 1993; Crawford 1995).  
The CH$^+$-like CH components are characterized by relatively low column densities (log[$N$(CH$^+$)] $\le$ 12.6), column density ratios $N$(CH)/$N$(CH$^+$) of order unity, and no detected CN.
The CN-like CH components have significantly larger $N$(CH)/$N$(CH$^+$) ratios, somewhat smaller $b$-values, and detections of CN.
For the CH$^+$-like CH components, Pan et al. found a strong correlation between $N$(CH) and $N$(CH$^+$), with log[$N$(CH)] = 0.6 $+$ 0.95~log[$N$(CH$^+$)], and proposed that the chemistry of CH may be tied primarily to that of CH$^+$ (see also Zsarg\'{o} \& Federman 2003).  
The data now available for CH and CH$^+$ in the SMC and LMC suggest that a similar relationship may apply in the Magellanic Clouds (Fig.~\ref{fig:chpcn} and discussions below).

The chemical models discussed in the following sections require estimates for the gas-phase interstellar abundances of carbon, nitrogen, and oxygen in the Magellanic Clouds.
While total abundances for those three elements have been estimated for both \ion{H}{2} regions and stars of various spectral types (e.g., references cited in the appendices to Welty et al. 1997, 1999a, and in prep.; Hill 2004), we have taken values only from studies of \ion{H}{2} regions (e.g., Garnett 1999) and main sequence O--B stars (e.g., Hunter et al. 2005; Korn et al. 2005), which should have abundances similar to the current total (gas + dust) interstellar values, unaffected by stellar CNO processing.
The adopted average fractional abundances of C, N, and O (relative to hydrogen, in parts per million) for the LMC (SMC) are $x$(C) = 82 ppm (24 ppm), $x$(N) = 14 ppm (3.4 ppm), and $x$(O) = 240 ppm (98 ppm), respectively.
Relative to the solar system values given by Lodders (2003), the adopted abundances for (C, N, O) are (0.33, 0.20, 0.49) times solar for the LMC and (0.10, 0.05, 0.20) times solar for the SMC.
The relative oxygen abundances are similar to those of most of the heavier elements in the LMC and SMC; carbon and nitrogen, however, appear to be somewhat more deficient.
While both carbon and oxygen are mildly (and apparently fairly uniformly) depleted into dust grains in the Galactic ISM (e.g., Sofia et al. 2004; Cartledge et al. 2004), the depletions of those elements are not yet known for the SMC or LMC.
Inclusion of depletion effects in the following analyses would yield correspondingly higher densities.

\subsubsection{H$_2$ Rotational Excitation:  Temperature, Density, and Radiation Field}
\label{sec-h2}

We first consider the physical conditions that can be extracted from the distribution of H$_2$ column densities in specific rotational levels of the ground vibrational state, under the assumptions that the lowest two levels ($J$ = 0 and 1) are thermalized via collisions with protons and that the higher $J$ levels are populated primarily via photon pumping (Jura 1974, 1975; Lee et al. 2002).  
For optically thick H$_2$ [i.e., $N$(H$_2$) $\ga$ 10$^{19}$ cm$^{-2}$], the local kinetic temperature may be approximated by the rotational temperature 
\begin{equation}
T_{01} = \frac{E_{01}/k}{ln[(g_1/g_0)N(0)/N(1)]},
\end{equation}
where $N$($J$) is the H$_2$ column density in rotational level $J$, the statistical weights $g_0$ = 1 and $g_1$ = 9, and the excitation energy $E_{01}/k$ = 171 K.
The total gas density [$n_{\rm H}$ $=$ $n$(\ion{H}{1}) $+$ 2~$n$(H$_2$)] may be obtained from
\begin{equation}
n_{\rm H} = 9.2\times10^8~\frac{N(4)}{N({\rm H}_{\rm tot})}
\left\{\frac{0.26~N(0)}{0.11~\left[N(0)+N(1)\right]}~+~0.19\right\}^{-1},
\end{equation}
where $N$(H$_{\rm tot}$) $=$ $N$(\ion{H}{1})~$+$~2~$N$(H$_2$) is the total hydrogen column along the line of sight.  
Eqn.~(2) assumes a rate coefficient for H$_2$ formation $R$ = $3 \times 10^{-18}$ cm$^3$ s$^{-1}$ for the Magellanic Clouds (Tumlinson et al. 2002) --- a factor of 10 lower than that usually adopted for Galactic clouds.
The enhancement of the ambient UV radiation field over the average Galactic field described by Draine (1978), $I_{uv}$, then may be inferred from the expression,
\begin{equation}
I_{uv} = \frac{7.1\times10^{-24}~n_{\rm H}}{0.11~\beta_0}
\left[\frac{N({\rm H}_{\rm tot})}{2~f({\rm H}_2)}\right]^{1/2},
\end{equation}
where $\beta_0$ = $5 \times 10^{-10}$ s$^{-1}$ is the H$_2$ photoabsorption rate corresponding to the average Galactic field, the factor 0.11 in front of $\beta_0$ represents the fraction of absorptions leading to dissociation, and $f$(H$_2$) = 2~$N$(H$_2$)/$N$(H$_{\rm tot}$) is the fraction of hydrogen in molecular form.
We also assume that the H$_2$ data pertain to all molecular components for a given line of sight --- which, in essence, suggests that each component is part of a larger complex that is characterized by fairly uniform local physical conditions ($I_{uv}$, $n_{\rm H}$, $T$).  
Substitution for $n_{\rm H}$ [from eqn.~(2)] and for $f$(H$_2$) in eqn.~(3) confirms that $I_{uv}$ depends only on the H$_2$ column density and relative rotational populations, and not on the assumed $R$.  
The uncertainties in $I_{uv}$ and $n_{\rm H}$ are typically dominated by the uncertainties in $N$($J$=4), which is often determined from lines on the flat part of the H$_2$ curve of growth; factors of 3 (or more) are not unusual for Magellanic Clouds sight lines with total $N$(H$_2$) $\ga$ 10$^{19}$ cm$^{-2}$ (Tumlinson et al. 2002).

Application of this analysis to {\it Copernicus} and {\it FUSE} H$_2$ data for 21 Galactic sight lines with $N$(H$_2$) $>$ 10$^{19}$ cm$^{-2}$ and data for the rotational populations (Spitzer, Cochran, \& Hirshfeld 1974; Jura 1975; Snow 1976, 1977; Savage et al. 1977; Frisch \& Jura 1980; Sonnentrucker et al. 2002, 2003) yields $I_{uv}$ from 0.5 to 40 and $n_{\rm H}$ from 15 to 800 cm$^{-3}$; all sight lines except one ($o$ Per) have $I_{uv}$ $<$ 8 and $n_{\rm H}$ $<$ 200 cm$^{-3}$, however.
Somewhat higher values for $I_{uv}$ and $n_{\rm H}$ are obtained toward HD~37903 (Lee et al. 2002), HD~73882 (Snow et al. 2000), and several stars in Cep OB2 (Pan et al. 2005).
Such elevated UV fluxes and moderately high densities (though not as high as for dense molecular clouds) can be associated with photon-dominated regions in star-forming clouds (e.g., Knauth et al. 2001; Lee et al. 2002).
The bulk of the material containing neutral atoms and CN molecules, however, is located sufficiently far from the illuminating source that both the flux and density are lower.  

The physical conditions inferred for the Magellanic Clouds sight lines are given in Table~\ref{tab:phys}, where the entries are based on the total H$_2$ and \ion{H}{1} column densities listed in Table~\ref{tab:coldens} and the detailed rotational populations given by Tumlinson et al. (2002), Cartledge et al. (2005), and Welty et al. (in prep.).  
For the SMC gas toward Sk~143, we have estimated $N$($J$=4) $\sim$ 10$^{14.8}$ cm$^{-2}$, based on the relative populations for other sight lines with similar $N$($J$=2)/$N$($J$=3) ratios.
The values for $T_{01}$ range from 46 to 91~K --- typical of the values found for the SMC and LMC (and similar to those found in the Galactic ISM) (Savage et al. 1977; Tumlinson et al. 2002).
The values for $I_{uv}$, however, are comparable to the values generally obtained for diffuse gas in our Galaxy only toward Sk~143, AV~476, and Sk~155 in the SMC and toward Sk$-67\arcdeg$2, Sk$-67\arcdeg$5, and Sk$-68\arcdeg$52 in the LMC. 
The other MC sight lines have ultraviolet fluxes between 8 and 900 times that of the average local interstellar field.  
In view of the large number of massive stars and the low dust-to-gas ratios in the Magellanic Clouds, such high fluxes are not unexpected (e.g., Tumlinson et al. 2002). 
The highest values in the LMC in our sample, for example, are for two sight lines near 30 Dor (Sk$-68\arcdeg$135, Sk$-69\arcdeg$246), while the highest values in the SMC are for two sight lines near the prominent star-forming regions at the southwest end of the main SMC ``bar'' (Sk~13, Sk~18).  

For values $N$(H$_{\rm tot}$) = 2 $\times$ 10$^{21}$ cm$^{-2}$ and $f$(H$_2$) = 0.05 typical of our Magellanic Clouds sample, eqn.~(3) yields $n_{\rm H}$ $\sim$ 55~$I_{uv}$.
For the sight lines with $I_{uv}$ $\la$ 10, the inferred densities thus range between 85 and 560 cm$^{-3}$.
These densities are larger than those found for Galactic sight lines with comparable $I_{uv}$ and molecular fraction, due primarily to the smaller H$_2$ formation rate ($R$) assumed for the LMC and SMC.
Because the density scales with flux, larger densities are obtained for the sight lines with higher $I_{uv}$.  

Bluhm \& de Boer (2001) and Andr\'{e} et al. (2004) reported comparable values for the total $N$(H$_2$) [but slightly different $N$($J$)] toward several of our LMC targets and, using similar analyses, derived physical conditions broadly consistent with those displayed in Table~\ref{tab:phys}.  
Andr\'{e} et al. derived $I_{uv}$ $\sim$ 3, 3000, and 20 toward Sk$-67\arcdeg$5, Sk$-68\arcdeg$135, and Sk$-69\arcdeg$246, respectively (vs. our 2.5, 860, 44); Bluhm \& de Boer estimated $I_{uv}$ $\sim$ 40 and $n_{\rm H}$ $\sim$ 5--280 cm$^{-3}$ in the main component(s) toward Sk$-69\arcdeg$246.
Under the assumption that higher $N$(\ion{Na}{1})/$N$(\ion{H}{1}) ratios result from stronger radiation fields, Cox et al. (2006) inferred relative $I_{uv}$ of 1/3/9 toward Sk$-67\arcdeg$5, Sk$-68\arcdeg$135, and Sk$-67\arcdeg$2 (vs. our 2.5/860/2.4), and low values toward two other stars (Sk$-69\arcdeg$223 and Sk$-69\arcdeg$243) in the vicinity of 30 Dor.
For the column density regime sampled in those Magellanic Clouds sight lines, however, strong radiation fields appear instead to be associated with lower $N$(\ion{Na}{1})/$N$(\ion{H}{1}) ratios (as toward the stars in the Sco-Oph and Orion Trapezium regions; e.g., Welty \& Hobbs 2001).

In view of the possibility (or likelihood) of differences in spatial distribution between the upper and lower H$_2$ rotational populations, however, these estimates for $I_{uv}$ and $n_{\rm H}$ should be taken with some caution (e.g., Browning, Tumlinson, \& Shull 2003).
As already noted, the higher resolution \ion{Na}{1} spectra indicate that most sight lines include multiple components, characterized by a range in column densities.
The observed general relationships between the line profiles and total column densities of \ion{Na}{1} and H$_2$ (e.g., Welty \& Hobbs 2001) suggest that while the bulk of the H$_2$ in any individual sight line is likely to reside in the (few) components with highest $N$(\ion{Na}{1}), the lower column density components also probably contain some H$_2$.
The effective $b$-values for H$_2$, derived from curve-of-growth analyses of the higher rotational levels, are generally much higher than the values that would be expected for a single component at $T_{01}$ (Tumlinson et al. 2002) --- also suggesting that multiple H$_2$ components are present in many of the sight lines.\footnotemark\
Even if a given sight line is dominated by a single cold cloud containing most of the total H$_2$, the upper $J$ populations thus may have significant contributions from more diffuse, lower column density gas --- which often is characterized by a significantly higher rotational temperature for all $J$.
Recent observations of both UV absorption and IR emission from H$_2$ rotational transitions in relatively diffuse gas do suggest that a few percent of the H$_2$ is warm and collisionally excited, perhaps due to MHD shocks and/or turbulence (e.g., Gry et al. 2002; Falgarone et al. 2005).
\footnotetext{While Lacour et al. (2005b) obtained progressively higher $b$-values for the higher H$_2$ $J$ levels in several moderately reddened Galactic lines of sight and argued that the column densities of those levels could be significantly overestimated under the usual assumption of constant $b$, Tumlinson et al. (2002) found no strong evidence for such an effect in their Magellanic Clouds sample.}

When there are additional contributions from such warmer, more diffuse H$_2$, the $I_{uv}$ and $n_{\rm H}$ estimated from integrated sight line rotational populations will be higher than the actual values characterizing the colder, denser cloud(s) containing the bulk of the H$_2$.
For example, $I_{uv}$ and $n$ could be overestimated by factors of order 2 for a line of sight containing one dominant cloud with $N$(H$_2$) $\sim$ 10$^{19}$ cm$^{-2}$, $T_{01}$ = 50~K, $T_{24}$ = 400~K, and $I_{uv}$ $\sim$ 1, plus a second cloud with $N$(H$_2$) $\sim$ 10$^{16}$ cm$^{-2}$ and $T_{04}$ = 400~K.
Conversely, if the higher $J$ levels (which are populated via radiative pumping) are preferentially located in the outer, relatively unshielded layers of the higher column density clouds, then the true $I_{uv}$ could be higher there (but lower in the cloud interiors where most of the lower $J$ gas resides).
Browning et al. (2003) estimated the ratios $N$(4)/$N$(2) and $N$(5)/$N$(3) (which are sensitive to $I_{uv}$) for both single clouds and various combinations of two clouds with different properties. 
For our small sample of sight lines, the observed relative H$_2$ rotational populations appear in most cases to be consistent with those predicted for either the single-cloud or 2-cloud models; only the values for Sk$-68\arcdeg$52 and Sk$-69\arcdeg$191 lie outside the ranges covered by the 2-cloud models. 

\subsubsection{Is CH Due to Steady-State Chemistry?}
\label{sec-chcn}

The nearly linear relationship between the column densities of CH and H$_2$ shown in Figure~\ref{fig:chh2} can be understood via models of steady-state gas-phase chemistry in moderate-density gas, in which the formation of CH is initiated by the radiative association reaction C$^+$~+~H$_2$ $\rightarrow$ CH$^{+}_{2}$ + $h\nu$ and the destruction of CH is due primarily to photodissociation. The range in $N$(CH) at any given $N$(H$_2$) may be due to differences in local density, as reactions with C$^+$ and O$^0$ contribute to the destruction of CH at higher $n_{\rm H}$ (Federman 1982; Danks et al. 1984; Federman \& Huntress 1989).  
The dotted line in Fig.~\ref{fig:chh2}, showing the column densities predicted by models T1--T6 of van Dishoeck \& Black (1989), is quite consistent with the observed column densities (at least for the higher values).
In steady state, we have
\begin{equation}
N({\rm CH}) \sim \frac{0.67~k({\rm C}^+,{\rm H}_2)~x({\rm C}^+)~N({\rm H}_2)~n_{\rm H}}
{I_{uv}~G_{0}({\rm CH})~+~[k({\rm C}^+,{\rm CH})~x({\rm C}^+)~+~k({\rm O}^0,{\rm CH})~x({\rm O}^0)]~n_{\rm H}},
\end{equation}
where the factor 0.67 represents the fraction of dissociative recombinations of CH$^{+}_{3}$ producing CH (Herbst 1978; Vejby-Christensen et al. 1997), $k$(C$^+$,H$_2$) = 1.0 $\times$ 10$^{-16}$ cm$^3$~s$^{-1}$ is the rate coefficient for the initiating reaction (Federman et al. 1997), $x$(C$^+$) and $x$(O$^0$) are the fractional abundances of C$^+$ and O$^0$ (which are assumed to be the dominant forms of carbon and oxygen in these diffuse clouds), $k$(C$^+$,CH) = 5.4 $\times$ 10$^{-10}$ cm$^3$~s$^{-1}$ and $k$(O$^0$,CH) = 9.5 $\times$ 10$^{-11}$ ($T$/300)$^{1/2}$ cm$^3$~s$^{-1}$ are the rate coefficients for CH destruction via C$^+$ and O$^0$ (Zsarg\'{o} \& Federman 2003), and $G_0$(CH) = 1.3 $\times$ 10$^{-9}$ e$^{-\tau_{uv}}$ is the CH photodissociation rate for the average Galactic field, reduced by dust extinction (Federman et al. 1997).  
We assume that the $I_{uv}$ and $T_{01}$ obtained from H$_2$ rotational excitation are appropriate for the gas containing the CH and that local densities can be approximated by column densities.  
The grain attenuation at 1000 \AA, $\tau_{uv}$, is taken to be 3 $\times$ $A_{\rm V}$, where the factor of 3 (instead of 2; Federman et al. 1994) allows for 50\% more attenuation compared to similar sight lines in our Galaxy, as suggested by the respective UV extinction curves (Fitzpatrick \& Savage 1984; Clayton \& Martin 1985; Misselt, Clayton, \& Gordon 1999; Hutchings \& Giasson 2001; Gordon et al. 2003; Cartledge et al. 2005).  
Since $E(B-V)$ $\sim$ 0.1 mag for the sight lines considered here, and since $R_{\rm V}$ = $A_{\rm V}$/$E(B-V)$ generally lies between 2 and 4 in the Magellanic Clouds (Gordon et al. 2003), $\tau_{uv}$ is typically of order unity. 
The values for $\tau_{uv}$ listed for each individual component in Table~\ref{tab:phys} were estimated based on the relative amounts of \ion{Na}{1} in each component, assuming that $N$(H$_{\rm tot}$) [and thus $E(B-V)$] is proportional to [$N$(\ion{Na}{1})]$^{1/2}$ (Welty \& Hobbs 2001).
If individual components are shielded from photodissociating radiation by other components along that sight line, then the effective $\tau_{uv}$ will be larger and the resulting density somewhat lower.

For Galactic sight lines [assuming $I_{uv}$ = 1, $T$ = 50 K, and gas-phase interstellar abundances $x$(C$^+$) = 1.6 $\times$ 10$^{-4}$ and $x$(O$^0$) = 3.0 $\times$ 10$^{-4}$], eqn.~(4) would be consistent with the observed average $N$(CH)/$N$(H$_2$) ratio (Table~\ref{tab:ratgal}) for $E(B-V)$ $\sim$ 0.3 mag and $n_{\rm H}$ $\sim$ 1300 cm$^{-3}$.
In the Magellanic Clouds, however, the lower carbon abundances and generally stronger radiation fields imply that somewhat higher densities would be needed to produce the observed $N$(CH)/$N$(H$_2$).
Assuming the LMC and SMC $x$(C$^+$) and $x$(O$^0$) given above, $E(B-V)$ $\sim$ 0.1 mag, and the average $N$(CH)/$N$(H$_2$) listed in Table~\ref{tab:ratgal}, eqn.~(4) yields $n_{\rm H}$ $\sim$ 4150~$I_{uv}$ for the LMC and $n_{\rm H}$ $\sim$ 3300~$I_{uv}$ for the SMC.
For our small sample of SMC and LMC sight lines, the densities obtained from eqn.~(4) thus are quite large --- typically factors of 10 to 100 higher than those obtained from H$_2$ rotational excitation (columns 8 and 7 of Table~\ref{tab:phys}, respectively).
Even the value for the +278 km~s$^{-1}$ cloud toward Sk$-67\arcdeg$2, where CN is detected, is a factor of $\sim$ 6 higher.
This consistent, significant discrepancy between the two density estimates suggests that the CH may not, in general, be a product of steady-state gas-phase chemistry in these Magellanic Clouds sight lines.

\subsubsection{Is CH Associated with CH$^+$?}
\label{sec-chchp}

We next compare our data, for clouds where at least one of the molecules was detected, to the relationship between CH and CH$^+$ found by Pan et al. (2005), in order to see whether the CH might instead be associated with CH$^+$.  
Tables~\ref{tab:comps},~\ref{tab:compf}, and~\ref{tab:phys} indicate that for most of the clouds in our sample with detectable amounts of both CH and CH$^+$, the ratio of column densities $N$(CH)/$N$(CH$^+$) lies between about 0.7 and 2.0.  
The only exceptions are the two clouds toward Sk$-67\arcdeg$2 
[the one at +278 (the only LMC cloud with significant CN absorption) has $N$(CH)/$N$(CH$^+$) $\sim$ 38, while 
the one at +267 (with relatively strong CH$^+$) has $N$(CH)/$N$(CH$^+$) $\sim$ 0.1]; the cloud toward Sk~143, with $N$(CH)/$N$(CH$^+$) $>$ 14; and the cloud toward AV~476, with $N$(CH)/$N$(CH$^+$) $\sim$ 3.5.  
The cloud toward LH~10-3061 (with strong CH$^+$ but no detectable CH) has $N$(CH)/$N$(CH$^+$) $<$ 0.3.  
For all the other clouds, the observed CH and CH$^+$ columns are consistent with the relationship found by Pan et al. within a factor of 2 (the approximate range allowed by our measurements).  
These results suggest that the chemistry involving CH and CH$^+$ in diffuse clouds in the Milky Way is relevant to gas in the Magellanic Clouds as well.

We now discuss what can be learned from the chemical scheme that leads to CH from CH$^+$.  
We do not explicitly consider the production of CH$^+$, which appears to rely on the endothermic reaction
C$^+$~$+$~H$_2$~$\rightarrow$~CH$^+$~$+$~H (Elitzur \& Watson 1978), but instead study what happens once observable amounts of CH$^+$ are produced.
The most significant pathway involves the reactions
\begin{center}
CH$^+$~$+$~H$_2$~$\rightarrow$~CH$_2^+$~$+$~H, \\
CH$_2^+$~$+$~H$_2$~$\rightarrow$~CH$_3^+$~$+$~H, and \\
CH$_3^+$~$+$~e~$\rightarrow$~CH~$+$~H$_2$~or~2H.
\end{center}
Assuming that steady-state conditions apply for these rapid reactions, we can write the rate equation for CH as
\begin{equation}
N({\rm CH}) \sim \frac{0.67~k({\rm CH}^+,{\rm H}_2)~N({\rm CH}^+)~f({\rm H}_2)~n_{\rm H}}{I_{uv}~G_{0}({\rm CH})},
\end{equation}
where the factor 0.67 (again) represents the fraction of dissociative recombinations producing CH and $k$(CH$^+$,H$_2$) = 1.2 $\times$ 10$^{-9}$ cm$^3$~s$^{-1}$ is the rate coefficient for the initial reaction (Draine \& Katz 1986).  
Photoionization of CH is not included because it recycles CH$^+$.    
Since CH$^+$ is both synthesized and destroyed through reactions involving H$_2$, a modest amount of H$_2$, $f$(H$_2$) $\approx$ 0.05 (comparable to the values found for most of these sight lines), must reside in the regions containing observable quantities of CH$^+$(e.g., Elitzur \& Watson 1978).
Note that eqn.~(5), which relates $N$(CH) to $N$(CH$^+$), does not depend on the elemental abundances for C, N, or O.

Use of the above values for $k$(CH$^+$,H$_2$), $f$(H$_2$), and $G_0$(CH) in eqn.~(5) leads to a simple prediction involving $n_{\rm H}$ and $I_{uv}$ for the diffuse clouds in the Magellanic Clouds.
Since $N$(CH)/$N$(CH$^+$) and $\tau_{uv}$ are both typically about 1, $n_{\rm H}$ $\sim$ 13~$I_{uv}$ cm$^{-3}$.  
If we assume that the UV flux inferred from $N$(4) of H$_2$ applies to the region containing CH$^+$ and use the actual values for $N$(CH)/$N$(CH$^+$) and $\tau_{uv}$, we obtain the gas densities shown in the last column of Table~\ref{tab:phys}.  
The resulting density estimates are somewhat closer to the values derived from H$_2$ rotational excitation than those obtained from steady-state CH chemistry, suggesting that the production of CH in our sample of diffuse clouds in the SMC and LMC may often occur together with that of CH$^+$; the cases with large $I_{uv}$ and $n_{\rm H}$ may be due to heated material in photon-dominated regions.  
In our Galaxy, such a source for CH$^+$ is relatively rare; the gas toward the Pleiades (White 1984) and toward IC~348-20 (Snow 1993) represent the best examples.

\subsubsection{CN Chemistry}
\label{sec:cn}
 
In diffuse clouds, CN is mainly a product of reactions between CH and C$_2$ with N (e.g., Federman et al. 1994).  We use the steady-state analytic expressions of Federman et al. (1994, 1997), with rate coefficients updated by Knauth et al. (2001) and Pan, Federman, \& Welty (2001).  
By reproducing the observed values for $N$(CN), using observed column densities for precursor species whenever possible, values for $n_{\rm H}$ and $I_{uv}$ may be derived.  
The kinetic temperature $T$ is assumed to be 50~K, based on the values for $T_{01}$ from the $J$ $=$ 0 and 1 rotational levels of H$_2$ (Table~\ref{tab:phys}; see also Tumlinson et al. 2002; Andr\'{e} et al. 2004; Cartledge et al. 2005); the chemical results are not very sensitive to $T$, however (Pan et al. 2001). 
The fractional abundances of C, N, and O adopted for the Magellanic Clouds are given above. 
In addition, the grains are again assumed to be somewhat more absorbing at UV wavelengths than those in Galactic diffuse clouds.  

We first consider the sight lines toward Sk~143 and Sk$-67\arcdeg$2, where CN is detected, and then discuss results for other sight lines to compare the values for $n_{\rm H}$ and $I_{uv}$ with those inferred from H$_2$ rotational excitation and from CH$^+$ chemistry.
For $I_{uv}$ $=$ 1, the steady-state chemical model suggests that the CN-containing components toward Sk~143 and Sk$-67\arcdeg$2 have densities of about 1600 cm$^{-3}$ and 1000 cm$^{-3}$, respectively.  
These densities are significantly higher than those inferred from both H$_2$ excitation and CH$^+$ chemistry --- a result reminiscent of those for diffuse clouds in our Galaxy.  
Such a situation arises because CN tends to trace high-density gas (e.g., Joseph et al. 1986), where under steady state conditions the destruction of CH$^+$ is enhanced (Cardelli et al. 1990; see also \S~\ref{sec-disc} below).  
The densities inferred toward Sk~143 and Sk$-67\arcdeg$2 are higher than those seen in our Galaxy (for clouds with comparable amounts of CH and CN) because the elemental abundances for C and N, which are important for CN synthesis, are lower in the SMC and LMC.  
Since photodissociation dominates the destruction of CN (also the result of low elemental abundances), the inferred gas density depends linearly on $I_{uv}$.  
If the values for $I_{uv}$ derived from H$_2$ excitation are applicable for these two sight lines, then the densities estimated from CN will be somewhat lower toward Sk~143 and several times higher toward Sk$-67\arcdeg$2 --- yielding reasonable agreement with the values obtained assuming steady-state formation of CH.
The models that reproduce the CN column densities also lead to predictions for the columns of C$_2$, $N$(C$_2$) $\sim$ 2.5 $\times$ 10$^{13}$ cm$^{-2}$ and 1.8 $\times$ 10$^{13}$ cm$^{-2}$, respectively.  

The sight lines with upper limits for $N$(CN) fall into two categories.  
For sight lines with $N$(CH) $\le$ $2 \times 10^{12}$ cm$^{-2}$ (most of our sample), the CN analysis indicates crude upper limits for $n_{\rm H}$ in excess of 1600 cm$^{-3}$ for $I_{uv}$ = 1.  
At such high densities, the chemical model is no longer appropriate because reactions pertaining to dark clouds are needed, but are not included in the model.  
The combination of lower elemental abundances and small amounts of grain attenuation leads to results that are not especially significant.  
For the two other sight lines in the SMC with CH detections (Sk~18 and AV~476), increasing the amount of attenuation by $\sim$ 30\% to account for the steeper far-UV extinction (e.g., Hutchings \& Giasson 2001) does not change our conclusion.  
More meaningful results are possible for the second set of sight lines, the $+$296 km s$^{-1}$ cloud toward Sk$-68\arcdeg$73 and the $+$277 km s$^{-1}$ cloud toward Sk$-68\arcdeg$135, where the larger amounts of grain attenuation yield upper limits for the gas density of 350 and 950 cm$^{-3}$, respectively.

\subsection{CN Rotational Excitation}
\label{sec-cnrot}

In diffuse clouds, excitation of the lowest rotational levels of CN is due primarily to absorption of cosmic microwave background (CMB) photons, with a possible minor contribution from collisions with electrons (Thaddeus 1972).
Precise observations of the CN rotational level populations thus can provide estimates for the CMB temperature (e.g., Roth \& Meyer 1995).
Alternatively, if that temperature is assumed to be equal to the value measured directly from the CMB spectrum, any excess CN excitation can yield constraints on the local electron density (Black \& van Dishoeck 1991).
Because CN typically traces fairly cold, dense gas, the CN lines can be quite narrow (e.g., Crawford 1995), and even relatively weak lines can be somewhat saturated.
It is therefore important to determine accurate $b$-values for the CN lines, so that accurate column densities may be obtained.

For the component at +278 km~s$^{-1}$ toward Sk$-67\arcdeg$2, $b$ $\sim$ 0.65$\pm$0.05 km~s$^{-1}$ is required to obtain consistent \ion{Na}{1} column densities from the two members of the $\lambda$3302 doublet and from the much stronger D lines; $b$ $\sim$ 0.6$\pm$0.2 km~s$^{-1}$ is obtained from fits to the CH $\lambda$4300 line (Table~\ref{tab:comps}).
If $b$ = 0.6 km~s$^{-1}$ is adopted for CN as well, then fits to the CN lines near 3875 \AA\ yield $N$($J$=0) = 1.31$\pm$0.06 $\times$ 10$^{12}$ cm$^{-2}$ from the R(0) line and $N$($J$=1) = 0.54$\pm$0.05 $\times$ 10$^{12}$ cm$^{-2}$ from the R(1) and P(1) lines.
These column densities are about 20\% and 10\% higher than the values estimated from integrating the ``apparent'' optical depths over the profiles of the lines from $J$=0 and $J$=1, respectively.
Substituting those column densities into the usual Boltzmann equation for the rotational excitation yields an excitation temperature $T_{01}$(CN) = 2.74$^{+0.16}_{-0.13}$~K --- in (perhaps fortuitously) excellent agreement both with the much more precise value 2.725$\pm$0.001 K measured directly from the CMB spectrum recorded by {\it COBE} (Fixsen \& Mather 2002) and with the values derived from CN absorption toward several Galactic stars (Roth \& Meyer 1995).
(Column densities obtained by assuming the CN lines to be optically thin yield a slightly larger $T_{01}$(CN) = 2.93~K.)
While the nominal agreement between $T_{01}$(CN) and the direct CMB value would seem to imply negligible collisional excitation (and thus a relatively low density), the relatively large uncertainties on $T_{01}$(CN) probably allow densities as large as those suggested by the analyses of H$_2$ rotational excitation and CH/CN chemistry (Black \& van Dishoeck 1991).

Fits to the lower S/N CN spectrum of Sk~143 yield $N$($J$=0) = 1.92$\pm$0.21 $\times$ 10$^{12}$ cm$^{-2}$ and $N$($J$=1) = 1.06$\pm$0.20 $\times$ 10$^{12}$ cm$^{-2}$, for an assumed $b$ = 0.5 km~s$^{-1}$.
The corresponding excitation temperature $T_{01}$(CN) = 3.2$^{+0.6}_{-0.3}$ K is slightly higher than the direct CMB value, but is consistent within the 2-$\sigma$ uncertainty.
The relatively low density obtained from the H$_2$ rotational excitation toward Sk~143 suggests that collisions should not make a significant contribution.

\section{Discussion}
\label{sec-disc}

As noted in the introduction, interstellar clouds in the Magellanic Clouds are subject to somewhat different environmental conditions than those typically found in our own local Galactic ISM:

\begin{enumerate}

\item{The overall metallicities, determined via studies of stellar and nebular abundances, are lower by factors of $\sim$ 2 in the LMC and $\sim$ 4--5 in the SMC, relative to solar system values; carbon and nitrogen are somewhat more deficient (e.g., Russell \& Dopita 1992; references in the appendices to Welty et al. 1997, 1999a, and in prep.; Hill 2004).
The interstellar metallicities, for the few sight lines with measured $N$(\ion{S}{2})/$N$(H$_{\rm tot}$) and/or $N$(\ion{Zn}{2})/$N$(H$_{\rm tot}$) ratios, are generally consistent with the stellar and nebular values, when allowance is made for a small amount of depletion of zinc into interstellar dust grains (Welty et al. 1999a, 2001, and in prep.).}

\item{The dust-to-gas ratios are lower by factors of $\sim$ 3 for the LMC and $\sim$ 8 for the SMC (due primarily to the lower metallicities there). 
The mean log[$N$(H$_{\rm tot}$)/$E(B-V)$] (cm$^{-2}$~mag$^{-1}$) is 21.71 for 166 Galactic sight lines, 22.15 for 49 LMC sight lines, and 22.62 for 43 SMC sight lines (Table~\ref{tab:ratgal}) --- where $N$(H$_{\rm tot}$) includes both \ion{H}{1} and H$_2$ (where available), the $N$(\ion{H}{1}) have been determined (mostly) from analyses of Lyman-$\alpha$ absorption, and the Galactic contributions to both $N$(H$_{\rm tot}$) and the color excesses have been removed from the LMC and SMC sight lines (Welty, in prep.). 
Similar values were found (for smaller samples) by Koornneef (1982) for the LMC and by Fitzpatrick (1985) for the SMC.
Note that the slightly higher log[$N$(H$_{\rm tot}$)/$E(B-V)$] listed in Table~\ref{tab:ratgal} for the Trapezium and Sco-Oph regions (relative to the mean Galactic value) are due largely to the higher average values of the ratio of total to selective extinction $R_{\rm V}$ = $A_{\rm V}$/$E(B-V)$ characterizing those regions.  
The average values of $R_{\rm V}$ in the LMC ($\sim$ 3.1) and in the SMC bar ($\sim$ 2.7) (Gordon et al. 2003) suggest that the dust-to-gas ratios in the Magellanic Clouds are actually slightly lower than can be attributed to the differences in overall metallicity alone.
Intriguingly, the relative dust-to-gas ratios in the Milky Way, LMC, and SMC (1.0/0.36/0.12) are very similar to the relative carbon abundances in the three galaxies (1.0/0.33/0.10).}

\item{Because of the lower dust-to-gas ratios and the higher rates of massive star formation per unit area, the ``typical'' ambient interstellar radiation fields are $\sim$ 5 times stronger in the LMC and SMC than in the local Galactic ISM (e.g., Lequeux 1989; see also \S~\ref{sec-h2} above); there are significant local variations, however (Tumlinson et al. 2002; Andr\'{e} et al. 2004; Table~\ref{tab:phys}).
The steeper far-UV extinction found for sight lines in the SMC bar and in the LMC2 region (e.g., Gordon et al. 2003) will affect the shape of the radiation field in the interiors of moderately thick interstellar clouds.}

\end{enumerate} 

\subsection{Structure and Chemistry of Interstellar Clouds}
\label{sec-dischem}

These differences in metallicity, dust-to-gas ratio, extinction, and radiation field are expected to affect the structure, properties, and composition of both atomic and molecular clouds in the Magellanic Clouds:

\begin{enumerate}

\item{In cool, diffuse, neutral atomic clouds, the heating is thought to be dominated by the photoejection of electrons from dust grains (with possible additional contributions from cosmic rays and soft X-rays), while the cooling is due primarily to the radiative decay of collisionally excited levels of \ion{C}{2} and \ion{O}{1}.
The models of Wolfire et al. (1995), which incorporate those processes, suggest that higher pressures are required for the stability of cold, neutral clouds when the metallicity is low and/or when the radiation field is enhanced --- as they are in the Magellanic Clouds.
Observations of the \ion{C}{1} fine-structure populations for the main neutral clouds in several sight lines in the SMC and LMC (Andr\'{e} et al. 2004; Welty et al., in prep.) do seem to imply somewhat higher thermal pressures ($n_{\rm H}T$) than are typical for diffuse Galactic clouds (e.g., Jenkins \& Tripp 2001).
The corresponding densities (assuming the $T_{01}$ from H$_2$) are similar to those obtained from the H$_2$ rotational populations in those sight lines (\S~\ref{sec-h2}).}

\item{The enhanced radiation fields and lower dust abundances in the Magellanic Clouds should also produce more extensive photon-dominated region (PDR) envelopes around molecular clouds (e.g., Johansson 1997; Pak et al. 1998) --- which could be responsible for observed differences in the emission from [\ion{C}{2}], CO, and H$_2$ in several regions of the SMC and LMC (Pak et al. 1998).}

\item{Because H$_2$ is formed on grain surfaces and destroyed primarily via photodissociation, the molecular fractions $f$(H$_2$) should generally be lower in the Magellanic Clouds; the stronger radiation fields there should also produce higher relative populations in the upper H$_2$ rotational levels ($J$ $>$ 2). 
Both effects have been observed by Tumlinson et al. (2002).}

\item{In general, the lower metallicities and stronger radiation fields might be expected to yield lower abundances of trace neutral species, diatomic molecules, and DIB carriers in the Magellanic Clouds.
On the other hand, in moderately reddened clouds, the steeper far-UV extinction could enhance the relative abundances of those atomic and molecular species (e.g., \ion{C}{1}, CN) whose photoionization or photodissociation depends on the strength of the far-UV radiation field (e.g., Jenkins \& Shaya 1979; Cardelli 1988; van Dishoeck \& Black 1989; Welty \& Fowler 1992).
On average, the column densities of both \ion{Na}{1} and \ion{K}{1} are lower, for a given $N$(H$_{\rm tot}$), in the LMC than in our Galaxy, and are even lower in the SMC (Welty, in prep.).
There does not appear to be a detailed correspondence, however, between the abundances of those trace neutral species and the $I_{uv}$ inferred from the H$_2$ rotational populations.
For example, while $N$(\ion{Na}{1}), $N$(\ion{K}{1}), and the equivalent widths of several of the DIBs do appear to be somewhat lower toward Sk~13 and Sk~18 (relative to the other SMC sight lines), they are not significantly lower toward the LMC stars Sk$-68\arcdeg$135 (with the highest inferred $I_{uv}$) and Sk$-69\arcdeg$246.
The column densities of \ion{Na}{1} and \ion{K}{1} might not be significantly reduced in such strong fields, however, if the local hydrogen densities $n_{\rm H}$ (and $n_e$?) are proportional to $I_{uv}$ and if the $N$(\ion{X}{1})/$N$(H$_{\rm tot}$) are proportional to [$I_{uv}$/$n_e$]$^{-1}$ [though Welty et al. (2003) found $n_e$ to be essentially independent of $n_{\rm H}$ in several Galactic sight lines].
The lower dust-to-gas ratios in the LMC and SMC may also contribute to the observed differences, if grain-assisted recombination can significantly enhance the trace neutral species (Lepp et al. 1988; Welty \& Hobbs 2001; Weingartner \& Draine 2001; Liszt 2003).}

\item{Studies of denser molecular material in the Magellanic Clouds (e.g., Johansson et al. 1994; Chin et al. 1998) suggest that differences in the {\it relative} abundances of carbon and oxygen (in particular), can have significant effects on the abundances of various molecular species.
For example, a lower C/O ratio (as found in both the LMC and the SMC) would result in more oxygen left after the formation of CO, and therefore (ultimately) less CN, since CN is destroyed in reactions with O$^{0}$ and O$_2$.
The abundance of any given molecular species thus cannot be assumed to just scale with the overall metallicity --- as it may depend strongly on more subtle differences in the relative abundances, on specific chemical reactions, and on the details of the local physical conditions.}

\item{Theoretical models constructed to interpret the observations of CO emission in the Magellanic Clouds suggest that the CO should be concentrated in relatively small, dense clumps, with little (if any) molecular gas in the lower density interclump regions (e.g., Lequeux et al. 1994; Israel et al. 2003).
The observations of absorption lines from H$_2$ and (in a few cases) CO toward stars outside the CO emission contours indicate, however, that there is some molecular gas outside the dense clumps (Tumlinson et al. 2002; Bluhm \& de Boer 2001; Andr\'{e} et al. 2004).
The highest molecular fractions in the Tumlinson et al. survey generally are in the range $f$(H$_2$) $\sim$ 0.01--0.1 --- most consistent with the values predicted in the low-density ($n_{\rm H}$ = 100 cm$^{-3}$) models.
The CO abundances, however, are closer to those predicted for the dense ($n_{\rm H}$ = 10$^4$ cm$^{-3}$) clumps (Andr\'{e} et al. 2004; though see the comments below).}

\end{enumerate}

While some differences in the properties of the diffuse molecular clouds in our (small) Magellanic Clouds sample may be ascribed to the differences in metallicity and radiation environment, the LMC and SMC clouds do exhibit some intriguing similarities to clouds observed in our Galaxy.  
For example, the Magellanic Clouds sight lines with no detected CN have molecular fractions and column densities of \ion{Na}{1}, \ion{K}{1}, and CH much like those observed toward several comparably reddened Sco-Oph stars ($\delta$, $\beta^1$, and $\omega^1$~Sco; Table~\ref{tab:coldens}), though they also have somewhat lower $N$(CH$^+$).
The two sight lines with detected CN (Sk~143 and Sk$-67\arcdeg$2), on the other hand, have molecular fractions and column densities of \ion{Na}{1}, CH, and CN very similar to those toward $o$ and $\zeta$~Per.
In both cases, the Magellanic Clouds sight lines have higher $N$(\ion{H}{1}) and $N$(H$_2$) (by factors of order 2), but weaker DIBs at 5780, 5797, and 6284 \AA\ (by factors of 2 to 4), than their Galactic counterparts (Table~\ref{tab:ewdib}).  
The C$_2$ DIBs toward Sk~143 and Sk$-67\arcdeg$2 are more comparable in strength to those toward $o$ and $\zeta$~Per, however (Table~\ref{tab:c2dibs}).
In the LMC, the average $N$(CH)/$N$(H$_2$) ratio is essentially identical to that found for diffuse molecular clouds in our Galaxy.
In the SMC, however, CH is less abundant by a factor of 4 to 5 (on average, for our small sample).

For directions in which no CN is detected, CH appears to be associated with the synthesis of CH$^+$, with inferred gas densities usually in the range from 100 to 600 cm$^{-3}$.
Those densities are much larger than the values from 20 to 30 cm$^{-3}$ derived from H$_2$ rotational excitation toward several Sco-Oph stars with similar $E(B-V)$, $f$(H$_2$), and $I_{uv}$ --- due primarily to the smaller H$_2$ formation rate assumed for the Magellanic Clouds.
Somewhat higher densities are inferred for the sight lines with highest $I_{uv}$, and for the components toward Sk~143 and Sk$-67\arcdeg$2 in which CN is detected.

In the LMC and SMC, more of the CH$^+$ (and CH) may be produced in photon-dominated regions.
Such regions are expected to be more extensive in the Magellanic Clouds, due to the lower metallicities and dust-to-gas ratios and to the generally stronger radiation fields there (e.g., Johansson 1997; Pak et al. 1998).  
The connection with photon dominated regions may explain why the $N$(CH)/$N$(H$_2$) ratio for diffuse clouds in the LMC is (on average) not lower than in Galactic clouds --- more extensive PDRs may compensate for the lower metallicity.
The smaller $N$(CH)/$N$(H$_2$) ratio in the SMC may be a consequence of the even smaller carbon abundance there.  

The enhanced H$_2$ excitation in the LMC and SMC provides additional evidence for the presence of warmer material.  
Higher temperatures yield more CH$^+$ (and subsequently CH) through the endothermic reaction C$^+$~$+$~H$_2$~$\rightarrow$~CH$^+$~$+$~H.
The high densities inferred from H$_2$ excitation and CH$^+$ chemistry for several directions may indicate the need to consider time-dependent effects, however, because CH$^+$ is quickly converted to other species when densities are high (Cardelli et al. 1990).  

The CO observed toward Sk$-68\arcdeg$135 and Sk$-69\arcdeg$246 (Bluhm \& de Boer 2001; Andr\'{e} et al. 2004) might also arise in the material containing CH$^+$.  
In a study of diffuse, relatively low-density molecular clouds with no detected CN, Zsarg\'{o} \& Federman (2003) found that the observed abundances of CH and CO could not be produced by steady-state chemistry alone. 
The CO abundances in the denser clouds ($n$ $\sim$ 100 to 200 cm$^{-3}$ --- toward several bright stars in the Sco OB2 Association) could be reproduced, however, through reactions involving the presence of CH$^+$, with the incorporation of a non-thermal (turbulent) component into the ion velocities.  
In view of the similarity in the abundances of CH$^+$, CH, and CO for the two sight lines in the LMC and toward $\delta$, $\beta^1$, and $\omega^1$ Sco, the same processes may be responsible for the CO in all those sight lines.
Toward Sk$-69\arcdeg$246, Andr\'{e} et al. (2004) derive a gas density of 300 to 600 cm$^{-3}$ from analysis of neutral carbon fine-structure excitation.  
That density, though lower than our estimate from H$_2$ for the photon-dominated region, is about what would be expected for CO production involving CH$^+$.
If CO formation via CH$^+$ is effective in moderate-density gas, then the very large densities ($\sim$ 10$^4$ cm$^{-3}$) suggested by Andr\'{e} et al. (2004) may not be required to produce the CO abundances observed in those diffuse molecular clouds. 
Additional observations of CH, CH$^+$, and CO in the Magellanic Clouds are needed to confirm and quantify these relationships.

\subsection{Behavior of Diffuse Interstellar Bands}
\label{sec-disdibs}

The general correlations of DIB strengths with the column densities of various other constituents of the ISM and the differences in DIB behavior seen for different regions suggest that the DIBs might (in principle) be used both as quantitative tracers of diffuse interstellar material and as diagnostics of the physical conditions characterizing that material (e.g., the strength of the ambient radiation field, the molecular fraction, extinction characteristics).
For the DIBs to be used in those ways, however, the specific dependences of DIB strength on the abundances of other species and on the local environmental conditions must be better understood.
For example, in the local Galactic ISM, the equivalent widths of the DIBs at 5780, 5797, and 6284 \AA\ generally seem to be closely related to both $N$(\ion{H}{1}) and $E(B-V)$ --- with similar correlation coefficients, slopes of order unity, and small scatter about the mean relationships (Fig.~\ref{fig:h1ebv}; Table~\ref{tab:corr}).
There appear to be real variations in the $N$(\ion{H}{1})/$E(B-V)$ ratio (both between our Galaxy and the Magellanic Clouds and within each galaxy), however, and it is not entirely clear whether the basic, primary correlations are with the amount of gas [$N$(\ion{H}{1})] or with the amount of dust [$E(B-V)$].
While it can be advantageous to compare ``normalized'' DIB equivalent widths (e.g., in attempts to identify environmental factors which affect the DIBs), it is thus not clear whether $W$(DIB) should be normalized by $N$(\ion{H}{1}) or by $E(B-V)$.
Resolving that issue might provide clues as to the origin of the DIBs (e.g., whether the DIB carriers are assembled in the gas-phase or on the surfaces of grains) and could also allow the 5780 \AA\ DIB to be used to estimate $N$(\ion{H}{1}) when that cannot be measured directly.

Because of the lower overall metallicities, the (typically) lower dust-to-gas ratios and molecular fractions, and the generally stronger radiation fields in the Magellanic Clouds, one might expect the DIBs to be weaker there.
The currently available data indicate that the DIBs at 5780, 5797, and 6284 \AA\ are generally weaker [on average and relative to $N$(\ion{H}{1})] than their counterparts in the Galactic ISM --- by factors of 7--9 in the LMC and factors of order 20 in the SMC (Ehrenfreund et al. 2002; Sollerman et al. 2004; Cox et al. 2006; this paper).
Those Magellanic Clouds DIBs are also weaker by factors of order 2 (on average), relative to $E(B-V)$, [$N$(\ion{Na}{1})]$^{1/2}$, and [$N$(\ion{K}{1})]$^{1/2}$.
The differences in $W$(DIB)/$N$(\ion{H}{1}) are larger than the differences in metallicity alone --- and also larger than the differences in carbon abundance --- so that one (or more) of the other environmental factors must play a role as well.
There is significant scatter, however, and there are sight lines in the LMC and SMC in which the DIBs are nearly as strong as they are in Galactic sight lines with comparable $N$(H$_{\rm tot}$) and $E(B-V)$.
Moreover, the Magellanic Clouds C$_2$ DIBs toward Sk~143 and Sk$-67\arcdeg$2 appear to be similar in strength to those in Galactic sight lines with comparable $E(B-V)$ and column densities of \ion{H}{1}, \ion{Na}{1}, H$_2$, CN, and CH --- even though the corresponding DIBs at 5780, 5797, and 6284 \AA\ are weaker than in the Galactic clouds.
As for the molecular abundances discussed above, there thus does not seem to be a simple, uniform relationship between DIB strength and metallicity that would apply to all DIBs in all Magellanic Clouds sight lines.

\begin{figure}
\epsscale{0.4}
\plotone{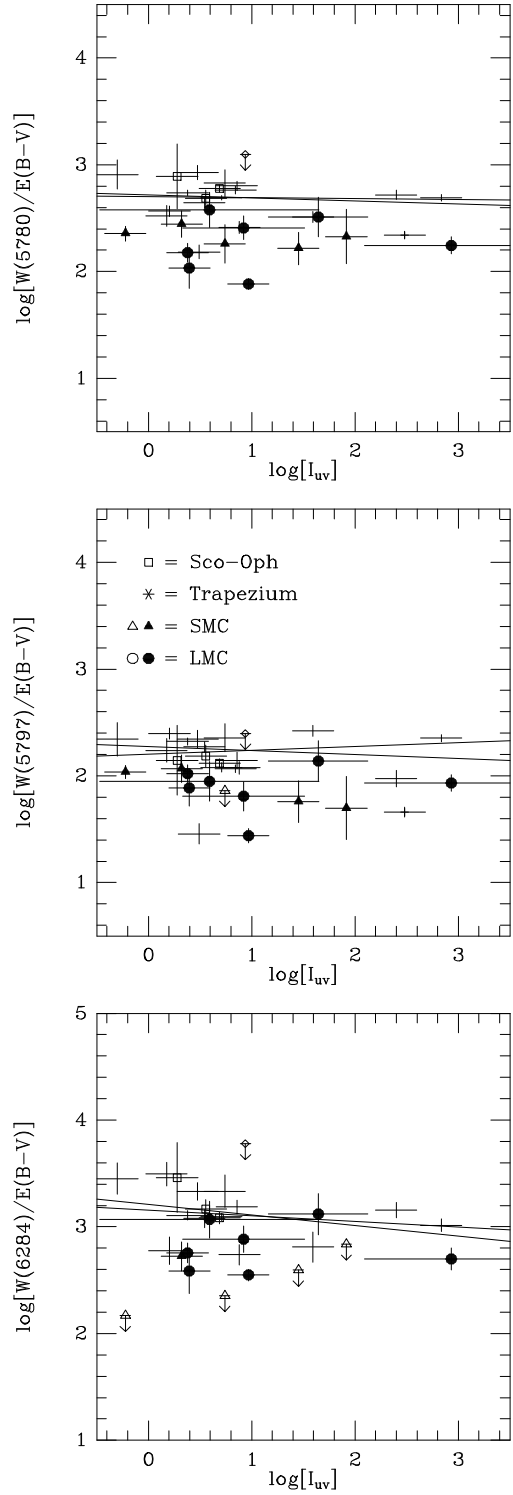}
\caption{Normalized equivalent widths of the DIBs at 5780, 5797, and 6284 \AA\, versus the intensity of the ambient radiation field ($I_{uv}$) inferred from the rotational excitation of H$_2$.
Circles denote LMC sight lines; triangles denote SMC sight lines; all others are Galactic sight lines (see caption to Fig.~7).
While the DIBs are generally weaker in the LMC and SMC, no significant trends with $I_{uv}$ are apparent in any of the three galaxies.}
\label{fig:dibiuv}
\end{figure}

Observations of the DIBs in the Galactic ISM seem to indicate that the DIB equivalent widths can be affected by the strength of the ambient radiation field, but the exact dependences and the corresponding implications for the DIB carriers are not yet understood.
The DIBs can be weak where the ambient fields are either weak or strong, and different DIBs appear to exhibit different dependences on various measures of field strength (e.g., Adamson, Whittet, \& Duley 1991; Snow et al. 1995; Cami et al. 1997; Sonnentrucker et al. 1997).
For example, relative to $N$(\ion{H}{1}), the DIBs are weaker (on average) in both the Sco-Oph and Orion Trapezium regions, where the presence of numerous early-type stars and lower than average abundances of both trace neutral species (e.g., \ion{Na}{1}, \ion{K}{1}) and molecules (e.g., H$_2$, CH) suggest that the radiation fields might be enhanced (Herbig 1993; Jenniskens et al. 1994; Welty \& Hobbs 2001).
Analysis of the H$_2$ rotational levels toward $\delta$~Sco and $\sigma$~Sco yields $I_{uv}$ $\sim$ 3--5 and fairly low densities [corresponding to the lower $f$(H$_2$) there; Table~\ref{tab:phys}]; H$_2$ data are not available for the three Trapezium sight lines.
In general, however, there does not seem to be an obvious detailed correspondence between the DIB strengths and the specific $I_{uv}$ estimated from the H$_2$ rotational excitation in individual lines of sight.
For example, the DIBs (and the trace neutral atomic and molecular species) do not appear to be weaker toward stars in the Cep OB2 association, even though analyses of the H$_2$ rotational level populations listed by Pan et al. (2005) (as in \S~\ref{sec-h2}) suggest quite strong radiation fields for several of those sight lines.
Figure~\ref{fig:dibiuv} plots the normalized DIB equivalent widths $W$(DIB)/$E(B-V)$ versus $I_{uv}$ for a relatively small set of sight lines in the Milky Way and Magellanic Clouds (references given in \S~\ref{sec-h2} above).
For all three DIBs (at 5780, 5797, and 6284 \AA), the slopes of the fits to the Galactic data (solid lines) are consistent (at the 1--2-$\sigma$ level) with 0.0; there also are no significant trends for the SMC or LMC sight lines.
At some level, the lack of detailed correspondence between the strength of those three DIBs and $I_{uv}$ (as estimated from H$_2$) might reflect the lack of association between those DIBs and the H$_2$, but it seems unlikely that $I_{uv}$ would be completely irrelevant to the \ion{H}{1} gas (and related DIBs) in a given sight line.

Various DIBs have now been observed in several extragalactic systems --- toward SN in other galaxies (D'Odorico et al. 1989; Sollerman et al. 2004), in the winds emanating from starburst galaxies (Heckman \& Lehnert 2000), and in at least one QSO absorption-line system (Junkkarinen et al. 2004).  
While comparisons of DIB behavior in those environments with the trends observed in the Milky Way and Magellanic Clouds might provide some clues as to the physical conditions in those more distant systems, the generally limited spectral resolution, S/N, and wavelength coverage of the existing spectra can make it difficult to determine accurate abundances for the other species (e.g., \ion{Na}{1}, \ion{H}{1}) needed for such comparisons.
For example, while the 5780 and 5797 \AA\ DIBs are about as strong in the ISM of NGC 5128 toward SN 1986G as they are in Galactic sight lines with comparable $E(B-V)$ (D'Odorico et al. 1989), the corresponding $N$(\ion{Na}{1}) $\sim$ 5.5 $\times$ 10$^{13}$ cm$^{-2}$, obtained from fits to the low resolution (FWHM $\sim$ 15--17 km~s$^{-1}$) spectrum of the D lines, should probably be viewed as a lower limit.
Toward SN 2001el (NGC 1448), the 5780, 5797, and 6284 \AA\ DIBs are quite comparable in strength to Galactic DIBs, relative to $N$(\ion{K}{1}) (Sollerman et al. 2004), but the relatively low $N$(\ion{Na}{1})/$N$(\ion{K}{1}) ratio ($\sim$ 30) suggests that the \ion{Na}{1} column density may be somewhat underestimated there as well.  
And although Heckman \& Lehnert (2000) found the 5780 and 6284 \AA\ DIBs in the galactic winds seen toward several starburst nuclei to be similar in strength to those in our Galaxy, relative to $N$(\ion{Na}{1}), the \ion{Na}{1} column densities appear to have been underestimated by factors of at least 2--3.\footnotemark\
In those cases, the DIBs likely are actually somewhat weaker, relative to the true $N$(\ion{Na}{1}) and $N$(\ion{K}{1}), than those in our Galaxy --- but more similar to those in the Magellanic Clouds --- perhaps as a result of enhanced radiation fields.
Finally, while there is currently only one reported detection of a DIB in a QSO absorption line system --- that of the 4430 \AA\ DIB in the $z$ = 0.524 damped Lyman-$\alpha$ system toward the BL Lac object AO~0235+164 (Junkkarinen et al. 2004), additional searches are underway (e.g., Lawton et al. 2005).
If the DIBs are due to large carbon-based molecules, it will be very interesting to see when they first appeared.
\footnotetext{The \ion{Na}{1} column densities were estimated via the doublet ratio method, using the D-line equivalent widths measured in the low resolution (FWHM $\sim$ 55--170 km~s$^{-1}$) spectra (Heckman et al. 2000).
In several cases, the $N$(\ion{Na}{1}) were ``checked'' by multiplying the column densities of \ion{K}{1} (determined from the $\lambda$7664 line) by 15 (the solar Na/K abundance ratio) --- which yielded values 2--3 times smaller than those obtained from the doublet ratio.
That scaling, however, appears to have omitted the differences in the photoionization and recombination rates for \ion{Na}{1} and \ion{K}{1} --- which typically add another factor $\sim$ 6 to the \ion{Na}{1}/\ion{K}{1} column density ratio in the Galactic ISM (e.g., Welty \& Hobbs 2001; though see also Kemp et al. 2002).
In addition, the \ion{K}{1} $\lambda$7664 line may well be somewhat saturated.}

\section{Summary / Conclusions}
\label{sec-summ}

We have discussed the molecular absorption lines and diffuse interstellar bands seen in high S/N, moderately high resolution optical spectra of 7 SMC and 13 LMC stars, most of which also exhibit relatively strong absorption from H$_2$ and \ion{Na}{1}.  
Absorption from CH $\lambda$4300 and/or CH$^+$ $\lambda$4232 has now been detected toward 3 SMC and 9 LMC stars, with typical equivalent widths of 1--5 m\AA; CN $\lambda$3874 is detected only toward Sk~143 (SMC) and Sk$-67\arcdeg$2 (LMC). 
These data represent nearly all the reported detections of optical molecular absorption lines in interstellar media beyond our Galaxy.
Column densities for CH, CH$^+$, and CN have been determined from the equivalent widths, by integrating over the ``apparent'' optical depth, and from multi-component fits to the line profiles.
Upper limits to $N$(CH) and the equivalent width of the DIB at 5780 \AA\ were obtained for an additional 25 sight lines, most of which have lower $E(B-V)$ and $N$(\ion{Na}{1}) than those characterizing our primary sample.

Column density relationships between various atomic and molecular species in the LMC and SMC have been compared with the corresponding trends found in the local Galactic ISM and in two more restricted regions (Sco-Oph and the Orion Trapezium) known to exhibit somewhat different abundances of trace neutral atomic and molecular species.
In the LMC, the relationships among the column densities of H$_2$, CH, \ion{Na}{1}, and \ion{K}{1} are similar to those found in Galactic diffuse clouds; the column densities of those species (and of CH$^+$, CN, and CO) in many cases resemble those seen toward several similarly reddened stars in the Sco-Oph region.
In the SMC, however, CH is weaker by about a factor of 4 to 5 (on average, but with a wide range), relative to H$_2$.
The abundance of CH in diffuse molecular material in the Magellanic Clouds thus appears to depend on local physical conditions --- and not just on metallicity.
In both the LMC and the SMC, the relationships between the total column density of CH and those of CN and CH$^+$ are similar to the trends found in the Galactic ISM --- though the $N$(CH)/$N$(CH$^+$) ratio can vary significantly for individual components (in all three galaxies).

Physical conditions in the diffuse molecular clouds in the LMC and SMC have been inferred both from the relative H$_2$ rotational populations and, using simple chemical models, from the abundances of CH, CH$^+$, and CN.
The H$_2$ rotational temperature $T_{01}$, obtained from the populations in the $J$ = 0 and 1 levels, ranges from 45 to 90 K.
The strength of the ambient radiation field, gauged from the relative H$_2$ population in $J$ = 4, is in most cases between 1 and 10 times the average local Galactic field; the local hydrogen densities $n_{\rm H}$ inferred for the diffuse molecular material in those sight lines range from 100 to 600 cm$^{-3}$.
Stronger radiation fields (30 to 900 times the local field) and correspondingly higher local densities are found for several sight lines near the prominent \ion{H}{2} region complexes in the southwestern part of the SMC and near 30 Dor in the LMC.
Much higher local densities are inferred if it is assumed that the observed CH is due to steady-state gas-phase chemical reactions; much better agreement with the values derived from H$_2$ is found if CH is instead primarily formed via the non-thermal process(es) responsible for the observed CH$^+$.
A significant fraction of the CH and CH$^+$ observed in diffuse molecular material in the Magellanic Clouds may arise in photon-dominated regions (which should be more extensive there than in our Galaxy).
Non-thermal chemistry may also contribute to the CO observed in several of these sight lines --- perhaps removing the need for very small clumps of high-density gas.

The diffuse interstellar bands at 5780, 5797, and/or 6284 \AA\ are detected, at velocities similar to those of the most prominent LMC and SMC components seen in \ion{Na}{1}, toward all 20 stars in our primary sample.
The three DIBs generally are weaker by factors of 7 to 9 (LMC) and about 20 (SMC), compared to those typically observed in Galactic sight lines with similar $N$(\ion{H}{1}).
The three DIBs are also weaker by factors of about 2 to 6, relative to $E(B-V)$ and to the column densities of \ion{Na}{1} and \ion{K}{1} in the Magellanic Clouds. 
While the relative weakness of those DIBs in the Magellanic Clouds is likely related to the lower metallicities and generally stronger radiation fields in the two galaxies, there is no apparent relationship between the strengths of the DIBs and the strength of the radiation field ($I_{uv}$) inferred from the relative rotational populations of H$_2$.
Moreover, the C$_2$ DIBs (relatively weak, narrow DIBs that are enhanced in sight lines with strong absorption from C$_2$ and CN) seen toward Sk~143 and Sk$-67\arcdeg$2 are similar in strength to those in Galactic sight lines with comparable $N$(\ion{H}{1}), $N$(H$_2$), and $E(B-V)$.
As for CH, local physical conditions (and not just metallicity) appear to play a significant role in determining the relative strengths of the various DIBs in individual lines of sight.

Within our relatively small sample, the lines of sight toward Sk~143 (SMC) and Sk$-67\arcdeg$2 (LMC) exhibit several unique characteristics.
They have
(1) the highest molecular fractions, with log[$f$(H$_2$)] = $-$0.28 and $-$0.19;
(2) the highest $N$(CH)/$N$(CH$^+$) ratios ($>$14 and 37.6);
(3) the lowest $I_{uv}$ in the SMC and LMC (0.6 and 2.4 --- inferred from the H$_2$ rotational populations);
(4) the sole SMC and LMC detections of CN $\lambda$3874 absorption; and
(5) the strongest SMC and LMC absorption from several of the ``C$_2$ DIBs''.  
The CN rotational excitation toward Sk$-67\arcdeg$2 corresponds to a temperature $T_{01}$(CN) = 2.74$^{+0.16}_{-0.13}$ K --- quite consistent with the temperature of the cosmic microwave background derived from {\it COBE} data and from observations of CN absorption in several Galactic sight lines.
This appears to be the first extragalactic measurement of the CMB temperature via CN absorption.

\acknowledgements

We thank P. Crowther for contributing the UVES spectra (runs C02 and C04) obtained for studies of SMC and LMC stellar properties.
D.E.W. thanks M. Rejkuba and M.-R. Cioni (ESO/Paranal) for providing essential guidance with the instrument and local software during run V03.
This work has been supported by NASA Long-Term Space Astrophysics grant NAG5-11413 to the University of Chicago.


\clearpage

\begin{deluxetable}{llcclclrc}
\tabletypesize{\scriptsize}
\tablecolumns{9}
\tablecaption{SMC and LMC Sight lines \label{tab:los}}
\tablewidth{0pt}

\tablehead{
\multicolumn{1}{l}{Star}& 
\multicolumn{1}{l}{ }& 
\multicolumn{2}{c}{RA (J2000) DEC}& 
\multicolumn{1}{c}{$V$\tablenotemark{a}}& 
\multicolumn{1}{c}{$E(B-V)$\tablenotemark{b}}& 
\multicolumn{1}{l}{Type\tablenotemark{a}}&
\multicolumn{1}{c}{t$_{\rm exp}$}&
\multicolumn{1}{l}{Runs}\\ 
\multicolumn{1}{c}{ }&  
\multicolumn{1}{c}{ }&  
\multicolumn{1}{c}{($^{h~m~s}$)}&  
\multicolumn{1}{c}{(\arcdeg~'~") }&  
\multicolumn{1}{c}{ }&  
\multicolumn{1}{c}{total/MC}&  
\multicolumn{1}{c}{ }&  
\multicolumn{1}{c}{(min)}}  

\startdata
Sk 13             & AV 18   & 00 47 12.0 & -73 06 31 & 12.45 & 0.18/0.14 & B2 Ia       & 160 & V03\\
Sk 18             & AV 26   & 00 47 50.1 & -73 08 21 & 12.51 & 0.12/0.08 & O7 III      & 160 & V03\\
Sk 40             & AV 78   & 00 50 38.4 & -73 28 19 & 11.05 & 0.11/0.07 & B1.5 Ia$^+$ &  20 & V03\\
                  & AV 80   & 00 50 43.8 & -72 47 41 & 13.33 & 0.16/0.12 & O4-6n(f)p   &  50 & C04\\
Sk~143            & AV~456  & 01 10 55.6 & -72 42 58 & 12.87 & 0.38/0.34 & O9.7 Ib     & 120 & E01\\
                  & AV~476  & 01 13 42.4 & -73 17 29 & 13.54 & 0.22/0.18 & O6.5 V      & 120 & V03\\
Sk~155            & AV~479  & 01 14 50.0 & -73 20 17 & 12.45 & 0.15/0.11 & O9 Ib       & 180 & V03\\
 & \\ 
LH~10-3061        &         & 04 56 42.5 & -66 25 18 & 13.68 & 0.29/0.26 & O2 III(f*)  &  67 & C04\\
Sk$-67\arcdeg$2   &HD270754 & 04 47 04.4 & -67 06 53 & 11.26 & 0.24/0.22 & B1 Ia$^+$   &  60 & V03\\
Sk$-67\arcdeg$5   &HD268605 & 04 50 19.0 & -67 39 38 & 11.34 & 0.16/0.13 & O9.7 Ib     &  60 & V03\\
                  &         &            &           &       &           &             &  30 & V04\\
BI~237            &         & 05 36 14.7 & -67 39 19 & 13.93 & 0.18/0.12 & O2 V((f*))  &  73 & C04\\
Sk$-68\arcdeg$52  &HD269050 & 05 07 20.6 & -68 32 10 & 11.54 & 0.15/0.09 & B0 Ia       &  90 & V03\\
                  &         &            &           &       &           &             &  50 & V04\\
Sk$-68\arcdeg$73  &HD269445 & 05 22 59.9 & -68 01 47 & 11.46 & 0.57/0.51 & Ofpe/WN9    & 120 & V03\\
Sk$-68\arcdeg$135 &HD269896 & 05 37 48.6 & -68 55 08 & 11.35 & 0.27/0.21 & ON9.7 Ia$^+$& 105 & V03\\
                  &         &            &           &       &           &             &  40 & V04\\
Sk$-69\arcdeg$191 &HD 37680 & 05 34 19.4 & -69 45 10 & 13.35 & 0.15/0.09 & WC4         &  80 & V04\\
Sk$-69\arcdeg$202 &SN 1987A & 05 35 26.7 & -69 16 14 &\nodata& 0.16/0.10 & (B3 Iab)    &     &    \\
BI~253            &         & 05 37 35.5 & -69 01 10 & 13.82 & 0.19/0.13 & O2 V((f*))  &  73 & C04\\
Melnick 42        &         & 05 38 42.1 & -69 05 55 & 10.96 & 0.42/0.36 & O3If*/WN6-A &  20 & C02\\
Sk$-69\arcdeg$246 &HD 38282 & 05 38 54.0 & -69 02 00 & 11.15 & 0.14/0.08 & WN6h        &  90 & V03\\
Sk$-70\arcdeg$115 &HD270145 & 05 48 49.8 & -70 03 58 & 12.24 & 0.20/0.14 & O6.5 Iaf    & 150 & V03\\
                  &         &            &           &       &           &             &  20 & C02\\
\enddata
\tablenotetext{a}{Most photometry and spectral types are from Danforth et al. 2002, Tumlinson et al. 2002, and Welty (in prep.), who list the original references.
Data for the other sight lines are from Brunet et al. 1975; Walborn et al. 2000, 2002; Massey 2002; and Massey et al. 2005.}
\tablenotetext{b}{Total $E(B-V)$ are based on the intrinsic colors of Fitzpatrick \& Garmany 1990.
MC $E(B-V)$ assume a Milky Way foreground of 0.04 mag for the SMC (Schlegel et al. 1998) and 0.02--0.06 mag for the LMC (Staveley-Smith et al. 2003).}
\end{deluxetable}

\clearpage

\begin{deluxetable}{lcrrrrrrc}
\tabletypesize{\scriptsize}
\tablecolumns{8}
\tablecaption{Equivalent Widths for CN, CH, and CH$^+$ \label{tab:ewmol}}
\tablewidth{0pt}

\tablehead{
\multicolumn{1}{l}{Star}&
\multicolumn{1}{c}{S/N\tablenotemark{a}}&
\multicolumn{2}{c}{CN B-X(0,0)}&
\multicolumn{2}{c}{CH }&
\multicolumn{2}{c}{CH$^+$}&
\multicolumn{1}{c}{Run/Ref\tablenotemark{b}}\\
\multicolumn{2}{c}{ }&
\multicolumn{1}{c}{R(1)}&
\multicolumn{1}{c}{R(0)}&
\multicolumn{1}{c}{A-X(0,0)}&
\multicolumn{1}{c}{B-X(0,0)}&
\multicolumn{1}{c}{A-X(0,0)}&
\multicolumn{1}{c}{A-X(1,0)}\\
\multicolumn{2}{c}{$\lambda$}&  
\multicolumn{1}{c}{3873.999}&
\multicolumn{1}{c}{3874.607}&
\multicolumn{1}{c}{4300.313}&
\multicolumn{1}{c}{3886.410}&
\multicolumn{1}{c}{4232.548}&
\multicolumn{1}{c}{3957.692}\\
\multicolumn{2}{c}{$f$}&
\multicolumn{1}{c}{0.02250}&
\multicolumn{1}{c}{0.03380}&
\multicolumn{1}{c}{0.00506}&
\multicolumn{1}{c}{0.00330}&
\multicolumn{1}{c}{0.00550}&
\multicolumn{1}{c}{0.00340}
}
\startdata
Sk~13                           & 245 & $<$1.3       & $<$1.3       &
                                        $<$1.8       & $<$1.1       &
                                        $<$1.2       & \nodata      & V03 \\
Sk~18                           & 255 & $<$1.3       & $<$1.3       &
                                         1.0$\pm$0.5 & $<$1.2       &
                                        $<$1.3       & \nodata      & V03 \\
Sk~40                           & 165 & $<$2.1       & $<$2.1       &
                                        $<$2.5       & $<$1.8       &
                                        $<$2.0       & \nodata      & V03 \\
AV~80                           & 130 & $<$2.9       & $<$2.9       &
                                        $<$2.5       & \nodata      &
                                        $<$2.7       & \nodata      & C04 \\
Sk~143\tablenotemark{c}         & 140 &  2.8$\pm$0.8 & 6.3$\pm$0.8  &
                                        20.7$\pm$0.6 & 6.6$\pm$0.6  &
                                        $<$2.6       & \nodata      & E01 \\
AV~476                          & 135 & $<$2.7       & $<$2.7       &
                                         3.0$\pm$0.8 & 1.0$\pm$0.6  &
                                         1.1$\pm$0.7 & \nodata      & V03 \\
Sk~155                          & 270 & $<$1.2       & $<$1.2       &
                                        $<$1.5       & $<$1.3       &
                                        $<$1.2       & \nodata      & V03 \\
 & \\
LH~10-3061\tablenotemark{d}     & 120 & $<$3.1       & $<$3.1       &
                                        $<$2.8       & \nodata      &
                                        13.0$\pm$1.2 & 9.9$\pm$1.1  & C04 \\
Sk$-67\arcdeg$2\tablenotemark{e}& 235 &  1.6$\pm$0.3 & 4.9$\pm$0.4  &
                                        10.5$\pm$0.8 & 2.9$\pm$0.3  &
                                         6.0$\pm$0.6 & 3.0$\pm$0.4  & V03 \\
Sk$-67\arcdeg$5                 & 265 & $<$1.3       & $<$1.3       &
                                        $<$1.6       & $<$1.2       &
                                        $<$1.3       & $<$1.1       & V03 \\
                                & 240 & $<$1.8       & $<$1.8       &
                                         0.9$\pm$0.4 & $<$1.6       &
                                        $<$1.5       & $<$1.4       & V04 \\
BI~237                          & 115 & $<$3.1       & $<$3.1       &
                                        $<$2.9       & \nodata      &
                                        $<$3.2       & \nodata      & C04 \\
Sk$-68\arcdeg$52                & 285 & $<$1.2       & $<$1.2       &
                                         1.2$\pm$0.5 & $<$1.0       &
                                         0.9$\pm$0.4 & 0.4$\pm$0.3  & V03 \\
                                & 205 & $<$1.9       & $<$1.9       &
                                         1.0$\pm$0.5 & $<$1.8       &
                                         0.6$\pm$0.4 & $<$1.6       & V04 \\
Sk$-68\arcdeg$73                & 285 & $<$1.5       & $<$1.5       &
                                         5.0$\pm$0.6 & 1.7$\pm$0.5  &
                                         2.5$\pm$0.4 & 1.3$\pm$0.3  & V03 \\
Sk$-68\arcdeg$135               & 270 & $<$1.1       & $<$1.1       &
                                         3.1$\pm$0.5 & 0.5$\pm$0.4  &
                                         2.7$\pm$0.5 & 1.5$\pm$0.4  & V03 \\
                                & 235 & $<$1.8       & $<$1.8       &
                                         2.4$\pm$0.7 & $<$1.6       &
                                         2.4$\pm$0.5 & 0.8$\pm$0.5  & V04 \\
Sk$-69\arcdeg$191               &  85 & $<$4.8       & $<$4.8       &
                                        $<$4.5       & $<$3.8       &
                                         2.7$\pm$1.2 & $<$3.8       & V04 \\
Sk$-69\arcdeg$202               & 435 & \nodata      & \nodata      &
                                         0.9$\pm$0.2 & \nodata      &
                                         0.4$\pm$0.2 & \nodata      & MG87 \\
BI~253                          & 130 & $<$2.8       & $<$2.8       &
                                        $<$2.6       & \nodata      &
                                        $<$2.9       & \nodata      & C04 \\
Melnick~42                      &  95 & $<$4.5       & $<$4.5       &
                                        $<$3.3       & \nodata      &
                                        $<$3.6       & \nodata      & C02 \\
Sk$-69\arcdeg$246               & 325 & $<$1.1       & $<$1.1       &
                                         1.7$\pm$0.4 & 0.7$\pm$0.3  &
                                         2.6$\pm$0.5 & 1.6$\pm$0.4  & V03 \\
Sk$-70\arcdeg$115               & 295 & $<$1.1       & $<$1.1       &
                                         0.8$\pm$0.5 & $<$1.1       &
                                         0.5$\pm$0.3 & $<$1.0       & V03 \\
\enddata
\tablecomments{Equivalent widths are in m\AA.  
Uncertainties are 1-$\sigma$ (photon noise + continuum uncertainty); limits are 3-$\sigma$.}
\tablenotetext{a}{Average S/N per half resolution element near 3875, 4232, and 4300 \AA.}
\tablenotetext{b}{V03, V04, C02, C04, E01 = UVES (this paper); MG97 = Magain \& Gillet 1987.}
\tablenotetext{c}{Other lines for Sk~143:
CN B-X(0,0) P(1) $\lambda$3875.764 ($f$ = 0.0113; $W$ = 1.9$\pm$0.7), CH B-X(0,0) $\lambda$3878.768 ($f$ = 0.0011; $W$ = 2.6$\pm$0.8), CH B-X(0,0) $\lambda$3890.213 ($f$=0.0022; $W$ = 4.9$\pm$0.8).}
\tablenotetext{d}{Other lines for LH~10-3061: 
CH$^+$ A-X(2,0) $\lambda$3745.310 ($f$ = 0.0014; $W$ = 5.0$\pm$1.4).}
\tablenotetext{e}{Other lines for Sk$-67\arcdeg$2: 
CN B-X(0,0) P(1) $\lambda$3875.764 ($W$ = 0.5$\pm$0.3), CH B-X(0,0) $\lambda$3878.768 ($W$ = 1.1$\pm$0.3).}
\end{deluxetable}

\clearpage

\begin{deluxetable}{lcccccc}
\tabletypesize{\scriptsize}
\tablecolumns{7}
\tablecaption{Diffuse Bands at 4963, 5780, 5797, and 6284 \AA\ \label{tab:ewdib}}
\tablewidth{0pt}

\tablehead{
\multicolumn{1}{l}{Star}&
\multicolumn{1}{c}{$E(B-V)$}&
\multicolumn{1}{c}{$\lambda$4963}&
\multicolumn{1}{c}{$\lambda$5780}&
\multicolumn{1}{c}{$\lambda$5797}&
\multicolumn{1}{c}{$\lambda$6284}&
\multicolumn{1}{c}{Run/Ref\tablenotemark{a}}}
\startdata
Sk~13             & 0.14 &  $<$4.6    &  23$\pm$ 5 &  8$\pm$ 2 &  $<$55      & V03 \\
Sk~18             & 0.08 &  $<$4.4    &  17$\pm$ 7 &  4$\pm$ 2 &  $<$55      & V03 \\
Sk~40             & 0.07 &  $<$7.7    &  20$\pm$ 8 &  $<$12    &  70$\pm$25  & V03 \\
AV~80             & 0.12 &  $<$11     &  22$\pm$14 &  $<$21    &  $<$80      & C04 \\
Sk~143            & 0.34 &  19$\pm$ 2 &  77$\pm$10 & 37$\pm$ 4 & \nodata     & E01 \\
                  &      &  \nodata   & (98$\pm$32)&(42$\pm$16)& ($<$50)     & E02 \\
AV~476            & 0.18 & 7.3$\pm$2.2&  50$\pm$11 & 21$\pm$ 5 &  95$\pm$25  & V03 \\
Sk~155            & 0.11 &  $<$4.3    &  20$\pm$ 6 &  $<$8     &  $<$60      & V03 \\
 & \\
LH~10-3061        & 0.26 &  $<$11     &  83$\pm$20 & 23$\pm$ 7 & 225$\pm$35  & C04 \\
Sk$-67\arcdeg$2   & 0.22 & 9.5$\pm$1.7&  33$\pm$ 5 & 23$\pm$ 3 & 125$\pm$20  & V03 \\
                  &      &  \nodata   &  \nodata   &  \nodata  & (93$\pm$70) & E02 \\
                  &      &  \nodata   &  \nodata   &  \nodata  &(155$\pm$35) & S05 \\
                  &      &  \nodata   &  \nodata   &  \nodata  &(150$\pm$29) & C06 \\
Sk$-67\arcdeg$5   & 0.13 &  $<$4.6    &  14$\pm$ 5 & 10$\pm$ 3 &  50$\pm$20  & V03 \\
                  &      &  \nodata   &  \nodata   &  \nodata  & ($<$163)    & C06 \\
BI~237            & 0.12 & 7.1$\pm$3.8& 111$\pm$16 &  \nodata  & 390$\pm$35  & C04 \\
Sk$-68\arcdeg$52  & 0.09 &  $<$4.4    &  34$\pm$ 5 &  8$\pm$ 2 & 105$\pm$20  & V03 \\
Sk$-68\arcdeg$73  & 0.51 & 3.3$\pm$1.7&  39$\pm$ 4 & 14$\pm$ 2 & 180$\pm$20  & V03 \\
Sk$-68\arcdeg$135 & 0.21 & 2.8$\pm$1.5&  37$\pm$ 4 & 18$\pm$ 2 & 105$\pm$20  & V03 \\
                  &      &  \nodata   &  \nodata   &  \nodata  & ($<$65)     & E02 \\
                  &      &  \nodata   &  \nodata   &  \nodata  & (50$\pm$25) & S05 \\
                  &      &  \nodata   &  \nodata   &  \nodata  & (30$\pm$30) & C06 \\
Sk$-69\arcdeg$202 & 0.10 &  \nodata   & (33)       & (7)       & (45)        & V87 \\
Sk$-69\arcdeg$223 & 0.20 &  \nodata   &(106$\pm$30)&(37$\pm$14)&(420$\pm$125)& E02 \\
                  &      &  \nodata   &(145$\pm$21)&(28$\pm$ 6)&(225$\pm$21) & S05 \\
                  &      &  \nodata   &(145$\pm$21)&(28$\pm$ 6)&(240$\pm$21) & C06 \\
BI~253            & 0.13 &  $<$11     &  75$\pm$16 &  $<$18    & 130$\pm$35  & C04 \\
Melnick~42        & 0.36 &  $<$10     &  94$\pm$17 & 12$\pm$ 4 & 210$\pm$35  & C02 \\
Sk$-69\arcdeg$243 & 0.38 &  \nodata   &  \nodata   &  \nodata  &(437$\pm$72) & E02 \\
                  &      &  \nodata   &  \nodata   &  \nodata  &(340$\pm$50) & S05 \\
                  &      &  \nodata   &  \nodata   &  \nodata  &(335$\pm$47) & C06 \\
Sk$-69\arcdeg$246 & 0.08 & 3.4$\pm$1.4&  26$\pm$ 4 & 11$\pm$ 2 & 105$\pm$20  & V03 \\
Sk$-70\arcdeg$115 & 0.14 &  $<$4.3    &  36$\pm$ 5 & 10$\pm$ 2 & 105$\pm$20  & V03 \\
                  &      &  \nodata   &  35$\pm$ 6 &  7$\pm$ 2 & 110$\pm$35  & C02 \\
 & \\
$o$ Per           & 0.31 &  16$\pm$ 2 & 101$\pm$ 7 & 82$\pm$ 6 & 200$\pm$60  & T03 \\
$\zeta$ Per       & 0.31 &   9$\pm$ 1 & 114$\pm$ 7 & 77$\pm$ 5 & 185$\pm$50  & T03 \\
$\delta$ Sco      & 0.17 &   3$\pm$ 1 &  82$\pm$ 5 & 26$\pm$ 4 & 250$\pm$25  & T03 \\
$\beta^1$ Sco     & 0.19 &   3$\pm$ 1 & 171$\pm$ 5 & 34$\pm$ 4 & 397$\pm$45  & T03 \\
$\omega^1$ Sco    & 0.22 &   3$\pm$ 1 & 192$\pm$ 5 & 40$\pm$ 4 & 403$\pm$40  & T03 \\
\enddata
\tablecomments{Equivalent widths are in m\AA.  
Uncertainties are 1-$\sigma$ (photon noise + continuum uncertainty); limits are 3-$\sigma$.
Magellanic Clouds values from other references are given in parentheses.}
\tablenotetext{a}{C02, C04, V03, V04, E01 = UVES runs (this paper); V87 = Vladilo et al. 1987; 
E02 = Ehrenfreund et al. 2002; S05 = Sollerman et al. 2005; C06 = Cox et al. 2006; T03 = Thorburn et al. 2003}
\end{deluxetable}

\clearpage

\begin{deluxetable}{lccccccccc}
\tabletypesize{\scriptsize}
\tablecolumns{10}
\tablecaption{DIBs and Molecules toward Sk$-$67 2, Sk~143, and Six Galactic Stars \label{tab:c2dibs}}
\tablewidth{0pt}

\tablehead{
\multicolumn{1}{l}{$\lambda$}&
\multicolumn{1}{c}{FWHM}&
\multicolumn{1}{c}{Sk$-67\arcdeg$2\tablenotemark{a}}&
\multicolumn{1}{c}{Sk~143}&
\multicolumn{1}{c}{o~Per}&
\multicolumn{1}{c}{HD~37061}&
\multicolumn{1}{c}{$\rho$~Oph~A}&
\multicolumn{1}{c}{HD~183143}&
\multicolumn{1}{c}{HD~204827}&
\multicolumn{1}{c}{HD~210121}}
\startdata
4963.96   & 0.62 & 9.5$\pm$1.7 & 18.9$\pm$2.3&
                  15.7$\pm$1.5 &  $<$2.5     &   23$\pm$ 1 &  26$\pm$ 1  &  55$\pm$ 1 &  11$\pm$ 1  \\
4984.73   & 0.53 & 3.9$\pm$1.8 &  7.7$\pm$1.8&
                   6.4$\pm$1.0 &  $<$2.5     & 10.3$\pm$0.8&  14$\pm$ 1  &  30$\pm$ 1 & 7.4$\pm$1.0 \\
5175.99   & 0.71 & 7.9$\pm$1.8 & 17.2$\pm$2.3&
                   2.7$\pm$0.8 &  $<$2.0     &  7.8$\pm$1.0& 2.2$\pm$0.7 &  37$\pm$ 1 & 6.2$\pm$1.0 \\
5418.91   & 0.80 & 7.4$\pm$2.0 & 13.6$\pm$1.8&
                  13.5$\pm$1.5 &  $<$4.0     &   21$\pm$ 1 & 9.7$\pm$1.0 &  48$\pm$ 1 &12.6$\pm$2.0 \\
5512.62   & 0.48 & 2.6$\pm$1.3 &  3.3$\pm$1.1&
                   8.6$\pm$0.8 &  $<$4.0     &   15$\pm$ 1 &  11$\pm$ 1  &  23$\pm$ 1 & 3.7$\pm$1.0 \\
C$_2$D-Sum&      &   31.3      &    60.7     &
                     46.9      & $<$15.0     &   77.1      &   62.9      &   193      &   40.9      \\
 & \\
5780.59   & 2.07 &  33$\pm$ 5  &   77$\pm$10 &
                   101$\pm$ 7  &  169$\pm$ 7 &  222$\pm$10 & 758$\pm$ 8  & 257$\pm$ 4 &   70$\pm$ 7 \\
5797.11   & 0.97 &  23$\pm$ 3  &   37$\pm$ 4 &
                    82$\pm$ 6  &   35$\pm$ 5 &   71$\pm$ 6 & 295$\pm$10  & 199$\pm$ 3 &   46$\pm$ 9 \\
6196.00   & 0.65 & 5.2$\pm$1.8 & 11.0$\pm$2.2&
                  12.7$\pm$1.0 & 12.6$\pm$1.5&   17$\pm$ 1 &  92$\pm$ 2  &  42$\pm$ 1 &  9.4$\pm$0.7\\
6284.31   & 2.58 & 125$\pm$20  &($<$50)      &
                   200$\pm$60  &  675$\pm$55 &  426$\pm$80 &1930$\pm$150 & 518$\pm$60 &  146$\pm$50 \\
6613.72   & 1.14 &15.0$\pm$2.6 &   53$\pm$ 4 &
                    51$\pm$ 3  & 34.4$\pm$3.0&   68$\pm$ 5 & 337$\pm$ 4  & 171$\pm$ 3 &   25$\pm$ 2 \\
 & \\
log[N(H I)]    & &    21.00    &    21.18    &
                      20.90    &    21.78    &    21.55    &   (21.54)   &   (21.17)  &   (20.63)   \\
log[N(H$_2$)]  & &    20.95    &    20.93    &
                      20.60    &  \nodata    &    20.57    &   (21.05)   &   (21.23)  &    20.75    \\
log[N(H$_{\rm tot}$)]& & 21.44 &    21.51    &
                      21.20    &    21.78    &    21.63    &   (21.76)   &   (21.69)  &    21.19    \\
log[N(Na I)]   & &    13.65    &    13.72    &
                      14.04    &    12.64    &    13.71    &   \nodata   &  \nodata   &   \nodata   \\
log[N(K I)]    & &    12.30    &    12.65    &
                      11.97    &    10.88    &    12.01    &     12.21   & $>$12.47   &    12.06    \\
log[N(CH)]     & &    13.12    &    13.60    &
                      13.33    & $<$12.34    &    13.33    &     13.70   &    13.90   &    13.47    \\
log[N(CH$^+$)] & &    12.85    & $<$12.47    &
                      12.80    &    12.54    &    13.20    &  $>$13.76   & $>$13.55   &    12.98    \\
log[N(CN)]     & &    12.20    &    12.47    &
                      12.17    & $<$11.73    &    12.32    &     12.33   &    13.74   &    13.09    \\
 & \\
E(B-V)    &      &     0.22    &     0.34    &
                       0.31    &     0.52    &     0.48    &      1.27   &     1.11   &     0.40    \\
log[$f$(H$_2$)]& &   $-0.19$   &  $-$0.28    &
                     $-0.30$   &  \nodata    &   $-0.76$   &   ($-0.41$) &  ($-0.16$) &  $-$0.14    \\
log[CN/CH]&      &   $-0.92$   &  $-$1.13    &
                     $-1.16$   &  \nodata    &   $-1.01$   &    $-1.37$  &   $-0.16$  &  $-$0.38    \\
5780/5797 &      &     1.4     &     2.1     &
                       1.2     &     4.8     &     3.1     &      2.6    &     1.3    &     1.5     \\
5780/C$_2$D-Sum& &     1.1     &     1.3     &
                       2.2     & $>$11.3     &     2.9     &     12.1    &     1.3    &     1.7     \\
6284/5780 &      &     3.8     &  $<$0.7     &
                       2.0     &     4.0     &     1.9     &      2.6    &     2.0    &     2.1     \\
\enddata
\tablecomments{DIB equivalent widths are in m\AA; uncertainties are 1-$\sigma$.  
Column densities are cm$^{-2}$.
Top five rows are (selected) C$_2$ DIBs; sixth row gives the sum of their equivalent widths.
Next five rows are other relatively strong, commonly measured DIBs. 
H~I values in parentheses are estimated from W(5780); H$_2$ values in parentheses are estimated from $N$(CH).
Diffuse band wavelengths, FWHM, and equivalent widths for Galactic stars are from Thorburn et al. 2003.}
\tablenotetext{a}{Cox et al. 2006 report equivalent widths of 5$\pm$3 m\AA, 150$\pm$29 m\AA, and 13$\pm$8 m\AA\ for the DIBs at 6196, 6284, and 6613 \AA, respectively.}
\end{deluxetable}

\setlength{\tabcolsep}{0.03in}

\clearpage

\begin{deluxetable}{lcrccrccrc} 
\tabletypesize{\scriptsize}   
\tablecolumns{10}   
\tablecaption{Components for \ion{Na}{1}, CH, and CH$^+$ (E01, V03) \label{tab:comps}}  
\tablewidth{0pt}   
  
\tablehead{  
\multicolumn{1}{l}{Star}&  
\multicolumn{3}{c}{-~-~-~-~-~-Na I U-~-~-~-~-~-}&
\multicolumn{3}{c}{-~-~-~-~-~-~-CH-~-~-~-~-~-~-}&
\multicolumn{3}{c}{-~-~-~-~-~-~-CH$^+$-~-~-~-~-~-~-}\\
\multicolumn{1}{c}{ }&
\multicolumn{1}{c}{$v$}&  
\multicolumn{1}{c}{$N_{12}$}&  
\multicolumn{1}{c}{$b$}&
\multicolumn{1}{c}{$v$}&  
\multicolumn{1}{c}{$N_{12}$}&  
\multicolumn{1}{c}{$b$}&
\multicolumn{1}{c}{$v$}&  
\multicolumn{1}{c}{$N_{12}$}&  
\multicolumn{1}{c}{$b$}\\  
\multicolumn{1}{c}{ }&   
\multicolumn{1}{c}{(km s$^{-1}$)}&   
\multicolumn{1}{c}{(cm$^{-2}$)}&   
\multicolumn{1}{c}{(km s$^{-1}$)}&
\multicolumn{1}{c}{(km s$^{-1}$)}&   
\multicolumn{1}{c}{(cm$^{-2}$)}&   
\multicolumn{1}{c}{(km s$^{-1}$)}&
\multicolumn{1}{c}{(km s$^{-1}$)}&   
\multicolumn{1}{c}{(cm$^{-2}$)}&   
\multicolumn{1}{c}{(km s$^{-1}$)}}   
\startdata
Sk~13             &   V03&$<$3.00&     &   V03&$<$2.1&     &    V03&$<$1.4&     \\
                  & 120.3&   1.00&(1.0)&      &      &     &       &      &     \\
                  & 148.6&   5.50& 2.7 &      &      &     &       &      &     \\
 & \\
Sk~18             &   V03&$<$2.30&     &   V03&$<$1.9&     &    V03&$<$1.5&     \\
                  & 124.3&   9.60& 1.1 & 123.3&  1.2 & 2.7 &       &      &     \\
 & \\
Sk~40             &   V03&$<$4.05&     &   V03&$<$3.0&     &    V03&$<$2.3&     \\
 & \\
Sk~143\tablenotemark{a}            &
                      E01&$<$9.20&     &   E01&$<$2.7&     &    E01&$<$3.0&     \\
                  & 132.7&  56.00& 0.7 & 132.8& 40.0 &(0.7)&       &      &     \\
 & \\
AV~476            &   V03&$<$6.00&     &   V03&$<$3.4&     &    V03&$<$2.9&     \\
                  & 168.8&  24.10& 1.8 & 168.4&  3.9 &(1.0)& 168.9 &  1.1 & 3.0 \\
 & \\
Sk~155            &   V03&$<$2.30&     &   V03&$<$1.8&     &    V03&$<$1.4&     \\
 & \\
Sk$-67\arcdeg$2\tablenotemark{b}   &
                      V03&$<$2.65&     &   V03&$<$2.3&     &    V03&$<$1.6&     \\
                  & 268.0&   3.80&(1.5)& 266.9&  0.73&(1.5)& 269.2 &  6.8 & 2.0 \\
                  & 278.2&  60.60& 0.6 & 278.0& 13.9 & 0.6 & 281.7 &  0.37&(1.5)\\
 & \\
Sk$-67\arcdeg$5   &   V03&$<$2.30&     &   V03&$<$1.9&     &    V03&$<$1.5&     \\
                  & 287.5&   3.10&(1.0)&      &      &     &       &      &     \\
                  & 294.0&   1.20&(1.0)&      &      &     &       &      &     \\
 & \\ 
Sk$-68\arcdeg$52  &   V03&$<$2.30&     &   V03&$<$1.8&     &    V03&$<$1.3&     \\
                  & 243.3&   4.90& 2.7 & 242.5&  1.3 &(3.5)& 242.1 &  1.1 &(3.5)\\
 & \\
Sk$-68\arcdeg$73  &   V03&$<$2.20&     &   V03&$<$1.6&     &    V03&$<$1.6&     \\
                  & 288.8&   2.40&(1.5)& 286.2&  0.29&(1.5)& 288.6 &  0.23&(2.0)\\
                  & 295.9&  20.80& 1.1 & 295.8&  6.0 & 2.2 & 295.5 &  3.0 & 1.9 \\
 & \\
Sk$-68\arcdeg$135 &   V03&$<$2.20&     &   V03&$<$1.5&     &    V03&$<$1.5&     \\
                  & 269.9&   3.20&(1.5)& 270.2&  0.66&(1.5)& 270.8 &  1.2 &(2.0)\\
                  & 277.2&  11.50& 3.6 & 277.5&  3.1 & 3.9 & 277.6 &  2.1 & 3.6 \\
 & \\
Sk$-69\arcdeg$246 &   V03&$<$2.10&     &   V03&$<$1.6&     &    V03&$<$1.2&     \\
                  & 279.2&   8.80& 1.7 & 279.2&  2.1 & 1.9 & 278.5 &  2.6 & 2.0 \\
                  & 286.0&   1.30&(1.5)&      &      &     &       &      &     \\
 & \\
Sk$-70\arcdeg$115 &   V03&$<$2.20&     &   V03&$<$1.7&     &    V03&$<$1.3&     \\
                  & 220.0&   4.80& 4.0 & 220.7&  0.91&(2.0)& 222.1 &  0.62&(2.0)\\
\enddata
\tablecomments{Column densities are in 10$^{12}$ cm$^{-2}$ for Na~I, CH and CH$^+$.
First line for each line of sight gives observing run and 3-$\sigma$ column density limit for a weak, unresolved component.
Velocities are heliocentric.
Line widths or velocities in parentheses were fixed in the fits to the line profiles.}
\tablenotetext{a}{ Fits to the CN B-X (0-0) $J$ = 0,1 lines toward Sk~143 yield $N$(0) = 1.92$\pm$0.21 $\times$ 10$^{12}$ cm$^{-2}$, $N$(1) = 1.06$\pm$0.20 $\times$ 10$^{12}$ cm$^{-1}$, and $v$ = 132.7 km~s$^{-1}$, for an assumed $b$ = 0.5 km~s$^{-1}$.}
\tablenotetext{b}{Fits to the CN B-X (0-0) $J$ = 0,1 lines toward Sk$-67\arcdeg$2 yield $N$(0) = 1.31$\pm$0.06 $\times$ 10$^{12}$ cm$^{-2}$, $N$(1) = 0.54$\pm$0.05 $\times$ 10$^{12}$ cm$^{-1}$, and $v$ = 277.6 km~s$^{-1}$, for an assumed $b$ = 0.6 km~s$^{-1}$.}
\end{deluxetable}

\clearpage

\begin{deluxetable}{lcrccrccrc}
\tabletypesize{\scriptsize}
\tablecolumns{10}
\tablecaption{Components for \ion{K}{1}, CH, and CH$^+$ (V04, C04) \label{tab:compf}}
\tablewidth{0pt}
 
\tablehead{
\multicolumn{1}{l}{Star}&
\multicolumn{3}{c}{-~-~-~-~-~-K I-~-~-~-~-~-}&
\multicolumn{3}{c}{-~-~-~-~-~-~-CH-~-~-~-~-~-~-}&
\multicolumn{3}{c}{-~-~-~-~-~-~-CH$^+$-~-~-~-~-~-~-}\\
\multicolumn{1}{c}{ }&
\multicolumn{1}{c}{$v$}&
\multicolumn{1}{c}{$N_{10}$}&
\multicolumn{1}{c}{$b$}&
\multicolumn{1}{c}{$v$}&
\multicolumn{1}{c}{$N_{12}$}&
\multicolumn{1}{c}{$b$}&
\multicolumn{1}{c}{$v$}&
\multicolumn{1}{c}{$N_{12}$}&
\multicolumn{1}{c}{$b$}\\
\multicolumn{1}{c}{ }&
\multicolumn{1}{c}{(km s$^{-1}$)}&
\multicolumn{1}{c}{(cm$^{-2}$)}&
\multicolumn{1}{c}{(km s$^{-1}$)}&
\multicolumn{1}{c}{(km s$^{-1}$)}&
\multicolumn{1}{c}{(cm$^{-2}$)}&
\multicolumn{1}{c}{(km s$^{-1}$)}&
\multicolumn{1}{c}{(km s$^{-1}$)}&
\multicolumn{1}{c}{(cm$^{-2}$)}&
\multicolumn{1}{c}{(km s$^{-1}$)}}
\startdata
LH~10-3061        &   C04 &  $<$2.9&     &   C04&$<$3.4&     &    C04&$<$3.5&     \\
                  & 262.1 &     4.3&(1.0)& \\
                  & 276.6 &     2.1&(1.0)& \\
                  & 286.5 &    19.5& 2.6 &      &      &     & 289.1 &  16.1& 2.8 \\
 & \\
Sk$-67\arcdeg$5   &   V04 &  $<$2.8&     &   V04&$<$1.8&     &    V04&$<$1.7&     \\  
                  & 287.8 &     8.8&(1.0)& 288.3&   1.0&(1.0)&       &      &     \\
                  & 294.3 &     2.0&(1.0)&      &      &     &       &      &     \\
 & \\
BI~237            &   C04 &  $<$3.5&     &   C04&$<$3.2&     &    C04&$<$3.7&     \\
                  & 277.2 &     3.0&(1.0)& \\
                  & 291.2 &    12.6&(1.0)& \\
                  & 296.0 &     5.0&(1.0)& \\
 & \\
Sk$-68\arcdeg$52  &   V04 &  $<$1.9&     &   V04&$<$2.3&     &    V04&$<$2.0&     \\  
                  & 243.0 &    12.2&(0.9)& 242.5&   1.2&(2.0)& 243.7 &  1.0 &(2.0)\\
 & \\
Sk$-68\arcdeg$135 &   V04 &  $<$1.6&     &   V04&$<$1.8&     &    V04&$<$1.6&     \\  
                  & 268.8 &     5.8&(1.0)& 270.5&   0.5&(1.5)&(271.0)&  0.2 &(2.0)\\
                  & 275.9 &    24.6& 3.7 & 278.0&   2.5&(4.5)& 278.4 &  2.7 & 2.5 \\
 & \\
Sk$-69\arcdeg$191 &   V04 &  $<$3.5&     &   V04&$<$5.4&     &    V04&$<$4.8&     \\  
                  & 237.6 &     3.4& 2.6 &      &      &     & 239.3 &  3.3 & 2.1 \\
 & \\
BI~253            &   C04 &  $<$6.1&     &   C04&$<$3.2&     &    C04&$<$3.3&     \\
                  &(265.5)&     3.6&(1.0)& \\
                  & 270.5 &    11.2&(1.0)& \\
                  & 276.9 &     3.5&(1.0)& \\
\enddata
\tablecomments{Column densities are in 10$^{10}$ cm$^{-2}$ for K~I and in 10$^{12}$ cm$^{-2}$ for CH and CH$^+$.
First line for each line of sight gives observing run and 3-$\sigma$ column density limit for a weak, unresolved component.
Velocities are heliocentric.
Line widths or velocities in parentheses were fixed in the fits to the line profiles.}
\end{deluxetable}

\clearpage

\setlength{\tabcolsep}{0.04in}

\begin{deluxetable}{lccccrrrrrr}
\tabletypesize{\scriptsize}
\tablecolumns{11}
\tablecaption{Magellanic Clouds Column Densities \label{tab:coldens}}
\tablewidth{0pt}

\tablehead{
\multicolumn{1}{l}{Star}&
\multicolumn{1}{c}{$E(B-V)_{\rm MC}$}&
\multicolumn{1}{c}{H~I}&
\multicolumn{1}{c}{H$_2$}&
\multicolumn{1}{c}{HD}&
\multicolumn{1}{c}{CO}&
\multicolumn{1}{c}{Na~I}&
\multicolumn{1}{c}{K~I}&
\multicolumn{1}{c}{CN}&
\multicolumn{1}{c}{CH}&
\multicolumn{1}{c}{CH$^+$}}
\startdata
Sk~13            & 0.14 &  22.04  &  20.36 & \nodata& \nodata&  12.81  &  11.20  &$<11.64$ &$<12.33$ &$<12.13$\\
Sk~18            & 0.08 &  21.82  &  20.63 & \nodata& \nodata&  12.98  &  11.20  &$<11.64$ &  12.08  &$<12.17$\\
Sk~40            & 0.07 &  21.54  & \nodata& \nodata& \nodata&  11.96  &$<10.75$ &$<11.85$ &$<12.48$ &$<12.35$\\
AV~80            & 0.12 &  21.81  & \nodata& \nodata& \nodata&  12.15  & \nodata &$<11.81$ &$<12.48$ &$<12.49$\\
Sk~143           & 0.34 &  21.18  &  20.93 & \nodata& \nodata&  13.72  &  12.65  &  12.47  &  13.60  &$<12.47$\\
AV~476           & 0.18 & \nodata &  20.95 & \nodata& \nodata&  13.38  &$<12.46$ &$<11.96$ &  12.59  &  12.04 \\
Sk~155           & 0.11 &  21.40  &  19.14 &  13.82 & \nodata&  11.92  & \nodata &$<11.62$ &$<12.26$ &$<12.14$\\
 & \\
LH~10-3061       & 0.26 & \nodata & \nodata& \nodata& \nodata&  12.86  &  11.26  &$<12.02$ &$<12.53$ &  13.20 \\
Sk$-67\arcdeg$2  & 0.22 &  21.00  &  20.95 & \nodata& \nodata&  13.79  &  12.30  &  12.27  &  13.17  &  12.86 \\
Sk$-67\arcdeg$5  & 0.13 &  21.00  &  19.46 &  13.62 &  13.88 &  12.50  &  11.03  &$<11.64$ &$<12.28$ &$<12.16$\\
BI~237           & 0.12 & \nodata & \nodata& \nodata& \nodata&  13.0:  &  11.3:  &$<12.02$ &$<12.55$ &$<12.57$\\
Sk$-68\arcdeg$52 & 0.09 & (21.23) &  19.47 & \nodata& \nodata&  12.69  &  11.09  &$<11.61$ &  12.11  &  12.04 \\
Sk$-68\arcdeg$73 & 0.51 & (21.70) &  20.09 & \nodata& \nodata&  13.37  &  11.76  &$<11.71$ &  12.80  &  12.51 \\
Sk$-68\arcdeg$135& 0.21 & (21.57) &  19.87 &  14.15 &  13.77 &  13.17  &  11.48  &$<11.57$ &  12.58  &  12.52 \\
Sk$-69\arcdeg$191& 0.09 &  21.02  &  18.94 & \nodata& \nodata& \nodata &  10.53  &$<11.81$ &$<12.76$ &  12.52 \\
Sk$-69\arcdeg$202& 0.10 & (21.40) & \nodata& \nodata& \nodata&  12.42  &  11.05  & \nodata &  12.04  &  11.66 \\
Sk$-69\arcdeg$223& 0.20 & \nodata & \nodata& \nodata& \nodata&  12.59  & \nodata & \nodata & \nodata & \nodata\\
BI~253           & 0.13 & \nodata & \nodata& \nodata& \nodata&  13.0:  &  11.2:  &$<11.97$ &$<12.50$ &$<12.52$\\
Melnick 42       & 0.36 & \nodata & \nodata& \nodata& \nodata&  12.55  & \nodata &$<12.18$ &$<12.60$ &$<12.62$\\
Sk$-69\arcdeg$243& 0.38 &  21.80  & \nodata& \nodata& \nodata&  12.84  &  11.24  & \nodata & \nodata & \nodata\\
Sk$-69\arcdeg$246& 0.08 &  21.30  &  19.71 &  13.62 &  13.57 &  13.01  &$<12.09$ &$<11.58$ &  12.32  &  12.41 \\
Sk$-70\arcdeg$115& 0.14 &  21.30  &  19.94 & \nodata& \nodata&  12.68  & \nodata &$<11.58$ &  11.96  &  11.79 \\
 & \\
$o$ Per          & 0.31 &  20.90  &  20.60 &  15.50 &  14.93 &  14.04  &  11.97  &  12.17  &  13.33  &  12.80 \\
$\zeta$ Per      & 0.31 &  20.81  &  20.67 &  15.60 &  14.86 &  13.89  &  11.85  &  12.51  &  13.34  &  12.50 \\
$\delta$ Sco     & 0.17 &  21.15  &  19.41 &  13.06 &  12.49 &  12.94  &  11.26  &$<11.12$ &  12.34  &  12.47 \\
$\beta^1$ Sco    & 0.19 &  21.10  &  19.83 & \nodata&  13.63 &  12.94  &  11.13  &$<11.12$ &  12.41  &  12.70 \\
$\omega^1$ Sco   & 0.22 &  21.18  &  20.05 & \nodata&  12.95 &  13.11  &  11.17  & \nodata &  12.51  &  12.79 \\
\enddata
\tablecomments{Entries for each species are log($N$) (cm$^{-2}$); limits are 3-$\sigma$, based on corresponding equivalent width limits in Table 2.  
For CN, limits assume $N$($J$=0)/$N$($J$=1) = 2.
Values for HD and CO are from Spitzer et al. 1974, Andr\`{e} et al. 2004, Lacour et al. 2005a, Sonnentrucker et al. 2006, or Welty et al. (in prep.).  
References for H~I and H$_2$ are given in Welty (in prep.); H~I values in parentheses are estimated from 21 cm emission or Zn~II absorption.
Values for Na~I and K~I are from Welty (in prep.), except those for Sk$-69\arcdeg$223 and Sk$-69\arcdeg$243, which are from Cox et al. 2006; values with a colon are less well determined.
Values for CH and CH$^+$ toward Sk$-69\arcdeg$202 (SN 1987A) are from Magain \& Gillet 1987.}
\end{deluxetable}

\setlength{\tabcolsep}{0.06in}

\begin{deluxetable}{llrrrrrr}
\tabletypesize{\scriptsize}
\tablecolumns{8}
\tablecaption{Correlations (Galactic Sight lines) \label{tab:corr}}
\tablewidth{0pt}

\tablehead{
\multicolumn{4}{c}{ }&
\multicolumn{2}{c}{Weighted}&
\multicolumn{2}{c}{Unweighted}\\
\multicolumn{1}{l}{Y}&
\multicolumn{1}{l}{X}&
\multicolumn{1}{c}{N}&
\multicolumn{1}{c}{r}&
\multicolumn{1}{c}{Slope}&
\multicolumn{1}{c}{rms}&
\multicolumn{1}{c}{Slope}&
\multicolumn{1}{c}{rms}}
\startdata
CH    & H$_2$   & 52 & 0.819 & 1.13$\pm$0.08 & 0.14 & 1.13$\pm$0.09 & 0.14  \\
      & Na~I    & 24 & 0.915 & 1.04$\pm$0.08 & 0.14 & 0.99$\pm$0.07 & 0.14  \\
      & K~I     & 74 & 0.855 & 1.06$\pm$0.07 & 0.14 & 1.02$\pm$0.06 & 0.15  \\
 & \\
CH$^+$& CH      & 92 & 0.437 & \nodata      &\nodata& 0.84$\pm$0.09 & 0.23  \\
 & \\
4963  & H~I     & 32 & 0.711 & \nodata      &\nodata& 1.04$\pm$0.14 & 0.15  \\
      & E(B-V)  & 54 & 0.805 & 1.04$\pm$0.11 & 0.12 & 1.11$\pm$0.09 & 0.12  \\
      & Na~I    & 14 & 0.629 & 0.50$\pm$0.16 & 0.15 & 0.73$\pm$0.19 & 0.18  \\
      & K~I     & 35 & 0.792 & 0.70$\pm$0.08 & 0.10 & 0.69$\pm$0.08 & 0.08  \\
      & H$_2$   & 26 & 0.514 & \nodata      &\nodata& 1.19$\pm$0.18 & 0.17  \\
      & CH      & 51 & 0.748 & 0.75$\pm$0.10 & 0.16 & 0.81$\pm$0.09 & 0.16  \\
 & \\
5780  & H~I     & 71 & 0.894 & 1.21$\pm$0.07 & 0.12 & 1.17$\pm$0.06 & 0.12  \\
      & E(B-V)  &101 & 0.889 & 0.99$\pm$0.06 & 0.14 & 1.10$\pm$0.05 & 0.14  \\
      & Na~I    & 39 & 0.769 & 0.49$\pm$0.04 & 0.24 & 0.48$\pm$0.05 & 0.23  \\
      & K~I     & 57 & 0.705 & 0.58$\pm$0.06 & 0.23 & 0.49$\pm$0.05 & 0.22  \\
      & C~I     & 24 & 0.725 & \nodata      &\nodata& 0.30$\pm$0.06 & 0.22  \\
      & H$_2$   & 50 & 0.515 & \nodata      &\nodata& 0.36$\pm$0.06 & 0.20  \\
      & CH      & 67 & 0.425 & \nodata      &\nodata& 0.53$\pm$0.08 & 0.23  \\
 & \\
5797  & H~I     & 67 & 0.875 & 1.05$\pm$0.08 & 0.13 & 1.29$\pm$0.07 & 0.12  \\
      & E(B-V)  & 93 & 0.883 & 0.99$\pm$0.06 & 0.13 & 1.12$\pm$0.05 & 0.12  \\
      & Na~I    & 34 & 0.828 & 0.52$\pm$0.04 & 0.18 & 0.55$\pm$0.05 & 0.20  \\
      & K~I     & 54 & 0.801 & 0.74$\pm$0.05 & 0.17 & 0.55$\pm$0.04 & 0.17  \\
      & C~I     & 21 & 0.814 & 0.42$\pm$0.06 & 0.19 & 0.35$\pm$0.05 & 0.18  \\
      & H$_2$   & 50 & 0.617 & 0.48$\pm$0.07 & 0.20 & 0.45$\pm$0.06 & 0.20  \\
      & CH      & 67 & 0.598 & 0.59$\pm$0.08 & 0.19 & 0.66$\pm$0.07 & 0.19  \\
 & \\
6284  & H~I     & 43 & 0.899 & 0.90$\pm$0.06 & 0.08 & 0.89$\pm$0.05 & 0.08  \\
      & E(B-V)  & 68 & 0.831 & 0.80$\pm$0.07 & 0.15 & 0.90$\pm$0.06 & 0.14  \\
      & Na~I    & 24 & 0.691 & 0.25$\pm$0.07 & 0.26 & 0.34$\pm$0.07 & 0.23  \\
      & K~I     & 45 & 0.594 & 0.36$\pm$0.07 & 0.25 & 0.37$\pm$0.06 & 0.24  \\
      & C~I     & 18 & 0.552 & 0.23$\pm$0.08 & 0.26 & 0.23$\pm$0.08 & 0.25  \\
      & H$_2$   & 34 & 0.218 & 0.30$\pm$0.08 & 0.24 & 0.13$\pm$0.08 & 0.22  \\
      & CH      & 58 & 0.365 & 0.49$\pm$0.08 & 0.25 & 0.45$\pm$0.08 & 0.25  \\
\enddata
\tablecomments{Correlations are for log[Y] vs. log[X]; N is the number of points for each included in the regression fits; r is the linear correlation coefficient.
The slopes and rms deviations are for weighted (w) and unweighted (n) fits, allowing for errors in both X and Y.
Correlations with H$_2$ are limited to N(H$_2$) $\ge$ 10$^{18.5}$ cm$^{-2}$ (where the H$_2$ is self-shielded).}
\end{deluxetable}

\setlength{\tabcolsep}{0.03in}

\begin{deluxetable}{lrrrrrrrrrr}
\tabletypesize{\scriptsize}
\tablecolumns{11}
\tablecaption{Ratios (Galactic, LMC, SMC) \label{tab:ratgal}}
\tablewidth{0pt}

\tablehead{
\multicolumn{1}{l}{Ratio}&
\multicolumn{2}{c}{Galactic}&
\multicolumn{2}{c}{Sco-Oph}&
\multicolumn{2}{c}{Trapezium}&
\multicolumn{2}{c}{LMC}&
\multicolumn{2}{c}{SMC}\\
\multicolumn{1}{c}{ }&
\multicolumn{1}{c}{Ratio}&
\multicolumn{1}{c}{N}&
\multicolumn{1}{c}{Ratio}&
\multicolumn{1}{c}{N}&
\multicolumn{1}{c}{Ratio}&
\multicolumn{1}{c}{N}&
\multicolumn{1}{c}{Ratio}&
\multicolumn{1}{c}{N}&
\multicolumn{1}{c}{Ratio}&
\multicolumn{1}{c}{N}}
\startdata
H$_{\rm tot}$/$E(B-V)$ & 
                      $21.71\pm0.23$ &166 & $21.89\pm0.09$ & 10 & $22.05\pm0.03$ &  3 & $22.16\pm0.22$ & 49 & $22.65\pm0.36$ & 42 \\
                    & \nodata        &    & $+0.18$         &  & $+0.34$           &  & $+0.44$        &    & $+0.91$        &    \\
 & \\
CH/H$_2$            & $-7.38\pm0.19$ & 52 & $-7.26\pm0.15$ &  7 & \nodata        &    & $-7.51\pm0.27$ &  6 & $-8.03\pm0.51$ &  3 \\
                    & \nodata        &    & $+0.12$        &    & \nodata        &    & $-0.13$        &    & $-0.65$        &    \\
CH/Na~I             & $-0.65\pm0.19$ & 24 & $-0.43\pm0.16$ &  6 & \nodata        &    & $-0.60\pm0.11$ &  7 & $-0.60\pm0.34$ &  3 \\
                    & \nodata        &    & $+0.22$        &    & \nodata        &    & $+0.05$        &    & $+0.05$        &    \\
CH/K~I              & $ 1.30\pm0.20$ & 74 & $ 1.38\pm0.15$ & 10 & \nodata        &    & $ 1.00\pm0.08$ &  5 & $ 0.92\pm0.03$ &  2 \\
                    & \nodata        &    & $+0.08$        &    & \nodata        &    & $-0.30$        &    & $-0.38$        &    \\
 & \\
CH/CH$^+$           & $ 0.06\pm0.33$ & 90 & $ 0.08\pm0.30$ & 10 & \nodata        &    & $ 0.17\pm0.15$ &  7 & $ 0.55$        &  1 \\
                    & \nodata        &    & $+0.02$        &    & \nodata        &    & $+0.11$        &    & $+0.49$        &    \\
 & \\
4963/H~I            &$-20.05\pm0.21$ & 32 &$-20.58\pm0.22$ &  6 & \nodata        &    &$-20.78\pm0.47$ &  4 & $-19.90$       &  1 \\
                    & \nodata        &    & $-0.53$        &    & \nodata        &    & $-0.73$        &    & $+0.15$        &    \\
4963/E(B-V)         & $ 1.48\pm0.17$ & 54 & $ 1.34\pm0.28$ &  8 & \nodata        &    & $ 1.39\pm0.37$ &  5 & $ 1.67\pm0.08$ &  2 \\
                    & \nodata        &    & $-0.14$        &    & \nodata        &    & $+0.09$        &    & $+0.19$        &    \\
4963/(Na~I)$^{1/2}$ & $-5.89\pm0.21$ & 14 & $-5.88\pm0.19$ &  6 & \nodata        &    & $-5.97\pm0.19$ &  5 & $-5.71\pm0.13$ &  2 \\
                    & \nodata        &    & $+0.01$        &    & \nodata        &    & $-0.08$        &    & $+0.16$        &    \\
4963/(K~I)$^{1/2}$  & $-4.87\pm0.12$ & 35 & $-4.95\pm0.23$ &  7 & \nodata        &    & $-5.16\pm0.22$ &  4 & $-5.05$        &  1 \\
                    & \nodata        &    & $-0.08$        &    & \nodata        &    & $-0.29$        &    & $-0.18$        &    \\
4963/H$_2$          &$-19.59\pm0.18$ & 26 &$-19.26\pm0.18$ &  7 & \nodata        &    &$-19.54\pm0.29$ &  4 &$-19.88\pm0.23$ &  2 \\
                    & \nodata        &    & $+0.33$        &    & \nodata        &    & $+0.05$        &    & $-0.29$        &    \\
4963/CH             &$-12.19\pm0.23$ & 51 &$-12.03\pm0.11$ &  8 & \nodata        &    &$-12.10\pm0.19$ &  4 &$-12.03\pm0.29$ &  2 \\
                    & \nodata        &    & $+0.16$        &    & \nodata        &    & $+0.09$        &    & $+0.16$        &    \\
 & \\
5780/H~I            &$-18.86\pm0.15$ & 71 &$-19.08\pm0.17$ &  9 &$-19.64\pm0.08$ &  3 &$-19.83\pm0.19$ &  8 & $-20.23\pm0.46$ &  6 \\
                    & \nodata        &    & $-0.22$        &    & $-0.78$        &    & $-0.97$        &    & $-1.37$        &    \\
5780/E(B-V)         & $ 2.68\pm0.20$ &101 & $ 2.76\pm0.13$ & 11 & $ 2.41\pm0.08$ &  3 & $ 2.45\pm0.30$ & 13 & $ 2.33\pm0.09$ &  7 \\
                    & \nodata        &    & $+0.08$        &    & $-0.27$        &    & $-0.23$        &    & $-0.35$        &    \\
5780/(Na~I)$^{1/2}$ & $-4.62\pm0.26$ & 39 & $-4.31\pm0.17$ &  8 & $-4.31\pm0.19$ &  3 & $-4.77\pm0.35$ & 13 & $-4.91\pm0.21$ &  7 \\
                    & \nodata        &    & $+0.31$        &    & $+0.31$        &    & $-0.15$        &    & $-0.29$        &    \\
5780/(K~I)$^{1/2}$  & $-3.62\pm0.24$ & 57 & $-3.49\pm0.18$ &  9 & $-3.27\pm0.05$ &  2 & $-4.06\pm0.32$ &  9 & $-4.35\pm0.08$ &  3 \\
                    & \nodata        &    & $+0.13$        &    & $+0.35$        &    & $-0.44$        &    & $-0.73$        &    \\
 & \\
5797/H~I            &$-19.33\pm0.19$ & 67 &$-19.71\pm0.17$ &  9 &$-20.37\pm0.17$ &  3 &$-20.25\pm0.28$ &  8 & $-20.66\pm0.76$ &  3 \\
                    & \nodata        &    & $-0.38$        &    & $-1.04$        &    & $-0.92$        &    & $-1.33$        &    \\
5797/E(B-V)         & $ 2.26\pm0.18$ & 93 & $ 2.16\pm0.10$ & 11 & $ 1.68\pm0.15$ &  3 & $ 1.88\pm0.21$ & 11 & $ 1.89\pm0.16$ &  4 \\
                    & \nodata        &    & $-0.10$        &    & $-0.58$        &    & $-0.38$        &    & $-0.37$        &    \\
5797/(Na~I)$^{1/2}$ & $-5.09\pm0.20$ & 34 & $-4.94\pm0.14$ &  8 & $-5.03\pm0.27$ &  3 & $-5.31\pm0.20$ & 11 & $-5.51\pm0.23$ &  4 \\
                    & \nodata        &    & $+0.15$        &    & $+0.06$        &    & $-0.22$        &    & $-0.42$        &    \\
5797/(K~I)$^{1/2}$  & $-4.04\pm0.18$ & 53 & $-4.08\pm0.13$ &  9 & $-3.93\pm0.03$ &  2 & $-4.59\pm0.17$ &  7 & $-4.82\pm0.13$ &  3 \\
                    & \nodata        &    & $-0.04$        &    & $+0.11$        &    & $-0.55$        &    & $-0.78$        &    \\
 & \\
6284/H~I            &$-18.46\pm0.13$ & 44 &$-18.65\pm0.13$ &  8 &$-18.91\pm0.04$ &  2 &$-19.33\pm0.23$ &  9 & $-19.69$        &  1 \\
                    & \nodata        &    & $-0.19$        &    & $-0.45$        &    & $-0.87$        &    & $-1.23$        &    \\
6284/E(B-V)         & $ 3.11\pm0.21$ & 68 & $ 3.12\pm0.21$ & 10 & $ 3.13\pm0.01$ &  2 & $ 2.84\pm0.18$ & 13 & $ 2.86\pm0.14$ &  2 \\
                    & \nodata        &    & $+0.01$        &    & $+0.02$        &    & $-0.27$        &    & $-0.25$        &    \\
6284/(Na~I)$^{1/2}$ & $-4.18\pm0.28$ & 24 & $-3.91\pm0.19$ &  7 & $-3.51\pm0.02$ &  2 & $-4.30\pm0.29$ & 14 & $-4.42\pm0.29$ &  2 \\
                    & \nodata        &    & $+0.27$        &    & $+0.67$        &    & $-0.12$        &    & $-0.24$        &    \\
6284/(K~I)$^{1/2}$  & $-3.20\pm0.28$ & 46 & $-3.13\pm0.22$ &  8 & $-2.59\pm0.02$ &  2 & $-3.55\pm0.31$ & 10 & \nodata        &    \\
                    & \nodata        &    & $+0.07$        &    & $+0.61$        &    & $-0.35$        &    & \nodata        &    \\
& \\
$\Delta$[DIBs/H~I]  & \nodata        &    & $-$0.4 to $-$0.2 &  & $-$1.0 to $-$0.5 &  & $-$1.0 to $-$0.9 &  & $-$1.4 to $-$1.2 &  \\
$\Delta$[DIBs/E(B-V)]&               &    & $-$0.1 to $+$0.1 &  & $-$0.6 to    0.0 &  & $-$0.4 to $-$0.3 &  & $-$0.4 to $-$0.2 &  \\
$\Delta$[DIBs/(Na~I)$^{1/2}$]&       &    & $+$0.2 to $+$0.3 &  & $+$0.1 to $+$0.7 &  & $-$0.2 to $-$0.1 &  & $-$0.4 to $-$0.2 &  \\
$\Delta$[DIBs/(K~I)$^{1/2}$]&        &    & $-$0.1 to $+$0.1 &  & $+$0.1 to $+$0.6 &  & $-$0.5 to $-$0.4 &  & $-$0.8 to $-$0.7 &  \\
\enddata

\tablecomments{Values are mean[log(ratio)], with standard deviations; N is the number of sight lines with values for both quantities.
Sco-Oph sight lines are 1~Sco, $\pi$~Sco, $\delta$~Sco, $\beta^1$~Sco, $\omega^1$~Sco, $\nu$~Sco, $\sigma$~Sco, $\rho$~Oph~D, HD147889, $\rho$~Oph~A, and 22~Sco.
Trapezium sight lines are $\theta^1$~Ori~C, $\theta^2$~Ori~A, and HD~37061.}
\end{deluxetable}

\setlength{\tabcolsep}{0.04in}

\begin{deluxetable}{lcccrcccc}
\tabletypesize{\scriptsize}
\tablecolumns{9}
\tablecaption{Physical Conditions \label{tab:phys}}
\tablewidth{0pt}

\tablehead{
\multicolumn{1}{l}{Cloud\tablenotemark{a}}&
\multicolumn{1}{c}{$T_{01}$(H$_2$)}& 
\multicolumn{1}{c}{log[$f$(H$_2$)]\tablenotemark{b}}&
\multicolumn{1}{c}{CH/CH$^+$}&
\multicolumn{1}{c}{$\tau_{uv}$}&
\multicolumn{1}{c}{$I_{uv}$\tablenotemark{c}}& 
\multicolumn{1}{c}{$n_{\rm H}$(H$_2$)\tablenotemark{d}}&
\multicolumn{1}{c}{$n_{\rm H}$(CH)\tablenotemark{e}}& 
\multicolumn{1}{c}{$n_{\rm H}$(CH$^+$)\tablenotemark{f}}\\
\multicolumn{1}{c}{ }&
\multicolumn{1}{c}{(K)}&
\multicolumn{4}{c}{ }& 
\multicolumn{1}{c}{(cm$^{-3}$)}&
\multicolumn{1}{c}{(cm$^{-3}$)}& 
\multicolumn{1}{c}{(cm$^{-3}$)}}
\startdata
Sk~13                    &  67 & $-$1.40 & \nodata & 1.30 &  28   &   580 &$<$60000 &   (250)\\
Sk~18                    &  54 & $-$0.93 &  $>$0.8 & 0.74 &  83   &  3500 &   99000 &$>$1000 \\
Sk~143                   & 45 & $-$0.28 & $>$14   & 3.16 &  (0.6)&    85 &    2000 &   $>$12 \\
AV~476                   & 54 & \nodata &     3.5 & 1.67 &   2.1 &\nodata&    1500 &      45 \\
Sk~155                   & 82 & $-$1.96 & \nodata & 1.02 &   5.5 &   125 &$<$2.2e5 &     (65)\\
 & \\
Sk$-67\arcdeg$2($+$268)  & 46 & $-$0.19 &     0.1 & 0.37 &   2.4 &   390 &    8500 &       6 \\
Sk$-67\arcdeg$2($+$278)  &    & \nodata &    37.6 & 1.67 &\nodata&   390 &    2200 &     550 \\
Sk$-67\arcdeg$5          & 57 & $-$1.26 & \nodata & 1.21 &   2.5 &   200 &$<$12000 &     (25)\\
Sk$-68\arcdeg$52         & 61 &($-$1.48)&     1.2 & 0.84 &   3.9 &  (190)&   33000 &      65 \\
Sk$-68\arcdeg$73(+288)   & 57 &($-$1.33)&     1.3 & 1.02 &   9.3 &  (300)&   29000 &     140 \\
Sk$-68\arcdeg$73(+296)   &    & \nodata &     2.0 & 3.72 &\nodata&  (300)&    6500 &      15 \\
Sk$-68\arcdeg$135($+$270)& 91 &($-$1.42)&     0.6 & 0.74 & 860   &(30000)&   5.8e6 &    7900 \\
Sk$-68\arcdeg$135($+$277)&    & \nodata &     1.5 & 1.21 &\nodata&(30000)&   8.3e6 &   13000 \\
Sk$-69\arcdeg$191        & 65 & $-$1.79 &  $<$1.7 & 0.84 &   9.4 &   400 &$<$6.5e5 &  $<$225 \\
Sk$-69\arcdeg$246        & 75 & $-$1.31 &     0.8 & 0.74 &  44   &  2300 &   3.6e5 &     550 \\
Sk$-70\arcdeg$115        & 54 & $-$1.10 &     1.5 & 1.30 &   8.4 &   560 &    6700 &     110 \\
 & \\
$o$ Per                  & 48 & $-$0.29 &     3.4 & 1.86 &  39   &   760 &   47000 &     630 \\
$\zeta$ Per              & 57 & $-$0.22 &     6.9 & 1.86 &   1.5 &    33 &    1500 &      50 \\
$\pi$ Sco                & 81 & $-$1.13 &  $<$0.3 & 0.30 &   1.9 &    23 & $<$1400 &   $<$12 \\
$\delta$ Sco             & 57 & $-$1.45 &     0.7 & 1.02 &   3.6 &    20 &   39000 &      30 \\
$\sigma$ Sco             & 65 & $-$1.28 &     0.6 & 2.44 &   4.9 &    26 &    3500 &       7 \\
\enddata
\tablenotetext{a}{If more than one cloud appears along a line of sight, the velocity is given in parentheses.}
\tablenotetext{b}{Values in parentheses are based on $N$(H~I) estimated from 21 cm emission or Zn~II absorption.}
\tablenotetext{c}{Value for Sk~143 is based on an estimated $N$($J$=4) = 10$^{14.8}$ cm$^{-2}$.}
\tablenotetext{d}{Density inferred from H$_2$ rotational populations (see text for references).  Values in parentheses are based on estimated $N$(H~I).}
\tablenotetext{e}{Density estimated assuming CH is produced via steady-state gas-phase chemistry.}
\tablenotetext{f}{Density estimated assuming CH is produced with CH$^+$, with $f$(H$_2$) = 0.05.  Values in parentheses further assume $N$(CH)/$N$(CH$^+$) = 1.0.}
\end{deluxetable}

\clearpage

\clearpage

\clearpage

\clearpage

\clearpage

\clearpage

\clearpage

\clearpage
 
\clearpage 
  
\clearpage

\clearpage
 
\clearpage
 
\end{document}